\documentclass[reqno,letterpaper,12pt]{amsart}
\usepackage{amsmath,amsthm,amscd,color}
\usepackage{hyperref}
\usepackage[dvips]{graphicx}

\theoremstyle{plain}

\theoremstyle{definition}

\theoremstyle{remark}

\numberwithin{equation}{section}
\numberwithin{theorem}{section}
\numberwithin{figure}{section}
\numberwithin{table}{section}

\DeclareMathOperator{\tr}{Tr}

\newcommand{\bino}[2]{\genfrac{(}{)}{0pt}{}{#1}{#2}}
\newcommand{\qbin}[2]{\genfrac{[}{]}{0pt}{}{#1}{#2}}

\newcommand{\cC}{{\mathcal C}}
\newcommand{\cD}{{\mathcal D}}
\newcommand{\cM}{{\mathcal M}}
\newcommand{\cO}{{\mathcal O}}

\newcommand{\bQ}{{\mathbf Q}}
\newcommand{\bL}{{\mathbf L}}
\newcommand{\bz}{{\mathbf z}}
\newcommand{\be}{{\mathbf e}}
\newcommand{\bu}{{\mathbf u}}
\newcommand{\bm}{{\mathbf m}}
\newcommand{\bn}{{\mathbf n}}
\newcommand{\bs}{{\mathbf s}}
\newcommand{\bt}{{\mathbf t}}
\newcommand{\bq}{{\mathbf q}}
\newcommand{\bx}{{\mathbf x}}
\newcommand{\by}{{\mathbf y}}
\newcommand{\ba}{{\mathbf a}}
\newcommand{\bb}{{\mathbf b}}

\newcommand{\bte}{\bt_{\text{e}}}
\newcommand{\btqp}{\bt_{\text{qp}}}
\newcommand{\bse}{\bs_{\text{e}}}
\newcommand{\bsqp}{\bs_{\text{qp}}}
\newcommand{\bqe}{\bq_{\text{e}}}
\newcommand{\bqqp}{\bq_{\text{qp}}}

\newcommand{\wbqe}{\widetilde{\bq}_{\text{e}}}
\newcommand{\wbqqp}{\widetilde{\bq}_{\text{qp}}}

\newcommand{\bbq}{\bar{\bq}}
\newcommand{\bbqe}{\bar{\bq}_{\text{e}}}
\newcommand{\bbqqp}{\bar{\bq}_{\text{qp}}}

\newcommand{\bbx}{\bar{\bx}}
\newcommand{\bby}{\bar{\by}}

\newcommand{\qee}{q_{\text{e},1}}
\newcommand{\qet}{q_{\text{e},2}}
\newcommand{\qqpe}{q_{\text{qp},1}}
\newcommand{\qqpt}{q_{\text{qp},2}}

\renewcommand{\AA}{{\mathbb A}}
\newcommand{\BB}{{\mathbb B}}
\newcommand{\CC}{{\mathbb C}}
\newcommand{\DD}{{\mathbb D}}
\newcommand{\EE}{{\mathbb E}}
\newcommand{\FF}{{\mathbb F}}
\newcommand{\GG}{{\mathbb G}}
\newcommand{\II}{{\mathbb I}}
\newcommand{\JJ}{{\mathbb J}}
\newcommand{\KK}{{\mathbb K}}
\newcommand{\MM}{{\mathbb M}}
\newcommand{\NN}{{\mathbb N}}

\newcommand{\PP}{{\mathbb P}}
\newcommand{\QQ}{{\mathbb Q}}

\newcommand{\WW}{{\mathbb W}}
\newcommand{\XX}{{\mathbb X}}
\newcommand{\ZZ}{{\mathbb Z}}

\newcommand{\id}{{\mathbf 1}}

\newcommand{\KKe}{\KK_{\text{e}}}
\newcommand{\KKqp}{\KK_{\text{qp}}}
\newcommand{\WWe}{\WW_{\text{e}}}
\newcommand{\WWqp}{\WW_{\text{qp}}}
\newcommand{\wKK}{\widetilde{\KK}}
\newcommand{\wKKe}{\wKK_{\text{e}}}
\newcommand{\wKKqp}{\wKK_{\text{qp}}}
\newcommand{\bKK}{\bar{\KK}}
\newcommand{\bKKe}{\bar{\KK}_{\text e}}
\newcommand{\bKKqp}{\bar{\KK}_{\text{qp}}}

\newcommand{\nonu}{\nonumber}
\newcommand{\txtfrac}[2]{{\textstyle{\frac{#1}{#2}}}}
\newcommand{\hp}{\hphantom{-}}
\newcommand{\eql}{=}

\newcommand{\qn}[1]{(q)_{#1}}

\newcommand{\qinf}{(q)_\infty}

\newcommand{\qp}{{\rm qp}}
\newcommand{\fg}{\mathfrak{g}}
\newcommand{\wfg}{\widehat{\fg}}
\makeatletter
\def\dddots{\mathinner{\mkern1mu\raise\p@
    \vbox{\kern7\p@\hbox{.}}\mkern2mu
    \raise4\p@\hbox{.}\mkern2mu\raise7\p@\hbox{.}\mkern1mu}}
\makeatother
\newcommand{\dCo}{{\text{h}^\vee}}

\newcommand{\la}{\lambda}

\newcommand{\latot}{\lambda_{\text{tot}}}
\newcommand{\mutot}{\mu_{\text{tot}}}

\newlength{\strutlengte}
\newlength{\strutdiepte}
\newlength{\strutbreedte} 
\newlength{\extraonderruimte}
\newlength{\extrabovenruimte}
\newcommand{\strt}[1]{
\settoheight{\strutlengte}{$#1$}
\addtolength{\strutlengte}{\extrabovenruimte}
\settodepth{\strutdiepte}{$#1$}
\addtolength{\strutdiepte}{\extraonderruimte}
\addtolength{\strutlengte}{\strutdiepte}
\rule[-\strutdiepte]{0mm}{\strutlengte}}

\def\dottedhline(#1,#2)#3#4#5{%
\multiput(#1,#2)(#3,0){#4}{\circle*{#5}}
}

\newlength{\kolom}
\newlength{\kolomm}

\begin{document}

\title[K-matrices for 2D conformal field theories]{K-matrices for
       2D conformal field theories}

\author[E. Ardonne]{Eddy Ardonne}
\address[Eddy Ardonne]{
Institute for Theoretical Physics\\
University of Amsterdam\\
Valckenierstraat 65\\
1018~XE~~Amsterdam \\
The Netherlands}
\address[]{
Department of Physics\\
University of Illinois at Urbana-Champaign\\
Loomis Lab of Physics\\
1110 W. Green Street\\
Urbana, IL 61801-3080\\
USA (Present address)}
\email{ardonne@uiuc.edu}

\author[P. Bouwknegt]{Peter Bouwknegt}
\address[Peter Bouwknegt]
{Department of Physics and Mathematical Physics
and Department of Pure Mathematics\\
University of Adelaide\\
Adelaide, SA 5005 \\
Australia}
\email{pbouwkne@physics.adelaide.edu.au, pbouwkne@maths.adelaide.edu.au}

\author[P. Dawson]{Peter Dawson}
\address[Peter Dawson]{
Department of Physics and Mathematical Physics\\
University of Adelaide\\
Adelaide, SA 5005 \\
Australia}
\email{pdawson@physics.adelaide.edu.au}




\thanks{EA would like to thank the Department of Physics and
Mathematical Physics at the University of Adelaide, where most
of this work was carried out, for hospitality.
PB acknowledges financial support from the Australian Research Council.
The research of EA was supported in part by the foundation FOM of the
Netherlands and by the National Science Foundation through the
grant DMR-01-32990.\\
ADP-01-30/M99, ITFA-02-33, [{\tt arXiv:hep-th/0212084}].}

\begin{abstract}
In this paper we examine fermionic type characters (Universal Chiral
Partition Functions) for general 2D conformal field theories
with a bilinear form given by a matrix of the form $\KK\oplus \KK^{-1}$.
We provide various techniques for determining these K-matrices,
and apply these to a variety of examples including (higher level)
WZW and coset conformal field theories.
Applications of our results to fractional quantum Hall
systems and (level restricted) Kostka polynomials are discussed.
\end{abstract}

\maketitle

\section{Introduction}

Two dimensional conformal field theories can be studied in a variety
of ways. In this paper, we will pursue the quasiparticle description,
which has attracted a lot of attention recently.
In a quasiparticle description, the characters of the conformal field
theories are of the fermionic sum type. It has been conjectured
that all these fermionic sums are of a form which goes under the name
of the `Universal Chiral Partition Function' (UCPF), see for instance
\cite{BM}, \cite{BCR} and \cite{BSb} (and references therein).
In general, the statistics of the quasiparticles is fractional and
interpolates between Fermi and Bose statistics. Moreover, to describe
general CFTs, we need to be able to incorporate the effect of the
non-trivial fusion rules of the fields, which can be done by allowing for
so called pseudoparticles. These pseudoparticles do not carry any energy
and are essential in describing the non-abelian statistics which is found in
the CFTs with non-trivial fusion rules.

Fractional statistics can be described in terms of the Haldane
`exclusion statistics' \cite{Hal}. If we allow for new types of particles,
such as the pseudoparticles, the same is true for the non-abelian
statistics, see \cite{GS} and \cite{BCR}.
The exclusion statistics is defined in terms of the
exclusion statistics parameters of the particles.
The parameters are intimately related to the Universal Chiral Partition
Functions, as it is these parameters which lie at heart of the UCPF, via
the so called K-matrix, which contains all the (mutual) statistics
parameters. In this paper, we will determine the K-matrices related
to the affine Lie algebra CFTs, in a particular basis. This basis
was first proposed in the context of the fractional quantum Hall states.

The topological properties of (fractional) quantum Hall states are
also encoded in matrices, which turned out to be the same as the
K-matrices alluded to in the above. In the abelian states, the
entries correspond to the coupling parameters of the Chern-Simons
fields which appear in the effective action of the quantum Hall
system (see, in particular \cite{Wen}, and references therein).
The Chern-Simons term effectively  changes the statistics
of the matter fields, making the relation between with the exclusion
statistics plausible. More details on this relation can be found in
\cite{ABS}.  The basis used in the description of certain classes of
non-abelian quantum Hall states is found to be useful in the context
of general affine Lie algebra CFTs as well.

One of the reasons that this basis is useful relates to the presence
of a duality, which relates the `electron-like' particles to the
quasiparticles (the notion of electron-like and quasiparticles will
be explained in Section~\ref{qhb}). Moreover, there is no mutual statistics
between these two types of particles. As this structure simplifies
the study of the conformal field theories, we will use this type
of basis throughout this paper.

One of the main themes in this paper will be the determination of the
K-matrices for the affine Lie algebra CFTs. We will develop a scheme
which is used to find the general K-matrices. The main idea is
to use `abelian coverings' of the (in general non-abelian) CFTs, and
project out some degrees of freedom. Having obtained the K-matrices,
we will propose a scheme to obtain the K-matrices for conformal
field theories which are of the coset form. We will address the diagonal
cosets, as well the parafermion CFTs, related to the affine Lie algebra
CFTs. Another application are the Kostka-polynomials
(see, e.g., \cite{Ki,Kib}, and references therein),
which can also be described in terms of the K-matrices.

In more detail, the outline of this paper is as follows. We start with a
general introduction to the role of the K-matrix in 2D conformal field
theories in Section \ref{kmatcft}. We will review some results concerning
the Universal Chiral Partition Function and the relation with exclusion
statistics. The structure of the basis of quasiparticles which will
be used throughout this paper is explained. We will end Section \ref{kmatcft}
by explaining the relation between the pseudoparticles and the fusion
rules of CFTs. In Section \ref{composites} we will explain the tools we
will use in determining the K-matrices for a general affine Lie algebra.
The idea is to embed the level-$k$ affine Lie algebra in $k$ copies
of the level-1 version, and project out certain degrees of freedom,
by using what we call a P-transformation. In Section \ref{kmatrices},
we will explicitly give the K-matrices for all the simple (untwisted)
affine Lie algebras. We will apply these results to obtain K-matrices
for cosets in Section \ref{cosets}. Finally, in Section \ref{secF}, we will
present some new results on level restricted Kostka polynomials related
to affine Lie algebras.
Some of the details are presented in the Appendices. Appendix~\ref{cartan}
deals with some notational issues, and explicitly gives all the Cartan
matrices and there inverses. Appendices~\ref{so5app} and \ref{g2app}
deal with the K-matrices for $\mathfrak{so}(5)_1$ and $G_{2,1}$
respectively, while Appendix~\ref{relbases} relates two different bases for
$\mathfrak{sl}(3)_k$.

\section{K-matrices for 2D conformal field theories}
\label{kmatcft}

\subsection{The UCPF and exclusion statistics}

Quasiparticles play an important role in the description of
$2$-dimensional conformal field theories (CFTs).
The exclusion statistics of these particles is closely related to
characters for CFTs, or more precisely, the `Universal Chiral Partition
Function' (UCPF).

\subsubsection{Quasiparticle basis}

We will start the discussion by introducing quasiparticle bases
for two dimensional conformal field theories, and in particular
(truncated) partition functions based on these bases.
In CFTs, the quasiparticles take the form of chiral vertex operators
$\phi^{(i)} (z)$ ($i=1,\ldots,n)$, which intertwine between
irreducible representations of the chiral algebra. By applying the modes
of these operators on a set of vacua $\lvert \omega \rangle$, one finds
(in general) an over complete basis, which, by using suitable restrictions
on the modes $(s_1, \ldots , s_N)$, can be turned into a maximal,
linearly independent set of states
\begin{equation}
\label{qpbasis}
\phi_{-s_N}^{i_N} \cdots \phi_{-s_2}^{i_2} \phi_{-s_1}^{i_1}
\lvert \omega \rangle \ .
\end{equation}
The grand canonical
partition function is obtained by taking the trace over this basis
\begin{equation}
\label{sec2:part}
  P(\bz;q) = \tr \biggl( \Bigl( \prod_i z_i^{N_i} \Bigr) q^{L_0} \biggr) \ .
\end{equation}
$N_i$ is the number operator for the quasiparticles $\phi^{(i)}$ and
$L_0 =\sum_i s_i$. Furthermore, $z_i = e^{\beta\mu_i}$ is a
(generalized) fugacity and
$q=e^{-\beta \varepsilon}$.
To find the `one particle grand canonical partition functions' $\lambda_i$,
we will use truncated partition functions, see \cite{Sca}.
In particular, one defines the truncated partition function
$P_\bL (\bz;q)$ by restricting the trace over the states
\eqref{sec2:part} in such a way that the modes $s$ of
the quasiparticles of species $i$ satisfy $s \leq L_i$
($\bL = (L_1,\ldots,L_n)$). In the limit of large $\bL$ one has
\begin{equation}  \label{largeL}
  \frac{P_{\bL+\be_i} (\bz;q)}{P_\bL (\bz;q)}  \ \sim\
  \lambda_i (z_i q^{L_i}) \ ,
\end{equation}
where $\be_i$ is the unit vector in the $i$-direction.
By using a recursion relation for the truncated partition function
$P_\bL (\bz;q)$ (which can be obtained from the basis \eqref{qpbasis})
and the limit \eqref{largeL}, one finds relations for the one particle
partition functions $\lambda_i$ (for more details, see \cite{BSb,BCR}).
For all the CFTs which were investigated by means of a quasiparticle
basis as discussed in this section, the equations determining $\lambda_i$
are of the form \eqref{iow}, and thus the quasiparticles satisfy
so-called `exclusion statistics', see Section \ref{exclstat}.

\subsubsection{The Universal Chiral Partition Function} \label{secbab}

It has been conjectured (see \cite{BM}, and references therein) that
the characters of all the irreducible representations of (rational)
conformal field theories can be written in the form
\begin{equation}  \label{sec2:ucpf}
  P(\bz;q) = \sideset{}{'} \sum_\bm
  \Bigl( \prod_i z_i^{m_i} \Bigr)
  q^{\frac{1}{2} \bm \cdot \KK  \cdot \bm + \bQ \cdot \bm }
  \prod_i \qbin{\bigl( (\II-\KK)\cdot\bm+\bu\bigr)_i}{m_i} \ ,
\end{equation}
which goes under the name of the `Universal Chiral Partition
Function' (UCPF) (or `fermionic-type character').
The matrix $\KK$ is a symmetric $n\times n$ matrix, $\II$ is the
$n\times n$ identity matrix
and $\bQ$ and $\bu$ are $n$-vectors. The sum is over
the non-negative integers $m_1,\ldots,m_n$. The restrictions denoted by
the prime are (in general) such that the coefficients of the $q$-binomials
are integers. These $q$-binomials are defined by
\begin{align}
  \qbin{M}{m} & = \frac{\qn{M}}{\qn{M-m}\qn{m}} \ , &
  \qn{m} = \prod_{k=1}^{m} (1-q^k) \ .
\end{align}
Depending on the parameters $u_i$, the associated particles
are of certain type. For {\em physical particles} $u_i = \infty$,
while {\em pseudoparticles} have $u_i < \infty$.
Note that in the limit $u_i \to\infty$ the $i$-th $q$-binomial reduces to
$1/(q)_{m_i}$ due to
\begin{equation} \label{mtoinf}
  \lim_{M \rightarrow \infty} \qbin{M}{m} = \frac{1}{\qn{m}} \ .
\end{equation}
As will become clear below, pseudoparticles do not carry
energy. They come about in theories with a non-abelian symmetry,
and in a sense they serve as bookkeeping devices for the internal
structure of the theory.

It was conjectured in \cite{GS,BCR} that the UCPF \eqref{sec2:ucpf}
is the partition function of a set of particles satisfying
exclusion statistics.
To be able to make this connection with exclusion statistics,
we will take a closer look at truncated versions of the
UCPF, and continue with a discussion on exclusion statistics
and the relation between the two.

Suppose that the truncated partition function $P_{\bL}(\bz;q)$ takes
the form of a `finitized' UCPF\footnote{While this is the case for
many examples, in general the finitized UCPF corresponding to a set
of (quasi)particles may differ from \eqref{eqPBaa} by terms $q^n$
with $n = \cO(L_i)$.  This will however not affect the conclusion.}
\begin{equation} \label{eqPBaa}
  P_{\bL}(\bz;q) = \sideset{}{'} \sum_\bm
 \Bigl( \prod_i z_i^{m_i} \Bigr)
  q^{\frac{1}{2} \bm \cdot \KK  \cdot \bm + \bQ \cdot \bm }
  \prod_i \qbin{\bigl( \bL + (\II-\KK)\cdot\bm+\bu\bigr)_i}{m_i} \ .
\end{equation}
One can then derive recursion relations for these truncated characters
by using the $q$-binomial relation
\begin{equation}
\qbin{M}{m} = \qbin{M-1}{m} + q^{M-m} \qbin{M-1}{m-1} \,.
\end{equation}
This leads to the recursion relations \cite{Bo2,ABS}
\begin{equation} \label{eqPBab}
  P_\bL (\bz;q) = P_{\bL-\be_i} (\bz;q) +
  z_i q^{-\frac{1}{2} \KK_{ii} + \bQ_i + \bu_i + \bL_i}
 P_{\bL-\KK\cdot\be_i} (\bz;q) \ .
\end{equation}
After dividing by $P_\bL (\bz;q)$, setting $q=1$, taking the large
$\bL$ limit and using relation \eqref{largeL}, one finds
\begin{equation} \label{iowtucpf}
  1 = \lambda_i^{-1} + z_i \prod_j \lambda_j^{-\KK_{ji}} \,,
\end{equation}
or equivalently
\begin{equation} \label{eqPBac}
  \left( \frac{\lambda_i -1}{\lambda_i} \right) \prod_j
  \lambda_j^{\KK_{ij}} = z_i \,.
\end{equation}
These relations are known as the Isakov-Ouvry-Wu (IOW) \eqref{iow}
equations, which give the one particle partition functions
for a system of particles which obey exclusion statistics;
this will be addressed in the next section.
For more details on this issue, we refer to \cite{ABS} and references
therein.

In the case of WZW Conformal Field Theories, i.e.\ CFTs with affine
Lie algebra symmetry, it is known that
in many cases (see \cite{NY,Yam,HKKOTY,BH},
and references therein) the (chiral) partition function
can be written in the form
\begin{equation}
  P(\bz;q) = \sum  M^{(k)}_{\lambda\mu}(q) M^{(\infty)}_\mu(\bz;q) \,,
\end{equation}
where $M^{(k)}_{\lambda\mu}(q)$ are the so-called
level-$k$ truncated Kostka polynomials,  $M^{(\infty)}_\mu(\bz;q)$
their $k\to\infty$ limit (with fugacity parameter $\bz$).
Having found an expression for the K-matrices of these CFTs
will thus give a natural guess for an explicit expression of these
level-$k$ truncated Kostka polynomials.  We will explore this
further in Section \ref{secF}.

For completeness, let us recall the value of the central charge
$c_{\text{ALA}}$,
of a CFT with affine Lie algebra symmetry $\wfg$ at level $k$,
\begin{equation} \label{eqcala}
  c_{\text{ALA}} = \frac{ k\ \text{dim}\,\fg }{k+\dCo} \,,
\end{equation}
where $\dCo$ is the dual Coxeter number corresponding to $\fg$.

For convenience, throughout this paper we will denote the (untwisted)
affine Lie algebra at level $k$, corresponding to a finite dimensional
Lie algebra $X_n$, by $X_{n,k}$, rather than by $(X_n^{(1)})_k$ which
is more common in the literature.

\subsubsection{Exclusion statistics}  \label{exclstat}

The starting point of the discussion on exclusion statistics will
be an ideal gas of particles which satisfy `fractional (exclusion)
statistics' \cite{Hal}.

The one particle grand canonical partition functions $\lambda_i$
for a set of quasiparticles obeying fractional exclusion statistics
can be obtained from the IOW equations \cite{IOW}
\begin{equation}
\label{iow}
\left( \frac{\lambda_i-1}{\lambda_i} \right)
\prod_j \lambda_j^{\KK_{ij}^{\rm st}} = x_i \ ,
\end{equation}
where $\KK^{\rm st}$ is the `statistics matrix' and
$x_i= z_i q = e^{\beta\mu_i}e^{-\beta\varepsilon}$ the fugacity.
Here, $\mu_i$ is the chemical potential of species $i$ and
$\varepsilon$ the energy.
Under the assumption of a symmetric matrix $\KK^{\rm st}$,
the one particle distribution functions follow
\begin{equation}
\label{sec2:2}
n_i (\varepsilon) = x_i \frac{\partial}{\partial x_i}
\log \prod_j \lambda_{j|_{x_i=e^{\beta(\mu_i-\varepsilon)}}}
= \sum_j x_j \frac{\partial}{\partial x_j} \log
\lambda_{i|_{x_i=e^{\beta(\mu_i-\varepsilon)}}} \ .
\end{equation}
These distribution functions are in general interpolations
between the Bose-Einstein and Fermi-Dirac distribution functions.

The discussion above holds in the case of abelian statistics,
but can be generalized to the non-abelian case
\cite{ABGS,ABS}. Non-abelian statistics arises when quasiparticle
operators (chiral vertex operators, see below)
in the underlying CFT have non-trivial fusion rules.
The effect of these fusion rules can be taken into account via
so-called `pseudoparticles', which do not carry any energy  (i.e.\
$q=1$). Note that for all the cases we consider, a formulation in
which the pseudoparticles have $x=1$ is possible. In fact, we only
consider formulations in which $x=1$ for the pseudoparticles.
More on the relation between fusion rules and
pseudoparticles can be found in Section \ref{pseudofus}.

We will now turn to the question of how to calculate the central
charge of a system of quasiparticles satisfying exclusion statistics with
statistics matrix $\KK^{\rm st}$ (and speak of the central charge
associated to the matrix $\KK^{\rm st}$).
First, we consider an abelian system, i.e.\ a system
without pseudoparticles.
In that case, the central charge is given by
\begin{equation}
\label{cclatot}
c_{\rm CFT} = \frac{6}{\pi^2} \int_0^1 \frac{dx}{x} \latot (x) \ ,
\end{equation}
where $\latot (z)$ denotes the product
\begin{equation}
\label{latotab}
\latot (x) = \prod_i \la_i (x_j=x) \ .
\end{equation}
By using the IOW-equations the central charge of Eqn.~\eqref{cclatot} can be
rewritten in the form (see, for instance, \cite{BSb})
\begin{equation}
\label{sec2:3}
c_{\rm CFT} = \frac{6}{\pi^2} \sum_i L(\xi_i) \,,
\end{equation}
where the $\xi_i$'s are solutions of the `central charge equations'
\begin{equation}
\label{sec2:4}
\xi_i = \prod_j (1-\xi_j)^{\KK_{ij}^{\rm st}} \,,
\end{equation}
and $L(z)$ is Rogers' dilogarithm
\begin{equation}
\label{sec2:5}
L(z) = -\frac{1}{2} \int_0^z dy \left(
\frac{\log y}{1-y} + \frac{\log(1-y)}{y}
\right) \ .
\end{equation}
The presence of pseudoparticles gives rise to a reduction of
the central charge. This reduction can be calculated in a similar
way, by considering the central charge equations restricted
to the pseudoparticles. For future convenience, we will denote the
statistics matrix restricted to the pseudoparticles by $\KK_{\psi\psi}$.
The central charge equations become (the prime denotes the restriction
to the pseudoparticles)
\begin{equation}
\label{sec2:6}
\xi'_i = \sideset{}{'} \prod_j (1-\xi'_j)^{(\KK_{\psi\psi})_{ij}} \ ,
\end{equation}
giving rise to a reduction $\tfrac{6}{\pi^2} \sum_j L(\xi'_j)$.
The central charge becomes
\begin{equation}
\label{sec2:7}
c_{\rm CFT} = \frac{6}{\pi^2} \Bigl( \sum_i L(\xi_i)
- \sideset{}{'} \sum_j L(\xi'_j) \Bigr) \,.
\end{equation}
This formula agrees with the central charge calculated from the
asymptotics of the UCPF \eqref{sec2:ucpf} (see, e.g., the discussion in
\cite{ABS}).

To summarize the above, we note that the truncated UCPFs in the large
$\bL$ limit give rise to one particle partition functions \eqref{eqPBac},
which are of the form of the IOW-equations \eqref{iow},
with statistics matrix $\KK^{\rm st} = \KK$.
Thus the K-matrix of the UCPF can be interpreted as a matrix
which describes the statistical interactions between the (quasi)particles.

The other important point was that in all the cases where conformal
field theories were studied by means of quasiparticle bases,
the equations \eqref{largeL} which determine $\lambda_i$, were shown to
be of the form of the IOW-equations.

We end this section by discussing the so-called quantum Hall
basis, which turns out to be very convenient for determining and
studying K-matrices for conformal field theories.

\subsubsection{The quantum Hall basis}  \label{qhb}

A convenient basis for WZW conformal field theories was first
proposed in the context of the quantum Hall effect \cite{ES}.
[This basis is also very natural from the mathematical point of
view as it is closely related to the existence of generalizations
of the Durfee square formula in combinatorics \cite{Bo2}.]
The `electron-like' particles (with unit charge and spin
$\txtfrac{1}{2}$ and
(fractionally) charged quasiparticles (sometimes called quasiholes)
are chosen to form a basis. It was found that a basis could be
chosen in such a way that the statistics matrix
$\KK_{\rm e}$ for the electron-like particles, and the matrix
$\KK_{\rm qp}$ for the quasiparticles are each others inverse
\begin{equation}
\label{dual}
\KK_{\rm qp}=\KK_{\rm e}^{-1} \,,
\end{equation}
while, furthermore, there is no mutual statistics between
the quasiparticles
and electrons, i.e.\
\begin{equation} \label{directsum}
  \KK = \KKe \oplus \KKqp \ .
\end{equation}
This is a very important observation, which will have many consequences.
Though this basis was first proposed in the context of the Laughlin
and Jain states \cite{ES}, it was soon realized that a basis
with a similar structure could be constructed
for the non-abelian generalizations of the abelian
quantum Hall states \cite{GS,ABGS,ABS}. These non-abelian generalizations are
based upon Wess-Zumino-Witten conformal field theories.
In this paper, we will determine bases for general WZW conformal
field theories. In the next section, we will review and develop some
techniques which are needed to perform this task. Here, we will first
explore some consequences of the `duality' between the electron and
quasiparticle sector.

In the description of the quantum Hall effect, the quantum numbers
of the particles play an important role, as they are used to calculate
physical properties. The most important are the charge and spin
quantum numbers, which are usually grouped in the so-called charge
and spin vectors, $\bt$ and $\bs$, respectively (see, for instance,
\cite{Wen}). Denoting a general vector for the
electron (quasiparticle) sector by $\bq_{\rm e}$ $(\bq_\qp)$ we have
\begin{equation}  \label{dualvec}
  \bq_\qp = - \KK_{\rm e}^{-1} \cdot \bq_{\rm e} \,.
\end{equation}
The  filling fraction $\nu$ and the spin filling $\sigma$ are
given by the expressions
\begin{align} \label{fillfracs}
  \nu &= \bte^T \cdot \KKe^{-1} \cdot\bte =
  \btqp^T \cdot \KKqp^{-1} \cdot \btqp \,, &
  \sigma &= \bse^T \cdot \KKe^{-1} \cdot \bse =
  \bsqp^T \cdot \KKqp^{-1} \cdot \bsqp \,.
\end{align}
These quantities are important physically; from a mathematical point
of view they are interesting, as they are conserved by the W- and
P-transformations of Section \ref{composites}. In a sense, these
transformations are constructed in such a way that they have this
property.

Let us explore some consequences of the duality, in particular
Eqns.~\eqref{dual} and \eqref{directsum}. We will focus on
the thermodynamic properties first and have a closer look at the
IOW-equations \eqref{iow}. We will denote the one particle distribution
functions for the electron-like particles
and quasiparticles by $\mu_i$ and
$\la_i$, respectively. The corresponding fugacities are given by
$y_i$ and $x_i$. Thus, the $\mu_i$ and $\la_i$ are the solutions
to the equations
\begin{align}
\left( \frac{\mu_i-1}{\mu_i} \right)
\prod_j \mu_j^{(\KKe)_{ij}} &= y_i \ , &
\left( \frac{\lambda_i-1}{\lambda_i} \right)
\prod_j \lambda_j^{(\KKqp)_{ij}} &= x_i \ .
\end{align}
Now Eqn.~\eqref{dual} leads to the following relations
\begin{align} \label{dualiow}
\la_i &= \frac{\mu_i-1}{\mu_i} \ , & x_i &=
\prod_j y_i^{-(\KKe)_{ij}^{-1}} \,.
\end{align}

Another important feature of the basis described in this section
is that the presence of pseudoparticles in the quasiparticle matrix
$\KK_{\rm qp}$ is accompanied by the presence of so-called `composite'
particles in the electron matrix $\KK_{\rm e}$.
The reason for this will become clear in Section \ref{composites}.
In general, the matrix $\KKe$ contains a few `electrons' (particles with
unit charge and spin up or down), with fugacities $y$.
In addition, there are composite particles, with fugacities
$y^{l_i}$, where the $l_i$ are positive integers.
The quantum numbers of the composites in the electron sector are
integer multiples of the quantum numbers of the electrons.
In the presence of composites in the electron sector, there will be
pseudoparticles in the quasiparticle sector. Pseudoparticles
have $x=1$, and as a consequence, pseudoparticles will have all quantum
numbers equal to zero. In principle, the fugacity of pseudoparticles
might be of the more general form $\tfrac{x_i}{x_j}$
(compare Eqns.~\eqref{dcompfug} and
\eqref{dcompfugsp}), but in all cases we will consider, this will not be the
case. Also, physical particles with all quantum numbers
trivial might occur, but again, we will not encounter such a
situation in this paper.

In the following, we will only encounter the situation where the electron
sector has composites, but no pseudoparticles, while the quasiparticle
sector does contain pseudoparticles, but no composites.
Thus, we will assume that the quasiparticle matrix has the following form
\begin{align}
\label{qpform}
\KK_{\rm qp} &=
\begin{pmatrix}
\KK_{\psi\psi} & \vdots & \KK_{\psi\phi} \\
\hdotsfor{3} \\
\KK_{\phi\psi} & \vdots & \KK_{\phi\phi}
\end{pmatrix} \ , & \KK_{\psi\psi}^T &= \KK_{\psi\psi}\,, &
\KK_{\phi\phi}^T &= \KK_{\phi\phi}\,, & \KK_{\psi\phi}^T &= \KK_{\phi\psi}\,,
\end{align}
where $\KK_{\phi\phi}$ denotes the statistic matrix for the physical
(as opposed to pseudo) quasi\-particles and $\KK_{\psi\phi}$ the mutual
statistics between the pseudo- and physical particles.

In the presence of composites and pseudoparticles, we have to
generalize the definition of $\latot$ (see Eqn.~\eqref{latotab}) to
\begin{equation}
\label{latotnab}
\latot (x) = \prod_i [\la_i (x_j=x^{l_j})]^{l_i} \ .
\end{equation}
With this definition, the central charge is still given by
Eqn.~\eqref{cclatot}.
In the absence of pseudoparticles, the central charge associated to
the system $\KKe\oplus\KKqp$, is simply given by the rank $n$ of the matrix
$\KKe$ (see, for instance, \cite{ABS}). To show this, we take a look at
the central charge equations
\begin{align}
\zeta_i &= \prod_j (1-\zeta_j)^{{\KKe}_{ij}} \ , &
\xi_i &= \prod_j (1-\xi_j)^{{\KKqp}_{ij}} \ .
\end{align}
Now because of the fact that $\KKqp = \KKe^{-1}$, the solutions to these
equations $\zeta_i$ and $\xi_i$ are simply related by
$\xi_i=1-\zeta_i$. We find the central charge to be
\begin{equation}
c_{\rm CFT} = \frac{6}{\pi^2} \sum_i \bigl( L(\xi_i) + L(1-\xi_i) \bigr) =
\frac{6}{\pi^2} n L(1) = n \ ,
\end{equation}
by using the dilogarithm relation
\begin{equation}
\label{dilogrel}
L (z) + L(1-z) = L(1) = \frac{\pi^2}{6} \ .
\end{equation}
In the case pseudoparticles are present, we again have a simple subtraction
(see Eqn.~\eqref{sec2:7}, the prime denotes the restriction to the
pseudoparticles)
\begin{equation} \label{eqPBad}
  c_{\rm CFT} = n - \frac{6}{\pi^2}
  \sideset{}{'} \sum_j L(\xi'_j) \ .
\end{equation}

It is important to note that the knowledge of the K-matrix is not
enough to specify the theory completely. In addition, one has to know,
or rather specify, which particles are pseudoparticles. So two theories can
have the same K-matrix, but differ in the `particle content' and
thereby (for instance) have different central charge. We will encounter
this situation frequently, namely as we discuss the K-matrices for
CFTs with affine Lie algebra symmetry, in cases the Lie algebra is
non simply-laced.

\subsection{Pseudoparticles and fusion rules}  \label{pseudofus}

There is an intimate connection between the pseudoparticle K-matrix
$\KK_{\psi\psi}$ and the fusion rules of a CFT, which can be used
as a consistency check or guiding principle on the construction of
K-matrices.
To explain this connection, consider a CFT with fusion rules
$N_{ij}{}^k$, $i,j,k=1,\ldots,\ell$.
The incidence matrix of the fusion graph $\Gamma_i$,
corresponding to taking consecutive fusions with the field $i$, is given
by the matrix $N_i$ with components $(N_i)_j{}^k = N_{ij}{}^k$.
Hence, if $P_{ij}{}^k(M)$ denotes the number of paths of length $M$ on
the fusion graph $\Gamma_i$ beginning at $j$ and ending at $k$ we
have
\begin{equation} \label{eqPBAa}
  P_{ij}{}^k(M) = \left( (N_i)^M \right)_j{}^k \,.
\end{equation}
Thus we find a recursion relation
\begin{equation}\label{eqPBAb}
  P_{ij}{}^k(M) = \sum_l P_{ij}{}^l(N) P_{il}{}^k(M-N) \,,
\end{equation}
for each $0\leq N\leq M$, with initial condition
$P_{ij}{}^k(0) = \delta_j{}^k$.
These recursion relations, however, involve
paths beginning and ending at arbitrary points. To derive
a recursion relation for fixed $j$ and $k$ we apply
the characteristic equation of $N_i$, i.e., the $\ell$-th
order polynomial equation for $N_i$ arising from the eigenvalue
equation, to $P_{ij}{}^k(M)$.  If the characteristic equation is
given as
\begin{equation}\label{eqPBAc}
  \sum_{n=0}^\ell a_n (N_i)^{\ell-n} =0\,,  \qquad a_0\equiv 1 \,,
\end{equation}
then, by using \eqref{eqPBAb} for $N=1$, we find the recursion relation
\begin{equation} \label{eqPBAd}
  \sum_{n=0}^\ell a_n P_{ij}{}^k(M-n) = 0 \,.
\end{equation}
That is, a recursion relation for fixed $j$ and $k$ and with coefficients
independent of $j$ and $k$.  Different solutions
of \eqref{eqPBAd}, determined by different initial conditions,
correspond to different choices of $j$ and $k$.\footnote{In fact,
for specific initial conditions, the solution might actually
satisfy simpler recursion relations obtained by factorizing the
characteristic equation and taking a subset of the factors.}
In particular,
asymptotically the number of paths is given by $(\lambda_{\text{max}})^M$,
where $\lambda_{\text{max}}$ is the largest eigenvalue of $N_i$.

On the other hand,
according to the UCPF assumption, the number of paths $P(M)$
of length $M$ on the
fusion graph $\Gamma_i$ is given in terms of the $q\to1$ limit
of the UCPF \eqref{sec2:ucpf}, i.e.,
\begin{equation} \label{eqPBAe}
  P_{\mathbf L} = \sum_{m_i}  \prod_i \bino{((\II-\KK_{\psi\psi})
  \cdot \mathbf m)_i   + L_i}{m_i} \,,
\end{equation}
where $L_i = a_i M + u_i$ and $\KK_{\psi\psi}$ the
pseudoparticle K-matrix.  The numbers $a_i$ are fixed (only
depend on the sector $i$),
in fact they arise as the part of the K-matrix describing the
coupling of the pseudoparticles to physical particles,
while $u_i$ is determined by begin and end point of the path.
[The $q$-analogue of Eqn.~\eqref{eqPBAe} is related to (level restricted)
Kostka polynomials and will be discussed in Section \ref{secF}.]
The numbers $P_{\mathbf L}$ satisfy the recursion relations
(cf.~\eqref{eqPBab})
\begin{equation} \label{eqPBAf}
  P_{\mathbf L} = P_{\mathbf{L} - \mathbf e_i} + P_{\mathbf{L} -
  \KK_{\psi\psi}\cdot  \mathbf e_i} \,,
\end{equation}
where $\mathbf e_i$ is the unit vector in the $i$-th direction.
In principle, the recursion relations \eqref{eqPBAf} can be manipulated
to yield a recursion relation for $P(M) \equiv P_{a_iM}$, the
quantity of interest.  Ideally, this recursion relation
should be the same as \eqref{eqPBAd}.  In practice, however, one finds
that one corresponds to a factor of the other
due to the fact we are dealing with specific initial conditions.
In practice, it is easier to study
the recursion relations \eqref{eqPBAf} in the large $M$ limit,
where they reduce to the IOW-equations \eqref{eqPBac}.  These can then
be used to derive an equation for $\mu=\prod_i \lambda_i^{a_i}$ which
should correspond to the characteristic equation for the eigenvalues
of $N_{ij}{}^k$, i.e.\ Eqn.\ \eqref{eqPBAc}.  In particular, the
largest root of the equation determining
$\mu$ should be equal to $\lambda_{\text{max}}$.

Moreover, note that while the recursion relations corresponding to graphs
on $\Gamma_i$ depend on the sector $i$, they should all derive from
one and the same pseudoparticle matrix $\KK_{\psi\psi}$
(they just differ in the
choice of $a_i$).  This puts extra constraints on the possible choices of
$\KK_{\psi\psi}$, given a set of fusion rules $N_{ij}{}^k$.  Unfortunately,
this still does not suffice to uniquely associate a pseudoparticle
$\KK_{\psi\psi}$ with a set of fusion rules $N_{ij}{}^k$ as is illustrated
for instance by the matrix
\begin{equation} \label{eqPBAg}
  \KK_{\psi\psi} = \begin{pmatrix}
  \frac{4}{3} & \frac{2}{3} \\
  \frac{2}{3} & \frac{4}{3} \end{pmatrix} \,,
\end{equation}
which arises both in $A_{2,2}$ and $F_{4,1}$
(see Sections \ref{sec:A} and \ref{secf4}), while
these two theories clearly have different fusion rules.  This is because
additional information is present in the coupling of pseudoparticles
to the physical particles (i.e.\ the numbers $a_i$).  Conversely, given a
pseudoparticle K-matrix leading to the correct fusion rules,
one can always construct other K-matrices giving rise to the
same recursion relations by extending the matrix `symmetrically'.
An example of this will be given in Section \ref{exampl}.

Finally, given a set of fusion rules $N_{ij}{}^k$,
we can compute the modular
$S$-matrix, since this is the matrix which simultaneously diagonalizes
all matrices $N_i$ \cite{Ver}.
Since the $T$-matrix acts diagonally on the characters
of the CFT with values $\exp( 2\pi i (h_i - c/24))$,
we can find constraints on the conformal dimensions
$h_i$ and the central charge $c$ from the
condition $(ST)^3=1$ (when $S^2=1$) or $(ST)^6=1$ (when $S^4=1$).

The central charge constraint in particular can be compared to
the central charge \eqref{eqPBad} arising from a particular choice of
pseudoparticle K-matrix.  Obviously, the constraints on which
fusion rules correspond to which pseudoparticle K-matrix derived this
way are much weaker than those arising from the comparison of the
above recursion relations.

\subsection{Simple examples}  \label{exampl}

Let us illustrate the considerations of the previous section in a few
examples.

Consider a CFT with two primary fields $1$ and $\phi$ and nontrivial
fusion rule $\phi\times\phi=1$, i.e.,
\begin{equation} \label{eqPBBa}
  N_\phi = \begin{pmatrix} 0 & 1 \\ 1 & 0 \end{pmatrix} \,,
\end{equation}
which has eigenvalues $\lambda=\pm1$ and is diagonalized by
\begin{equation}  \label{eqPBBb}
  S = \txtfrac{1}{\sqrt2} \begin{pmatrix} 1 & 1 \\ 1 & -1 \end{pmatrix} \,,
\end{equation}
which satisfies $S^2=1$. We find that $(ST)^3=1$
yields the condition
\begin{equation}  \label{eqPBBc}
  h_\phi =  \txtfrac{1}{4} \mod \txtfrac{1}{2} \,,
\end{equation}
while
\begin{equation}  \label{eqPBBd}
c = \begin{cases} 1 \mod 8 & \text{for}\ h_\phi =  \frac{1}{4} \mod 1 \,,\\
              7 \mod 8 & \text{for}\ h_\phi = \frac{3}{4} \mod 1\,.
  \end{cases}
\end{equation}
Clearly, $A_{1,1}$    
is an example of the first possibility,
while $E_{7,1}$ 
is an example of the second.

Since $c$ is necessarily an integer, one would conclude
that as far as this calculation is concerned no pseudoparticles
are necessary.
The characteristic equation for $N_\phi$ is given by $\lambda^2-1=0$
and leads to the recursion
\begin{equation}\label{eqPBBe}
  P(M) = P(M-2) \,,
\end{equation}
which is trivially solved by $P(2M)=P(0)$ and $P(2M+1)=P(1)$.  Again,
this does not require pseudoparticles, since the fusion paths are
obviously unique.

Now consider $A_{1,k}$ 
for generic level $k$.
The fusion matrix of the generating field
$\phi_2$ is given by the incidence matrix of the Dynkin diagram
of $A_{k+1}$
(see, for example, \cite{dFMS}).
The characteristic equation is thus given by
\begin{equation} \label{eqPBBea}
  \sum_{j=1}^{[(k+1)/2]} (-1)^j \bino{k+1-j}{j} \lambda^{k+1-2j} = 0 \,,
\end{equation}
and has roots (see, e.g., \cite{dFMS})
\begin{equation}  \label{eqPBBf}
   \lambda_j = 2 \cos \left( \frac{\pi j}{k+2} \right)\,,
   \qquad j=1,\ldots,k+1\,.
\end{equation}
For example, the characteristic equation at the first few levels is
given by
\begin{align} \label{eqPBBg}
k=1 & \qquad \lambda^2-1 = 0 \,, \nonumber \\
k=2 & \qquad \lambda(\lambda^2-1) = 0 \,, \nonumber \\
k=3 & \qquad \lambda^4 - 3\lambda^2 +1 = (\lambda^2 +\lambda -1)
(\lambda^2 -\lambda -1) = 0 \,, \nonumber \\
k=4 & \qquad \lambda^5-4 \lambda^3+3 \lambda = \lambda(\lambda^2-3)(
\lambda^2-1)  =0 \,.
\end{align}
On the other hand, the pseudoparticle K-matrix for $A_{1,k}$,
is known to be $\KK_{\psi\psi} = \frac{1}{2} \AA_{k-1}$, while
$\mathbf a = (\frac{1}{2},0,\ldots,0)$.  This leads to, e.g.,
\begin{align}  \label{eqPBBh}
  k=2 & \qquad \mu^2-1 = 0\,, \nonumber \\
  k=3 & \qquad \mu^2 -\mu -1 = 0 \,.
\end{align}
which, in general, corresponds to a factor of \eqref{eqPBBg} as discussed
in Section \ref{pseudofus}.

As a final example consider a CFT with
two primary fields $1$ and $\phi$ and fusion rule
$\phi\times\phi = 1 + \phi$, i.e.
\begin{equation} \label{eqPBBi}
  N_\phi = \begin{pmatrix} 0 & 1 \\ 1 & 1 \end{pmatrix} \,.
\end{equation}
The characteristic equation is given by
\begin{equation} \label{eqPBBj}
  \lambda^2 - \lambda -1 = 0 \,,
\end{equation}
with roots $\lambda_\pm= \frac{1}{2}(1\pm\sqrt5)$.
The constraints on $h$ and $c$, arising from the modular matrices, are
(see, e.g., \cite{dFMS}, Exercise 10.16)
\begin{equation} \label{eqPBBk}
  c - 12  h = -2 \mod 8 \,,
\end{equation}
while
\begin{equation}\label{eqPBBl}
  h = \frac{m}{5}\mod 1  \,,\qquad m=1,2,3,4\,.
\end{equation}
$G_{2,1}$ 
is an example of a solution for $m=2$
($c=\frac{14}{5},\ h=\frac{2}{5}$), while
$F_{4,1}$ 
is an example of a solution for $m=3$
($c=\frac{26}{5},\ h=\frac{3}{5}$).  Examples of $m=1,4$ solutions
can be found among the minimal (non-unitary) models.

The characteristic equation \eqref{eqPBBj} leads to the recursion relation
\begin{equation}  \label{eqPBBm}
  P(M) = P(M-1) + P(M-2)\,,
\end{equation}
the solutions of which are (generalized) Fibonacci numbers.
Clearly, the recursion relation \eqref{eqPBBm} arises from the
pseudoparticle matrix (cf.~\eqref{eqPBAf})
\begin{equation}  \label{eqPBBn}
  \KK = \begin{pmatrix} 2 \end{pmatrix} \,,
\end{equation}
with $\mathbf a = (1)$.

The central charge subtraction corresponding to \eqref{eqPBBn} is,
according to \eqref{eqPBad}, given by
\begin{equation} \label{eqPBBo}
 \frac{6}{\pi^2}\ L\left( \txtfrac{3}{2} - \txtfrac{1}{2}\sqrt5\right) =
  \txtfrac{2}{5} \,,
\end{equation}
which is not the correct subtraction for either $G_2$ or $F_4$.
We can however double the subtraction while, at the same time,
keeping the recursion relation, by a `symmetric doubling' of
\eqref{eqPBBn}, i.e., by making a $2\times2$ matrix with entries that
sum to $2$ in all columns and rows and which is such that the solution
to the IOW-equation is identical for all components, e.g.,
\begin{equation} \label{eqPBBp}
\KK = \begin{pmatrix}  \frac{4}{3} & \frac{2}{3} \\
                       \frac{2}{3} & \frac{4}{3}  \end{pmatrix} \,,
\end{equation}
with, $\mathbf a = (a_1,a_2)$ where $a_1+a_2=1$.  This case is relevant
for $(F_4^{(1)})_{k=1}$ (see Section \ref{secf4}).
To get a subtraction of $\frac{6}{5}$,
as needed for $(G_2^{(1)})_{k=1}$, we need to do a `symmetric tripling'
such as
\begin{equation} \label{eqPBBq}
\KK = \begin{pmatrix}  1  & \frac{1}{2} & \frac{1}{2} \\
                         \frac{1}{2} & 1  & \frac{1}{2} \\
                         \frac{1}{2} & \frac{1}{2} & 1 \end{pmatrix} \,.
\end{equation}
Cf.~Section \ref{g2oneqp}.

\section{Composite and dual composite construction}
\label{composites}

As is well known in the context of the quantum Hall effect, the K-matrices
describing the abelian quantum Hall states are not unique, but are in fact
determined up to similarity transformations. These similarity transformations
can be thought of as changing the basis for the description.
Moreover, the physical properties such as the filling fraction are not
changed by this transformation. Also the central charge is left unchanged.

A similar situation occurs when we want to view the K-matrices as the data
for a general (i.e.\ non-abelian)
CFT. There exist transformations of the K-matrices, which leave the
corresponding characters unchanged. Therefore, the K-matrices related by
such a transformation correspond to the same theory. A prime example will
be described in Section \ref{compcons} and the dual version in Section
\ref{dualcomp}. At first sight, this might be a
disturbing observation because we would like to have a unique description
of the theory. However, the situation can be used in our advantage, for
instance, in the construction of the K-matrices for general affine Lie
algebra CFTs, as will be pointed out in Section \ref{Ptrans}.

\subsection{W-transformations} \label{Wtrans}

To describe the well-known W-transformations (see, for instance
\cite{Wen}), we will use the notation
of the fqH basis (as we will do in the rest of this section). Of course,
it is applicable to all abelian quantum Hall systems. So we have a K-matrix
$\KK_{\text{e}}$ and the quantum number vectors $\bqe$ (the dual data
is obtained by applying Eqns.~\eqref{dual} and \eqref{dualvec}).
Let $\WW$ be an $SL(n,\ZZ)$ matrix, where $n$ is the rank of $\KK$.
The W-transformation takes the form
\begin{equation} \label{wtrans}
  \wKK_{\text{e}} = \WW\cdot {\KK}_{\text{e}} \cdot \WW^T\,,\qquad
  \widetilde{\KK}_{\text{qp}} = (\WW^{-1})^T \cdot {\KK}_{\text{qp}}
   \cdot \WW^{-1}\,,
\end{equation}
while
\begin{equation}
  \widetilde{\bq}_{\text{e}} = \WW \cdot \bq_{\text{e}} \,, \qquad
  \widetilde{\bq}_{\text{qp}} =  (\WW^{-1})^T \cdot \bq_{\text{qp}} \,.
\end{equation}
Indeed, physical quantities of the form $\bqe^T
\cdot \KKe^{-1} \cdot \bqe$,
such as the filling fraction are invariant under this transformation.
Also, the central charge, which is given by $n$ for the abelian states,
is not changed. In the non-abelian case, we can also apply these
W-transformations, however, to conserve the central charge,
we can only use those transformations which do not change
the pseudoparticle part of the K-matrix.

In the following, we will concentrate on constructions based on
character identities (so we view the K-matrices as matrices
containing CFT data). In addition, we will show that extended matrices
obtained in this way can be used to make a reduction of the theory,
which turns out to be closely related to the W-transformations
described above. We will use the results of this section extensively
in the remainder of this paper, in particular in Section \ref{kmatrices},
where we will obtain the K-matrices for general affine Lie algebra
CFTs.

\subsection{Composite construction}
\label{compcons}

The basic `transformation' one can do on a K-matrix, leaving the
theory invariant, is the composite construction \cite{ABS}.
The effect of this transformation is to add a particle, which is the
composite of two particles already present in the theory.
The quantum numbers of this composite particle
are just the sum of the quantum numbers of the two constituent
particles. In order to keep the theory unchanged, one has to increase
the mutual exclusion statistics of the two constituent particles. In a
sense, they avoid one another more, while the gap is filled by the
composite particle.

To make this more precise, consider the IOW-equations \eqref{iow} with
a symmetric matrix $\KKe$ (i.e. $a_{12}=a_{21}$ and $\bKKe=\bKKe^T$),
fugacities $\by$ and quantum numbers $\bqe$
\begin{equation} \label{kdef}
  \KKe \eql
  \begin{pmatrix}  a_{11} & \ldots & a_{1n}  \\
  \vdots &  & \vdots   \\
  a_{n1}  & \ldots & a_{nn} \end{pmatrix}
  = \begin{pmatrix}
  a_{11} & a_{12} & \ba_1^T \\
  a_{21} & a_{22} & \ba_2^T \\
  \ba_1 & \ba_2 & \bKKe
  \end{pmatrix} \,, \qquad
  \by \eql \begin{pmatrix} y_1 \\ y_2 \\ \bby \end{pmatrix} \,, \qquad
  \bqe \eql \begin{pmatrix} \qee \\ \qet \\ \bbqe \end{pmatrix} \,.
\end{equation}
If we define the operation $\cC_{12}$, corresponding to adding
a composite of the quasiparticles $1$ and $2$ to the system,
by
\begin{equation} \label{eqDDb}
\cC_{12} \KKe = \begin{pmatrix}
 a_{11} & a_{12} + 1 & a_{11}+a_{12} & \ba_1^T \\
 a_{21} + 1 & a_{22} & a_{21}+a_{22} & \ba_2^T \\
 a_{11}+a_{21} & a_{12}+a_{22} & a_{11}+2a_{12} +a_{22} & \ba_1^T
  + \ba_2^T \\
 \ba_1 & \ba_2 & \ba_1 + \ba_2 & \bKK \end{pmatrix} \,,
\end{equation}
and
\begin{equation} \label{eqDDc}
 \cC_{12} \by =
  \begin{pmatrix} y_1 \\ y_2 \\ y_1 y_2 \\ \bby \end{pmatrix} \,, \qquad
 \cC_{12} \bqe =
   \begin{pmatrix} \qee \\ \qet \\ \qee+\qet \\ \bbqe \end{pmatrix} \,,
\end{equation}
then the two systems are equivalent, at least at the level of
thermodynamics.
The action of the general $\cC_{ij}$ is defined, as above,
by a suitable permutation of the rows and columns.
The solutions $\{\mu_i\}$ to the IOW-equations
defined by $(\KKe,\by)$  and $\{\mu_i'\}$ defined by
$(\KKe',\by') = (\cC_{ij}\KKe,\cC_{ij}\by)$
are simply related by
\begin{align} \label{compiow}
  \mu_i' & \eql \frac{\mu_i+\mu_j-1}{\mu_j} \,, &
  \mu_j' & \eql \frac{\mu_i+\mu_j-1}{\mu_i} \,,\nonumber\\
  \mu_{n+1}' & \eql \frac{\mu_i\mu_j}{\mu_i+\mu_j-1} \,, &
  \mu_k' & \eql \mu_k\,,\qquad(k\neq i,j,n+1) \,.
\end{align}
Note that, in particular, it follows $\mu_i=\mu_i'\mu_{n+1}'$ and
$\mu_j=\mu_j'\mu_{n+1}'$ such that $\mutot = \mutot'$.
Also, from $\mu_i=\mu_i'\mu_{n+1}'$ and $\mu_j=\mu_j'\mu_{n+1}'$ one
sees that the original one particle partition functions for $i$ and
$j$, receive contributions from the new particles
$i$ and $j$, respectively, as well as from the composite
particle $n+1$. The operation $\cC_{ij}$ has the effect that
states in the spectrum containing both particles $i$ and $j$ get less
dense (their mutual exclusion statistics is bumped up by $1$), while
the resulting `gaps' are now filled by the new composite particle.

A consistency check on the equivalence of the systems described by
$(\KKe,\by)$ and $(\KKe',\by')=(\cC_{ij}\KKe,\cC_{ij}\by)$
is the fact that both lead to the same central charge.
It was shown in \cite{BCR} that this is in fact a consequence of
the five-term identity for Rogers' dilogarithm.
For completeness, we repeat the argument here. It is not hard to check
that the solutions to the Eqns.~\eqref{sec2:4}, with $\KKe$ and
$\cC_{ij} \KKe$, which we will denote by $\zeta_i$ and $\zeta_i'$
respectively, are related by
\begin{align}
  \zeta_i' &= \frac{\zeta_i(1-\zeta_j)}{1-\zeta_i \zeta_j} \,, &
    \zeta_j' &= \frac{\zeta_j(1-\zeta_i)}{1-\zeta_i \zeta_j} \,,\nonumber \\
  \zeta_{n+1}' &=\zeta_i \zeta_j \ , & \zeta_k' &= \zeta_k \,,
  \qquad (k\neq i,j,n+1) \,.
\end{align}
The equivalence of the central charge for both matrices follows from
\begin{equation}
\label{dilogvt}
  L(x) + L(y) = L\left(\frac{x(1-y)}{1-xy}\right) +
  L\left(\frac{y(1-x)}{1-xy}\right) + L(xy) \,,
\end{equation}
which is the five-term identity for Rogers' dilogarithm.

Finally, we note that the composite
transformation \eqref{eqDDb} can be derived from the following
character identity, which is a special case of the
$q$-Pfaff-Saalsch\"utz sum (see \cite{GR}) 
\begin{equation} \label{compid}
  \qbin{M_1}{m_1} \qbin{M_2}{m_2} = \sum_{m\geq 0} q^{(m_1-m)(m_2-m)}
  \qbin{M_1-m_2}{m_1-m} \qbin{M_2-m_1}{m_2-m}
  \qbin{M_1+M_2-(m_1+m_2)+m}{m} \,.
\end{equation}
If one inserts this identity at the $(i,j)$-th entry in the UCPF of
Eqn.~\eqref{sec2:ucpf}, one finds, after shifting the summation variables
$m_i\mapsto m_i-m$ and $m_i\mapsto m_i-m$, another UCPF, based on
the data $(\cC_{ij} \KK, \cC_{ij} \by)$.  

The form \eqref{compid} is used for the composite construction on two
pseudoparticles. Taking the limit $M_1 \rightarrow \infty$
($M_1,M_2 \rightarrow \infty$) by using Eqn.~\eqref{mtoinf}, gives the
appropriate identity for the composite construction applied to a physical
and a pseudoparticle (two physical particles) respectively.

\subsection{Dual composite construction}
\label{dualcomp}

Using the logic of the fqH basis, one might expect that
upon inverting the extended matrix $\cC_{ij} \KKe$, one should
find a matrix, which is related to $\KKqp=\KKe^{-1}$ by a character identity
as well. This turns out to be the case.

We will denote this transformation by $\cD_{ij}$, thus we define
$\cD_{ij} \KKqp = (\cC_{ij}\KKqp^{-1})^{-1}$.
After performing this transformation, the quasiparticles
corresponding to $i$ and $j$ have become pseudoparticles. This is
necessary, because otherwise the central charge of the transformed
system $\widetilde\KK = \cC_{ij}\KKe \oplus \cD_{ij}\KKqp$
would have been increased by one with respect to $\KKe\oplus\KKqp$,
because the rank of the K-matrices is increased by one.
The presence of the extra pseudoparticles reduces the central charge
by precisely the right amount, to keep the total central charge the same
(see below).

The action of $\cD_{ij}$ on a symmetric matrix $\KKqp$,
in the case of two (physical) particles, can be described in the following
way.
\begin{align} \label{eqPBda}
  \KKqp &= \begin{pmatrix}
  a & b \\
  b & c \\ \end{pmatrix} \ , &
  \cD_{12} \KKqp = \frac{1}{\Delta}
 \begin{pmatrix}
  1 & \Delta-1 & a-b-1\\
 \Delta-1 & 1 & c-b-1\\
  a-b-1 & c-b-1 & (1+b)^2-ac\\
 \end{pmatrix} \,,
\end{align}
where $\Delta = 2-(a-2b+c)$.
In addition, in the transformed formulation, the particles $1$ and $2$
are pseudoparticles. When, in the original formulation, the particles $i$ and
$j$ are physical, it is easily verified that the reduction of the central
charge, in the transformed formulation, due to the particles $i$ and $j$
is in fact equal to one. This is precisely the value needed to give
the transformed system the same central charge as the original formulation,
as was to be expected.

The action of $\cD_{12}$ on the fugacity and quantum number vectors
$\bx^T=(x_1,x_2)$ and $\bqqp^T=(\qqpe,\qqpt)$ is
given by
\begin{align}
  \cD_{12} \bx &= \begin{pmatrix}
  (\frac{x_1}{x_2})^\frac{1}{\Delta} \\
  (\frac{x_2}{x_1})^\frac{1}{\Delta} \\
  x_1^{\frac{1+b-c}{\Delta}} x_2^{\frac{1+b-a}{\Delta}}
  \end{pmatrix} \,, &
  \cD_{12} \bqqp &= \frac{1}{\Delta} \begin{pmatrix}
  \qqpe-\qqpt \\ \qqpt-\qqpe \\ (1+b-c)\qqpe+(1+b-a)\qqpt
  \end{pmatrix} \,.
\end{align}
If we have $x_1=x_2=x$ and hence, $\qqpe=\qqpt=\tilde{q}_{\text{qp}}$,
as will always be the case in this paper, we find
\begin{align}
  \cD_{12} \bx & = \begin{pmatrix}
  1 \\ 1 \\ x \end{pmatrix} \,, &
 \cD_{12} \bqqp & = \begin{pmatrix}
  0 \\ 0 \\ \tilde{q}_{\text{qp}} \end{pmatrix} \,.
\end{align}

{}From a character identity point of view, the
transformation \eqref{eqPBda} is based on the
$q$-binomial doubling formula
\begin{equation} \label{eqPBdb}
  \qbin{M+N}{n}
  = \sum_{p-q=M-n} q^{(M-p)(N-q)} \qbin{M}{p} \qbin{N}{q} \,.
\end{equation}
Indeed, considering the UCPF for two physical particles
with $\KKqp$ as in Eqn.~\eqref{eqPBda}, i.e.
\begin{equation} \label{eqPBdc}
  Z = \sum \frac{q^{\frac{1}{2} (am_1^2+cm_2^2+2bm_1m_2)}}{(q)_{m_1}
  (q)_{m_2}} =
  \sum \frac{q^{\frac{1}{2} (am_1^2+cm_2^2+2bm_1m_2)}}{(q)_{m_1+m_2}}
  \qbin{m_1+m_2}{m_1} \,,
\end{equation}
and then applying \eqref{eqPBdb} with
\begin{align}
  M & = - (b-c)m_1 + (1+b-a)m_2 \,, \nonumber \\
  N & = (1+b-c) m_1 - (b-a) m_2 \,,\nonumber \\
  n & = m_1 \,,
\end{align}
to the $q$-binomial in \eqref{eqPBdc}, results in the UCPF
based on $\cD_{12}\KKqp$ of \eqref{eqPBda}, with the identifications
\begin{equation}
  m_1' = p\,,\qquad m_2' = q\,,\qquad m_3' = m_1+m_2\,,
\end{equation}
and where the first two particles in $\cD_{12}\KKqp$ are pseudo.

The general case can be derived from \eqref{eqPBdb} as well, and is
described in the following way.  Again, we will focus on the case
where we let $\cD$ work on the first two particles.  In addition
we will assume that both those particles are physical.
For ease of presentation, we now define $\Delta =2-(b_{11}-2b_{12}+b_{22})$,
$\delta_1 = 1+b_{12}-b_{11}$ and $\delta_2=1+b_{12}-b_{22}$.

Using similar notation as in Eqn.~\eqref{kdef}, we take
(the symmetric) $\KKqp$, the fugacities and quantum numbers
\begin{equation} \KKqp = \begin{pmatrix}
  b_{11} & b_{12} & \bb_1^T \\
  b_{21} & b_{22} & \bb_2^T \\
  \bb_1 & \bb_2 & \bKKqp
  \end{pmatrix} \,, \qquad
  \bx = \begin{pmatrix} x_1 \\ x_2 \\ \bbx \end{pmatrix} \,, \qquad
  \bqqp = \begin{pmatrix} \qqpe \\ \qqpt \\ \bbqqp \end{pmatrix} \,.
\end{equation}
The dual composite construction, applied on the first two particles
is given by
\begin{multline} \label{dualcompact}
  \cD_{12} \KKqp = \\ \frac{1}{\Delta}
  \begin{pmatrix}
  1 & \Delta-1 & \vdots & -\delta_1 & \vdots & \bb_1^T - \bb_2^T \\
  \Delta-1 & 1 & \vdots & -\delta_2 & \vdots & \bb_2^T - \bb_1^T \\
  \hdotsfor[1.5]{6} \\
  -\delta_1 & -\delta_2 & \vdots & (1+b_{12})^2-b_{11}b_{22} & \vdots &
  \delta_2 \bb_1^T  + \delta_1 \bb_2^T \\
  \hdotsfor[1.5]{6} \\
  \bb_1 -\bb_2 & \bb_2 - \bb_1 & \vdots &
  \delta_2 \bb_1 + \delta_1 \bb_2 & \vdots &
  \Delta (\bKKqp)_{ij} + (\bb_1-\bb_2)_i
  (\bb_1-\bb_2)_j
  \end{pmatrix} \,.
\end{multline}
The first two particles have become pseudoparticles, while the extra
particle is a physical particle. Note that this construction based on the
character identity Eqn.~\eqref{eqPBdb} only
works in the case that the particles on which it is applied are
physical particles. We have not found a character identity for the case
where the dual composite construction is applied to two pseudoparticles.
However, we will show below that also in that case the central charge
works out alright, so we suspect that there is indeed a character identity
relating the two systems.

The action of the dual composite construction on the fugacities
and quantum number vectors is given by
\begin{align} \label{dcompfug}
  \cD_{12} \bx &= \begin{pmatrix}
  (\frac{x_1}{x_2})^{\frac{1}{\Delta}} \\
 (\frac{x_2}{x_1})^{\frac{1}{\Delta}} \\
  x_1^{\frac{\delta_2}{\Delta}} x_2^{\frac{\delta_1}{\Delta}} \\
  \bar x_i \bigl( \frac{x_1}{x_2} \bigr)^\frac{(\vec b_1 -
  \vec b_2)_i}{\Delta}  \end{pmatrix} \,, &
  \cD_{12} \bqqp &= \frac{1}{\Delta} \begin{pmatrix}
  \qqpe-\qqpt \\ \qqpt-\qqpe \\ \delta_2 \qqpe +\delta_1 \qqpt \\
  \Delta \bbqqp + (\bb_1 -\bb_2) (\qqpe-\qqpt) \end{pmatrix} \,.
\end{align}
Again, specifying to the situation where $x_1=x_2=x$ and
$\qqpe=\qqpt=\tilde{q}_{\text{qp}}$,
as holds in all the cases we consider, we find
\begin{align} \label{dcompfugsp}
\cD_{12} \bx &= \begin{pmatrix} 1\\1\\ x \\ \bbx \end{pmatrix} \ , &
\cD_{12} \bqqp &=
\begin{pmatrix} 0\\0\\ \tilde{q}_{\text{qp}} \\ \bbqqp \end{pmatrix}\ .
\end{align}
The solutions $\{\la_i\}$ to the IOW-equations
defined by $(\KKqp,\bx)$  and $\{\la_i'\}$ defined by
$(\KKqp',\bx') = (\cD_{ij}\KKqp,\cD_{ij}\bx)$ are, as was the case
for the composite construction (compare \eqref{compiow}), related in a
simple way
\begin{align} \label{dualcompiow}
\la_i' & \eql \frac{\la_i \la_j-1}{\la_j-1} \,, &
\la_j' & \eql \frac{\la_i \la_j-1}{\la_i-1} \,,\nonumber\\
\la_{n+1}' & \eql \la_i\la_j \,, &
\la_k' & \eql \la_k\,,\qquad(k\neq i,j,n+1) \,.
\end{align}
Using the relations \eqref{dualcompiow} it is not hard to show that
the IOW-equations based the two systems $(\KKqp,\bx)$ and
$(\cD_{ij}\KKqp, \cD_{ij} \bx)$ are in fact equivalent.
We also find that $\latot=\latot'$ by using the fact that the
particles $i$ and $j$ are pseudoparticles after the dual composite
construction has been applied. The composite particle which is
created is a physical (pseudo) particle if particles $i$ and $j$ are
physical (pseudo) in the original description.

From Eqn.~\eqref{dualcompiow} it follows that the dual composite
construction can not be applied on a physical and pseudoparticle.
In that case, $\latot'$ can not be made equal to $\latot$. Note that
such a restriction does not apply to the composite construction of
Section \ref{compcons}. Though we do not quite understand this
difference, it will not affect any results in this paper.

Let us now focus on the central charge, and look at the case in which
all the particles are physical particles first. Because the rank of the
transformed matrices is increased by one, we need that the two created
pseudoparticles reduce the central charge by one. This is easily verified.
Also, because the central charge of the matrix $\cC_{ij}\KKe$ equals
the central charge of $\KKe$, we need to find the result that the
central charge related to $\cD_{ij}\KKqp$ {\em without the pseudoparticle
subtraction} equals the central charge related to $\KKqp$ plus one.
To show this, we need to relate the solutions to the Eqns.~\eqref{sec2:4},
which we denote by $\xi_i$ and $\xi_i'$ for $\KKqp$ and $\cD_{ij} \KKqp$,
respectively. The relations are given by
\begin{align}
  \xi_i'&=\frac{\xi_i}{\xi_i+\xi_j - \xi_i \xi_j} \,, &
  \xi_j'&=\frac{\xi_j}{\xi_i+\xi_j - \xi_i \xi_j} \,, \nonumber\\
  \xi_{n+1}' &= \xi_i+\xi_j-\xi_i \xi_j \ , &
 \xi'_k &= \xi_k \,, \qquad (k\neq i,j,n+1) \,.
\end{align}
Because of the relation between the central charges, we require the
following dilogarithm identity
\begin{equation}
  L(x)+L(y) = L\left( \frac{x}{x+y-xy} \right) +
  L\left( \frac{y}{x+y-xy} \right) + L(x+y-xy) - L(1) \,,
\end{equation}
which is easily derived from Eqn.~\eqref{dilogvt} by applying
Eqn.~\eqref{dilogrel} to each term, and making the change of variables
$(x\mapsto 1-x, y\mapsto 1-y)$.

The argument above not only shows that the central charge works out
correctly in the absence of pseudoparticles. It can also be used to
show that the reduction of the central charge increases by one if we
apply the dual composite construction on pseudoparticles.

What remains to be checked is the central charge if we apply the
composite construction to physical particles, while pseudoparticles
are present.  For this, we need to compare the central charge
equations for the original pseudoparticles with the ones where the
additional two pseudoparticles are present. Though non-trivial, one
can convince oneself that the solutions to the central charge
equations of the original pseudoparticles do not change, while the
solutions for the two pseudoparticles which are introduced add up
to one and therefore increase the reduction by one, which gives the
correct result.

\subsection{P-transformations} \label{Ptrans}

In this section, we will discuss a transformation which is based on
the (dual) composite construction. This construction is very useful
in determining K-matrices for general affine Lie algebra CFTs. We
will motivate this construction by using a simple example, which
captures the essence of the method. In the end, this P-transformation
is very similar to the W-transformations described in Section
\ref{Wtrans}, with one
important difference. After applying a P-transformation, some of the
physical quasiparticles have transformed into pseudoparticles. One of the
consequences of this is a reduction of the central charge.

As we will use the P-transformations mainly as a tool to obtain
K-matrices for level-$k$ affine Lie algebras from the direct sum
of $k$ level-1 algebras, we will explain the construction using the
simplest case. Afterwards, we will present the general case.
In the next section, we will use the results obtained here to
find the K-matrices we are after.

\subsubsection{The case $\mathfrak{sl}(2)_2$} \label{mrcase}
The goal in this section is to obtain the K-matrices for the
$\mathfrak{sl}(2)_2$ affine CFT, which describes the Moore-Read
(or pfaffian) quantum Hall state. The corresponding matrices are
known, see for instance \cite{Scb,ABGS,ABS}. Let us recall the
K-matrices for the (bosonic) $\nu=1$ case, which corresponds
to $\mathfrak{sl}(2)_2$
\begin{align} \label{mrke}
  \KKe^{\mathfrak{sl}(2)_2} & = \begin{pmatrix}
  2 & 2 \\ 2 & 4 \\ \end{pmatrix} \ , &
  \bte & = -  \begin{pmatrix} 1 \\ 2 \end{pmatrix}\,.
\end{align}
The first particle can be identified with the (bosonic) electron,
while the second is a composite of two electrons.  In the
quasiparticle sector
\begin{align} \label{mrkp}
  \KKqp^{\mathfrak{sl}(2)_2} = \KKe^{-1} & = \begin{pmatrix}
 \hphantom{-}1 & -\frac{1}{2} \\ -\frac{1}{2} &
  \hphantom{-}\frac{1}{2} \\ \end{pmatrix} \,, &
  \btqp & = - \KKe^{-1}\cdot \bte =
  \begin{pmatrix} 0 \\ \frac{1}{2} \end{pmatrix}  \,,
\end{align}
where the first particle is a pseudoparticle. The K-matrices for the general
Moore-Read state, at filling fraction $\nu=\frac{1}{M+1}$, are
obtained by applying the so-called shift map, which is described in
detail in \cite{ABS}. Though the theory for general $M$ has the
same central charge, the theory does not have the underlying
$\mathfrak{sl}(2)_2$ structure anymore,
but rather a deformation (along the charge direction) of this.
In this paper, we concentrate on the $M=0$ case throughout;
the K-matrices for
general $M$ are obtained by applying the shift map as indicated above.
Note that the pseudoparticle matrices $\KK_{\psi\psi}$ are unchanged
under this shift map.

The main idea is now to obtain these $\mathfrak{sl}(2)_2$ matrices
via an embedding of $\mathfrak{sl}(2)_2$ in
$\mathfrak{sl}(2)_1 \oplus \mathfrak{sl}(2)_1$
(which we will call an abelian covering, see also \cite{CGT}).
By introducing a composite,
and projecting out some degrees of freedom, we obtain the K-matrices for
$\mathfrak{sl}(2)_2$. In physical terms, we start from two, uncoupled,
quantum Hall layers with filling $\nu=\frac{1}{2}$
(these are in fact bosonic Laughlin states). In a sense, this state is a
covering state for the Moore-Read state at filling $\nu=1$.
By increasing the interactions between the two layers, one might encounter
a phase transition to the Moore-Read state, as described in \cite{Ho}.
The bosons form pairs, and condense. In the terminology of an effective
Landau-Ginzburg theory, (see \cite{FNS}), the difference of the gauge fields
describing the bosons acquires a mass, and decouples from the spectrum.
This is the Meissner effect.

On the level of the K-matrices, we can describe this in the following
way. We first introduce the composite of the two bosonic particles,
and afterwards simply delete
(or `project out') one of the original bosons. So we actually
reduced the theory, as required.
We start with the direct sum of two $\mathfrak{sl}(2)_1$ K-matrices
\begin{align}
\label{mrcoverke}
\KKe^{\rm cover} &= \begin{pmatrix}
    2 & 0 \\ 0 & 2 \\ \end{pmatrix} \,, &
\bte &= - \begin{pmatrix}  1 \\ 1 \end{pmatrix} \,,
\end{align}
\begin{align} \label{mrcoverkp}
\KKqp^{\rm cover} & = \begin{pmatrix}
\frac{1}{2} & 0 \\ 0 & \frac{1}{2} \\ \end{pmatrix} \,, &
\btqp  & = \begin{pmatrix} \frac{1}{2} \\ \frac{1}{2} \end{pmatrix}  \,.
\end{align}
Now, applying the composite and dual composite constructions
(Eqns.~\eqref{eqDDb} and \eqref{dualcompact}) on
these matrices gives the following, equivalent description
\begin{align}
\label{mrcoverketr}
\widetilde{\KK}_{\text{e}} & = \begin{pmatrix}
2 & 1 & 2 \\ 1 & 2 & 2 \\
2 & 2 & 4 \\ \end{pmatrix} \ , &
\widetilde{\bt}_{\text{e}} &= - \begin{pmatrix} 1\\ 1 \\ 2 \end{pmatrix}\,,
\end{align}
\begin{align} \label{mrcoverkptr}
   \widetilde{\KK}_{\text{qp}} &= \begin{pmatrix}
  \hp1 & \hp0 & -\frac{1}{2} \\
  \hp0 & \hp1 & -\frac{1}{2} \\
  -\frac{1}{2} & -\frac{1}{2} & \hp\frac{3}{4} \\
  \end{pmatrix} \,, &
\widetilde{\bt}_{\text{qp}} & = \begin{pmatrix} 0 \\ 0 \\ \frac{1}{2}
  \end{pmatrix}\,.
\end{align}
Note that the first two particles of the quasiparticle matrix are
pseudoparticles. To obtain the $\mathfrak{sl}(2)_2$ matrices, we have
to project out one of these pseudoparticles, by putting it into the
vacuum state. In addition, we discard one of the original bosons.

However, while projecting out one of the bosons in the electron
sector simply corresponds to deleting the respective row and
column in $\wKKe$, projecting out one of the pseudoparticles is
more subtle, due to the negative coupling between the pseudoparticles
and the physical particle in $\wKKqp$.

For explicitness, consider the UCPF corresponding to $\wKKqp$ of
\eqref{mrcoverkptr}
\begin{equation} \label{eqPBCa}
  \sum  \frac{q^{\frac12 (m_1^2 + m_2^2 - (m_1+m_2)m_3 +
  \frac34 m_3^2 )}}{(q)_{m_3}}
  \qbin{\frac12 m_3}{m_1}\qbin{\frac12 m_3}{m_2} \,.
\end{equation}
Due to the minus-sign in the coupling between particles
$2$ and $3$ in $\wKKqp$, the vacuum state for particle $2$ is
not achieved for $m_2=0$, but rather for $m_2=\frac12 m_3$.
Hence, rather than just omitting particle $2$ from $\wKKqp$,
we need to set $m_2=\frac12 m_3$ in the bilinear form.  This results
in
\begin{multline}
  \bm^T \cdot \wKKqp \cdot \bm  =
  m_1^2 + m_2^2 - (m_1+m_2)m_3 +  \txtfrac{3}{4} m_3^2  \\ \to
  m_1^2 + (\txtfrac12 m_3)^2 - (m_1+ \txtfrac12 m_3)m_3 +
  \txtfrac{3}{4} m_3^2
  = m_1^2 - m_1m_3 + \txtfrac12 m_3^2 \,,
\end{multline}
which precisely corresponds to the matrix $\KKqp$ of \eqref{mrkp}.

To summarize, the results of projecting out degrees of freedom in
Eqns.~\eqref{mrcoverketr} and \eqref{mrcoverkptr},
gives rise to the K-matrices of
Eqns.~\eqref{mrke} and \eqref{mrkp}. One of the key points of this section
is that there is an elegant way of going from K-matrices for
the (abelian) coverings (Eqns. \eqref{mrcoverke} and \eqref{mrcoverkp})
to the K-matrices of $\mathfrak{sl}(2)_2$, by what we call a
`P-transformation'. This also hold for the general case, as we will
show below.  We find
\begin{align} \label{ptrans}
  \KKe^{\mathfrak{sl}(2)_2} & =
  \PP \cdot \KKe^{\rm cover} \cdot \PP^T\,, &
  \KKqp^{\mathfrak{sl}(2)_2} & =
  (\PP^{-1})^T \cdot \KKqp^{\rm cover} \cdot \PP^{-1} \ .
\end{align}
The vectors containing the quantum numbers (denoted by $\bq_{\rm e}$ and
$\bq_{\rm qp}$) transform as
\begin{align}\label{qvtrans}
  \tilde{\bq}_{\rm e} & = \PP \cdot \bq_{\rm e}\,, &
  \tilde{\bq}_{\rm qp} & = (\PP^{-1})^T \cdot \bq_{\rm qp} \ .
\end{align}
In the above, we have to take
$\PP=\bigl(\begin{smallmatrix}1&0\\1&1\end{smallmatrix}\bigr)$, and
hence
$(\PP^{-1})^T=\bigl(\begin{smallmatrix}1&-1\\0&\hp1\end{smallmatrix}\bigr)$.
A few remarks need to be made here. First of all, the P-transformation
described by Eqns.~\eqref{ptrans} and \eqref{qvtrans} closely
resembles the W-transformation, as they act on the K-matrices in
the same way (compare \eqref{wtrans}). However, there are a few
important differences.
As we explained above, upon applying a P-transformation, we
introduced a pseudoparticle in the quasiparticle sector. This is
important, as the presence of a pseudoparticle changes the theory.
For instance, the central charge is reduced, in the case at hand by
$\frac{1}{2}$, which is precisely the difference in central charge
between $\mathfrak{sl}(2)_1 \oplus \mathfrak{sl}(2)_1$ and
$\mathfrak{sl}(2)_2$ (given by $c=2$ and $c=\tfrac{3}{2}$ respectively).
So the P-transformation actually changes the
theory, while the W-transformation is a basis transformation,
which does not change the theory.

In the remainder of this section, we will show how a P-transformation
works on a general K-matrix. These results are used in the next
section to find the K-matrices for the general affine Lie algebra CFTs,
in a similar way as we constructed the $\mathfrak{sl}(2)_2$ matrices above.

\subsubsection{The general case}

In this section, we will relate the introduction of a composite
(in the electron sector), and the corresponding transformation
in the quasiparticle sector to a general P-transformation.
For notational simplicity, consider introducing a composite
of particles $1$ and $2$ in a general symmetric K-matrix
as given by Eqn.~\eqref{eqDDb}.
Now, suppose we delete particle $2$ from the
resulting matrix $\cC_{12} \KKe$, we then find a new K-matrix
system $(\wKKe,\wbqe)$ given by
\begin{equation}
  \wKKe = \begin{pmatrix}
  a_{11} & a_{11}+a_{12} & \ba_1^T \\
  a_{11}+a_{21} & a_{11}+2 a_{12} +a_{22} & \ba_1^T
   + \ba_2^T \\
  \ba_1 & \ba_1 + \ba_2 & \bKK \end{pmatrix} \,,\qquad
  \wbqe =
   \begin{pmatrix} q_1 \\ q_1+q_2 \\ \bbq \end{pmatrix} \,.
\end{equation}
Notice that we can write the relation
between $(\wKKe,\wbqe)$ and $(\KKe,\bqe)$ as
\begin{equation}
  \wKKe = \PP \cdot \KKe \cdot \PP^T\,,\qquad
  \wbqe = \PP \cdot \bqe \,,
\end{equation}
with
\begin{equation}  \label{eqPBCf}
\PP = \begin{pmatrix}
  1 & 0 &   \boldsymbol{0}^T \\
  1 & 1 &  \boldsymbol{0}^T   \\
  \boldsymbol{0}  & \boldsymbol{0}  & \II  \end{pmatrix} \,.
\end{equation}

Now consider the dual composite construction $\cD_{12} \KKqp$ (see
Eqn.~\eqref{dualcompact}).
In analogy with Eqn.~\eqref{eqPBCa}, putting the second
pseudoparticle in its vacuum state amounts to setting
\begin{equation}
  m_2 = - (\Delta-1) m_1 + \delta_2 m_3 - (\bb_2 - \bb_1)\cdot
  \overline{\bm} \,.
\end{equation}
Substituting this in the quadratic form yields,
after a lengthy calculation,
\begin{equation}
  \bm^T \cdot (\cD_{12}\KKqp) \cdot\bm
  \quad \to \quad  \bm^T \cdot \wKKqp \cdot\bm \,,
\end{equation}
where $\wKKqp$ is given by
\begin{equation} \label{eqAcd}
  \wKKqp = \begin{pmatrix}
  b_{11}-2b_{12}+b_{22} & b_{12}-b_{22} & \bb_1^T-\bb_2^T  \\
  b_{12}-b_{22} &  b_{22} & \bb_2^T        \\
  \bb_1 -\bb_2 & \bb_2 & \bKK
\end{pmatrix} \,,
\end{equation}
which is related to $\KKqp$ by
\begin{equation}
   \wKKqp = (\PP^{-1})^T \cdot {\KK}_{\text{qp}}
   \cdot \PP^{-1}\,,
\end{equation}
with
\begin{equation}
(\PP^{-1})^T = \begin{pmatrix}
  1 & -1 &   \boldsymbol{0}^T \\
  0 & \hp1 &  \boldsymbol{0}^T   \\
  \boldsymbol{0}  & \hp\boldsymbol{0}  & \II  \end{pmatrix} \,,
\end{equation}
in accordance with Eqn.~\eqref{eqPBCf}. It is important to note that
the first particle of $\wKKqp$ in Eqn.~\eqref{eqAcd} is a pseudoparticle.
The presence of this pseudoparticle causes the reduction of the central
charge of the system $\wKKe\oplus\wKKqp$ with respect to
$\KKe\oplus\KKqp$. Of course, this is to be expected when
degrees of freedom are projected out.

Summarizing, a P-transformation acts on the K-matrices and
quantum number vectors (denoted by $\bq_{\text{e}}$ and $\bq_{\text{qp}}$) as
follows
\begin{equation} \label{psum}
  \wKKe = \PP\cdot {\KK}_{\text{e}} \cdot \PP^T\,,\qquad
  \wKKqp = (\PP^{-1})^T \cdot {\KK}_{\text{qp}}
   \cdot \PP^{-1}\,,
\end{equation}
and
\begin{equation}\label{pvsum}
  \wbqe = \PP \cdot \bq_{\text{e}} \,, \qquad
  \wbqqp =  (\PP^{-1})^T \cdot \bq_{\text{qp}} \,,
\end{equation}
where in addition, some of the quasiparticles have been transformed into
pseudoparticles.

In Section \ref{construction} we will repeatedly use the
(dual) composite construction combined with the projecting
out of degrees of freedom to determine K-matrices for a variety
of CFTs.  Rather than specifying the particles to which we
consecutively apply
this construction we will simply state the required
resulting P-transformation, and specify which
quasiparticles have become pseudoparticles.

Because the P-transformations take the form \eqref{psum},
properties such as the filling fraction
(see \eqref{fillfracs}), are not changed upon performing the
P-transformation. Of course, the statistics properties are changed
in a profound way, because the induced pseudoparticles lead to
non-trivial fusion rules as described in Section \ref{pseudofus}.
In turn, this leads to the non-abelian statistics of the physical
quasiparticles (see, for instance, \cite{MR}).

One important remark needs to be made before closing this section.
In the construction of the K-matrices, we will use the (dual)
composite construction via the P-transformation. We will always apply
the dual composite construction
to identical (quasi) particles.  Hence, the quantum numbers of the
quasiparticles (and also their electronic equivalents) are the same.
Moreover, we will
always have $a_{ii} = a_{jj}$ and $b_{ii} = b_{jj}$. As a result, it
does not matter which of the electron-like particles is projected out.
If $a_{ii} \neq a_{jj}$, the two different projections are related by
$\PP'=\PP^T$. The general form for $\PP$ we use in this paper will be
discussed in the next section (see, in particular, Eqn.~\eqref{genp}).

\section{K-matrices for affine Lie algebras}
\label{kmatrices}

One of the main themes of this paper is the identification of the K-matrices
for general affine Lie algebra CFTs. We will work in the so-called
quantum Hall basis, as described above. In \cite{ABS} (see also \cite{ABGS}),
the K-matrices corresponding to the $\mathfrak{sl}(2)_k$ and
$\mathfrak{sl}(3)_k$ CFTs were derived. Here, we will give an alternative
construction of the $k>1$ cases directly from the $k=1$ cases,
which can be found in \cite{ABGS}.  This construction is based
on the embedding of the level-$k$ theory in the direct sum of $k$ level-1
theories. By applying composite and dual composite constructions, we
introduce pseudoparticles. After projecting out some of these, we have
reduce the theory to the level-$k$ theory. We will phrase all of this in
terms of the P-transformations of the previous section. Apart from the
$\mathfrak{sl}(2)_k$ and $\mathfrak{sl}(3)_k$ theories, we will also use
this construction for the other (simply-laced) affine Lie algebra cases,
and provide a few non-trivial checks to show that we indeed found
the correct K-matrices. The non simply-laced cases can be obtained by
embedding the level-1 affine algebras into simply-laced algebras,
and performing a similar construction as outlined above.

\subsection{Constructing the matrices}  \label{construction}

We will use the techniques described in the previous section to construct
the K-matrices for general affine Lie algebras.

In the this section, we will describe how
this works in detail for the simplest examples, which have all the
characteristics of the general case. Motivation of this construction
can be found in the previous section.
In Sections \ref{kel} and \ref{kqp} we will present the results
for the K-matrices for general affine Lie algebra CFTs.


\subsubsection{Example: the case $\mathfrak{sl}(2)_k$}
Let us illustrate the construction for the level $k>2$ generalizations
of the Moore-Read states, the so-called Read-Rezayi states \cite{RR}.
The covering state in this case is the direct sum of $k$ level-$1$ theories
(instead of just 2 for the MR case). So we have
\begin{align}
\label{sl2cover}
  \KK_{\rm e}^{\rm cover} &= \begin{pmatrix}
  2\\& 2 \\ & & \ddots \\& & & 2 \end{pmatrix} \, , &
\bte^{\rm cover} &= -\begin{pmatrix} 1 \\ 1 \\ \vdots \\ 1 \end{pmatrix} \,.
\end{align}
[Here, and in the following
we use the convention that `empty' entries contain zeroes, if not implied
otherwise by `dots'.] We also indicated the charge vector, containing
the charge quantum numbers of the particles, as the transformation behavior
of the quantum numbers under the P-transformation clearly shows that
composites are introduced.
To obtain the K-matrices for $\mathfrak{sl}(2)_k$, describing the
Read-Rezayi states, we need to introduce all types of composites,
from a pair up to a cluster made out of the $k$ original particles.
Thus $\PP$ takes the following form
\begin{equation}
\PP = \begin{pmatrix}
  1 \\
  1 & 1  \\
  \vdots & \ddots & \ddots \\
  1 & \cdots & 1 & 1\\
  \end{pmatrix} \,.
\end{equation}
This leads to following matrix $\KKe$ and charge vector $\bte$
(by using Eqns.~\eqref{psum} and \eqref{pvsum})
\begin{align}
  \KKe &=
  \begin{pmatrix}
  2 & 2 & 2 & \cdots & 2\\
  2 & 4 & 4 & & 4 \\
  2 & 4 & 6 & & 6 \\
  \vdots & & & \ddots & \vdots \\
  2 & 4 & 6 & \cdots & 2k \\
  \end{pmatrix} \,, &
  \bte &= -\begin{pmatrix} 1 \\ 2 \\ \vdots \\ k \end{pmatrix} \,,
\end{align}
which are indeed the correct for the $\mathfrak{sl}(2)_k$
theory. The dual sector is simply obtained by using the duality relations
\eqref{dual}, \eqref{dualvec}. Alternatively, we can apply the dual
P-transformation on the dual (i.e. the inverse) of the covering matrix
Eqn.~\eqref{sl2cover}. The corresponding P-matrix is
\begin{equation}
\label{pitsl2}
  (\PP^{-1})^T = \begin{pmatrix}
  1 & -1 \\
  & 1 & \ddots \\
  & & \ddots & -1 \\
  & & & 1\\
  \end{pmatrix} \,,
\end{equation}
from which we find
\begin{align}
  \KKqp &= \begin{pmatrix}
  1 & -\frac12 & & & \vdots\\
  -\frac12 & 1 & & & \vdots \\
  & & \ddots & -\frac12 & \vdots \\
  & & -\frac12 & 1 & \vdots & -\frac12 \\
  \hdotsfor{6} \\
  & & & -\frac12 & \vdots &  \frac12\\
  \end{pmatrix} \,, &
  \btqp &= \begin{pmatrix} 0 \\ \vdots \\ 0 \\ \frac12 \end{pmatrix} \,.
\end{align}
From the matrix Eqn.~\eqref{pitsl2} we read of that the first $k-1$ particles
are pseudoparticles. These results are in perfect agreement with the
results of \cite{GS,ABS}.

\subsubsection{Example: the case $\mathfrak{sl}(3)_k$}

As an example of a case where the rank $n$ of the affine Lie algebra
is greater than $1$,
we show that a similar construction can be carried out to obtain
the K-matrices related to the $\mathfrak{sl}(3)_k$ CFT.
This is the underlying theory of
the `non-abelian spin-singlet' quantum Hall states as defined in \cite{AS}.
Finding the K-matrices when the rank $n>1$ is somewhat more complicated
than for $n=1$. The K-matrices for the $\mathfrak{sl}(3)_k$ CFT were
obtained in \cite{ABS}. There, the basis was chosen in such a way
that all the particles in the electron sector had the same sign for
the charge. The reason for this choice was that the electron
operators (for spin up and spin down) appearing in the construction
of the quantum Hall state have the same sign of the charge.
These electron operators are associated to the roots $\alpha_1$ and
$-\alpha_2$ of $\mathfrak{sl}(3)$. From mathematical point of view,
it is more natural to work with $\alpha_1$ and $\alpha_2$, as the
resulting K-matrices have
a simpler structure. So here we will present the results using the
(mathematically) more natural formulation, based on the positive roots.
In Appendix \ref{relbases}, we will explain the precise relationship
between the two descriptions.
Essentially, the relation is a W-transformation on the physical particles,
which leaves the pseudoparticles unchanged. This is required, because the
pseudoparticles are related to the fusion rules of the affine Lie algebra
and they also determine the central charge.
The K-matrix for the electron sector at level 1 takes the form
in the representation chosen here
\begin{align}  \label{kesl31}
  \KKe &= \begin{pmatrix}
  \hp 2 & -1 \\ -1 & \hp 2\\
  \end{pmatrix} \,, & \bte &= \begin{pmatrix}\hp1 \\ -1 \end{pmatrix} \,, &
  \bse &= \begin{pmatrix} 1 \\ 1 \end{pmatrix}  \,.
\end{align}
In the other formulation, used in \cite{ABS}, the off-diagonal elements
of $\KKe$ are $1$, while the role of $\bte$ and $\bse$ is interchanged.

The K-matrix in Eqn.~\eqref{kesl31} is the building block of the covering
matrix, from which we construct the level-$k$ K-matrices
\begin{align} \label{ascover}
  \KKe^{\rm cover} &=
  \begin{pmatrix}
  \hp 2 & -1  \\
  -1 & \hp 2  \\
  & & \hp 2 & -1 \\
  & & -1 & \hp 2 \\
  & & & & \ddots \\
  & & & & & \hp 2 & -1\\
  & & & & & -1 & \hp 2\\
  \end{pmatrix} \,, &
  \bte^{\rm cover} &= \begin{pmatrix}
  \hp 1 \\ -1\\ \hp1 \\ -1 \\ \hp\vdots \\ \hp1 \\ -1
  \end{pmatrix} \,, &
  \bse^{\rm cover} &= \begin{pmatrix} 1 \\ 1 \\ 1\\ 1\\ \vdots \\
  1 \\ 1 \end{pmatrix} \,.
\end{align}
At this point, we need to specify the matrix $\PP$, which is used
to project to the K-matrix for the $\mathfrak{sl}(3)_k$
theory. However, because we have $n=2$ in this case, we can construct
the composites (up to order $k$) in different ways.
We will first state the form which gives the correct result, and comment
on the other possibilities afterwards.
The P-transformation which gives the correct central charge is given
by
\begin{equation}  \label{psl3}
  \PP=\begin{pmatrix}
  \II_2 \\
  \II_2 & \II_2 \\
  \vdots & \ddots & \ddots \\
  \II_2 & \cdots & \II_2 & \II_2 \\
  \end{pmatrix} \,,
\end{equation}
where $\II_2$ is the $2\times2$ identity matrix.
The resulting K-matrix has the following form (explicit forms of
the Cartan matrix $\AA_2$ of $A_2$ and the symmetrized Cartan
matrix $\MM_k^{-1}$ of $B_k$ can be found in Appendix~\ref{cartan})
\begin{equation}
\label{kesl3}
  \KK_e = \AA_2 \otimes \MM_k = \begin{pmatrix}
  \hp2 & -1 & \hp2 & -1 & \cdots & \hp2 & -1 & \hp2 & -1\\
  -1 & \hp2 & -1 & \hp2 & \cdots & -1 & \hp2 & -1 & \hp2 \\
  \hp2 & -1 & \hp4 & -2 & & \hp4 & -2 & \hp4 & -2 \\
  -1 & \hp2 & -2 & \hp4 & & -2 & \hp4 & -2 & \hp4 \\
  \vdots & \vdots & & & \ddots & & & \vdots & \vdots \\
  \hp2 & -1 & \hp4 & -2 & & 2(k-1) & -(k-1) & 2(k-1) & -(k-1) \\
  -1 & \hp2 & -2 & \hp4 & & -(k-1) & 2(k-1) & -(k-1) & 2(k-1) \\
  \hp2 & -1 & \hp4 & -2 & \cdots & 2(k-1) & -(k-1) & 2k & -k \\
  -1 & \hp2 & -2 & \hp4 & \cdots & -(k-1) & 2(k-1) & -k & 2k \\
  \end{pmatrix} \,,
\end{equation}
while the charge and spin quantum numbers are given by
\begin{align}
\bte &= \begin{pmatrix}
 \hp 1 \\ -1\\ \hp2 \\ -2 \\ \hp\vdots \\ \hp k \\ -k
 \end{pmatrix} \,, &
\bse &= \begin{pmatrix} 1 \\ 1 \\ 2\\ 2\\ \vdots \\
 k \\ k \end{pmatrix} \,.
\end{align}
It is not to hard to see that introducing the composites can be done
in different ways.
For instance, we could move some of the $1$'s in the lower-triangular
part of the matrix $\PP$ of Eqn.~\eqref{psl3}
to the corresponding place in the upper-triangular
part. If done systematically, we still would introduce all the
composites, so the resulting quantum numbers would be the same.
Luckily, all the essentially different possibilities result in
different K-matrices, which have different central charge associated to
them. So we can pick the, presumably, correct description by looking at
the central charge and perform further checks to assure the validity of
the chosen matrices. In all the cases we encountered, only one
P-transformation gave rise to a rational central charge (as far as the
numerical checks could tell), which indeed was the central charge corresponding
to the affine Lie algebra CFT. We refer to Section~\ref{kqp} for more
details on the checks of the central charge associated to the K-matrices.
Whether or not the other possibilities correspond to (non-rational) CFTs
is not clear at the moment.

The K-matrices and quantum numbers for the quasiparticle sector are
obtained similarly as in the $\mathfrak{sl}(2)_k$ case, by applying
the dual P-transformation to the dual of the covering. Now, the
transformation matrix becomes the inverse transpose of Eqn.~\eqref{psl3}
\begin{equation}
  (\PP^{-1})^T =\begin{pmatrix}
  \II_2 & -\II_2 \\
   & \hp\II_2 & \ddots \\
   & & \ddots & -\II_2 \\
   &  &  & \hp\II_2 \\
  \end{pmatrix} \,,
\end{equation}
with the results
\begin{align} \label{kqpsl3}
  \KKqp &= \AA_2^{-1} \otimes \MM_k^{-1} = \begin{pmatrix}
 \AA^{-1}_{2} \otimes \AA_{k-1} &
  \begin{matrix}
  \vdots \\ \vdots  \\ \vdots & -\AA^{-1}_2 \\
  \end{matrix} \\
  \hdotsfor{2} \\
  \begin{matrix}
  \hphantom{\cdots} & \hphantom{\cdots} &  -\AA_2^{-1}
  \end{matrix} &
  \begin{matrix}\vdots & \hp \AA_2^{-1}\end{matrix}
  \end{pmatrix} \,, &
  \btqp &=
  \begin{pmatrix} \hp 0 \\ \hp \vdots \\ \hp 0 \\ -\frac13 \\ \hp\frac13
  \end{pmatrix} \,, &
  \bsqp &=
  \begin{pmatrix} \hp 0 \\ \hp \vdots \\ \hp 0 \\ -1 \\ -1 \end{pmatrix} \,.
\end{align}
The K-matrix is to be compared with the matrix (7.23) in \cite{ABS}.
Note that part of the K-matrix corresponding to the $2(k-1)$ pseudoparticles
is the same in both cases. So,
because we know the two descriptions are related (see Appendix~\ref{relbases}),
we can say that by using the method of the P-transformations, we were able
to obtain correct K-matrices for the $\mathfrak{sl}(3)_k$ theory. One
important check is the central charge. Because the pseudoparticles are the
same in both formulations, the central charge is also equal.
In Section \ref{kqp}, the quasiparticle matrices for all simple
affine Lie algebra CFTs will be given. The electron matrices are
specified in Section \ref{kel}. Before we come to that, we will first
describe in detail how to construct the general K-matrices, using the
P-transformations and suitable coverings.

\subsubsection{The general case}
\label{gencase}

Using the knowledge obtained in the previous section, we go on and
propose a scheme to obtain the K-matrices for general affine Lie
algebra CFTs. We will first concentrate on the simply-laced cases,
and discuss the non simply-laced cases afterwards.
As we discussed the case of $\mathfrak{sl}(3)$, which has all the essential
ingredients, in detail in the previous section, we will be brief here.
We saw that in
the case of $\mathfrak{sl}(3)_1$, we could use the particles related to
the simple roots as the basis of the electron sector. Simple roots are
roots which can not be written as a sum of two positive roots. A Lie
algebra of rank $n$ has $n$ simple roots, and their scalar products define
the Cartan matrix. So we found that the K-matrix for the electron
sector of $\mathfrak{sl}(3)_1$  was the Cartan matrix. In the following,
we will assume that this is the case for all the simply-laced affine Lie
algebras. What we need to do further to obtain the level-$k$ K-matrices is
construct the covering theory, which is just the direct sum of $k$ level-1
theories, and apply the correct P-transformation. The form of the
P-transformation is similar to the $\mathfrak{sl}(3)$ case, where the
rank is the only thing which needs to be changed. So we find $\PP$
for the simply-laced cases
\begin{align}  \label{genp}
  \PP &= \begin{pmatrix}
  \II_n \\
  \II_n & \II_n \\
  \vdots & \ddots & \ddots \\
  \II_n & \cdots & \II_n & \II_n \\
  \end{pmatrix} \,, &
  (\PP^{-1})^T & =  \begin{pmatrix}
  \II_n & -\II_n\\
  & \hp\II_n & \ddots \\
  & &\ddots & -\II_n \\
  &  &  & \hp\II_n \\
  \end{pmatrix} \,.
\end{align}
Applied to the covering matrix we find the result
$\KKe = \PP \cdot (\AA_n \otimes \II_k) \cdot
\PP^T = \AA_n \otimes \MM_k$.  See Section \ref{kel}
for an explicit example. Of course, $\AA_n$ can be replaced by the
Cartan matrix of any other simply-laced algebra, $\DD_n$ or $\EE_n$.
The K-matrix for the quasiparticle sector is obtained by applying
$(\PP^{-1})^T$ to the  dual covering $\AA_n^{-1}\otimes\II_k$, resulting in
$\KKqp=\AA_n^{-1} \otimes \MM_k^{-1}$. From the form of $(\PP^{-1})^T$ we
find that the first $n(k-1)$ particles are in fact pseudoparticles.
These matrices will be given
explicitly in Section \ref{kqp}. For now, we note that the central charge
associated to these systems does indeed have the correct value. More on this
can be found in Section \ref{kqp}.

Let us now focus our attention to the non simply-laced case.
The idea is to apply the same construction as for the simply-laced cases.
However, we need to find the correct starting point, that is, the
level $k=1$ formulation. The non simply-laced affine algebras have non-trivial
fusion rules already at level-1, so we already need pseudoparticles
at level-1. This is also reflected in the central charge, which is
non-integer. To find the K-matrices, we embed the non simply-laced algebra in a
simply-laced one, and basically do the same construction before: project out
some degrees of freedom by introducing pseudoparticles.
As an example,
we quote the case for $\mathfrak{so}(5)_1$, which is related to the
spin-charge separated quantum Hall states of \cite{ALLS}
(see also \cite{BCR,BSa}). There, the
K-matrices for the $\mathfrak{so}(5)_1$ were obtained from the
$\mathfrak{so}(6)_1$ K-matrices using the construction outlined above.
It turns out that in general, the matrices for the non-simply laced
affine Lie algebras are equal to the (simply-laced)
affine Lie algebra in which they are embedded. The difference is
the presence of pseudoparticles in the non-simply laced cases,
as described above.  Alternative descriptions are possible, e.g.,
for $G_{2,k}$  we have an alternative
description (which is used in the connection with the corresponding
parafermion CFT), where the $k=1$ K-matrix has a couple of sign changes
in comparison to the Cartan matrix of the algebra used for the embedding,
see Appendix~\ref{g2app}.

To check that we indeed found the correct matrices, we will provide
another way to obtain the K-matrix for non simply-laced CFTs at level one.
This time, we directly use the exclusion statistics parameters of
the electron-like operators, corresponding to the root lattice of the
algebra. It is important to know the exclusion statistics of the corresponding
parafermions (which are part of the electron operators, see Section
\ref{pfsec} and also \cite{Gep}),
but we can borrow results from the literature here. We
will show how this works for the case $\mathfrak{so}(5)_1$ in
Appendix \ref{so5app}, while $G_2$ at level-1 can be found in
Appendix \ref{g2app}. The other non simply-laced cases can be obtained
in a similar way.

Having identified the $k=1$ K-matrices for the non simply-laced algebras,
we can go on, and take the direct sum of $k$ of the level-1 matrices,
and do exactly the same P-transformations as in the simply-laced case.
Because the covering matrices for the non simply-laced cases are identical
to the ones used for the corresponding simply-laced cases, the resulting
K-matrices will be identical as well.
The only difference is the number of pseudoparticles, as there will be
more pseudoparticles in the non simply-laced case. So, specifying the
nature of the particles is the only way to tell the difference between
the two. It is important to note that in the P-transformation,
(dual) composites are made only out of identical particles. We never
have the situation where a physical particle is paired with a
pseudoparticle, in accordance with the results of Section~\ref{dualcomp}.

\subsection{The matrices $\boldsymbol{\KKe}$}
\label{kel}

The building blocks of all the K-matrices are the Cartan matrices
$\AA_n$, $\DD_n$, $\EE_n$ and their inverses. In addition, we need
the symmetrized Cartan matrix of $B_n$, which we denote by
$\MM_k$, and its inverse. All these matrices can be found explicitly
in Appendix~\ref{cartan}.

From Section~\ref{gencase}, we have the results that for the simply-laced
cases $A_{n,k}$, $D_{n,k}$ and $E_{n,k}$ the matrices $\KKe$
take the form $\AA_n \otimes \MM_k$, $\DD_n \otimes \MM_k$ and
$\EE_n \otimes \MM_k$, respectively.
As an example, we will give the result for $D_{4,2}$ explicitly
\begin{equation}
  \KKe = \DD_4 \otimes \MM_2 =
  \begin{pmatrix}
  \hp2 & -1 & \hp0 & \hp0 & \vdots & \hp2 & -1 & \hp0 & \hp0 \\
  -1 & \hp2 & -1 & -1 & \vdots & -1 & \hp2 & -1 & -1 \\
  \hp0 & -1 & \hp2 & \hp0 & \vdots & \hp0 & -1 & \hp2 & \hp0 \\
  \hp0 & -1 & \hp0 & \hp2 & \vdots & \hp0 & -1 & \hp0 & \hp2 \\
  \hdotsfor{9}\\
  \hp2 & -1 & \hp0 & \hp0 & \vdots & \hp4 & -2 & \hp0 & \hp0 \\
  -1 & \hp2 & -1 & -1 & \vdots & -2 & \hp4 & -2 & -2 \\
  \hp0 & -1 & \hp2 & \hp0 & \vdots & \hp0 & -2 & \hp4 & \hp0 \\
  \hp0 & -1 & \hp0 & \hp2 & \vdots & \hp0 & -2 & \hp0 & \hp4 \\
  \end{pmatrix} \,.
\end{equation}
For the non-simply laced cases, we have to take the Cartan matrix
corresponding to affine Lie algebra which we used for the embedding.
We find that the matrices $\KKe$ are
$\DD_{n+1}\otimes\MM_k$,
$\AA_{2n-1}\otimes\MM_k$, $\EE_{6}\otimes\MM_k$
and $\DD_{4}\otimes\MM_k$ for $B_{n,k}$, $C_{n,k}$, $F_{4,k}$ and
$G_{2,k}$, respectively.

\subsection{The matrices $\boldsymbol{\KKqp}$}
\label{kqp}

The matrices $\KK_{\rm qp}$ can be obtained from $\KK_{\rm e}$ by a simple
inversion (see \eqref{dual}). In the following, we will explicitly
give these matrices, and indicate which particles are in fact the
pseudoparticles. With this knowledge, one can calculate the central
charge corresponding to $\KKe\oplus\KKqp$ by using Eqn.~\eqref{eqPBad}.
As this is hard to do analytically in general, we determined the central
charge numerically for some low values of $(n,k)$. All the cases up to
rank $n=10$  have been checked up to level $k=20$. We found that
the central charge corresponding to the matrices was equal to the central
charge of the CFTs up to $10^{-20}$ or better. The central charge of an
affine Lie algebra CFT is given by (cf.\ \eqref{eqcala})
\begin{equation} \label{ccala}
c_{\rm ALA} = \frac{k\  \text{dim} X_n}{k+\dCo} \ ,
\end{equation}
where $\text{dim} X_n$ is the dimension and $\dCo$ the dual Coxeter
number of the Lie algebra $X_n$. Both can be found in Appendix~\ref{cartan}
for every simple Lie algebra.

In the following, we will denote the $i$-th column of the matrix
$M$ by $(M)_i$. Recall that the quasiparticle matrices are of the
form (see Eqn.~\eqref{qpform})
\begin{align}
  \KK_{\rm qp} &=
  \begin{pmatrix}
  \KK_{\psi\psi} & \vdots & \KK_{\psi\phi} \\
  \hdotsfor{3} \\
  \KK_{\phi\psi} & \vdots & \KK_{\phi\phi}
  \end{pmatrix} \ , & \KK_{\psi\psi}^T &= \KK_{\psi\psi}\,, &
  \KK_{\phi\phi}^T &= \KK_{\phi\phi}\,, & \KK_{\psi\phi}^T &=
  \KK_{\phi\psi}\,,
\end{align}
where $\psi$ denotes the pseudoparticles, and $\phi$ the physical
quasiparticles.

\subsubsection{The case $A_{n,k}$.}  \label{sec:A}

The quasiparticle matrix $\KK_{\rm qp}$ for $\mathfrak{sl}(n+1)_k$
is given by
\begin{equation}
  \KKqp = \AA^{-1}_n \otimes \MM^{-1}_k = \begin{pmatrix}
  \AA^{-1}_{n} \otimes \AA_{k-1} &
  \begin{matrix}
  \vdots \\ \vdots  \\ \vdots & -\AA^{-1}_n \\
  \end{matrix} \\
  \hdotsfor{2} \\
  \begin{matrix}
  \hphantom{\cdots} & \hphantom{\cdots} &  -\AA_n^{-1}
  \end{matrix} &
  \begin{matrix}\vdots & \hp \AA_n^{-1}\end{matrix}
  \end{pmatrix} \,.
\end{equation}
In particular, the pseudoparticle matrix is given by
$\KK_{\psi\psi} = \AA_n^{-1}\otimes \AA_{k-1}$.

\subsubsection{The case $B_{n,k}$} \label{secbnk}

As already pointed out, we need to an embedding
to obtain the $B_{n,1}$ description first. This is done
for $\mathfrak{so}(5)$ in Appendix B, where we used $D_{3,1}$ for the
embedding. In general, we need $D_{n+1,1}$.
We find that we need one extra pseudoparticle, which corresponds to the
first node of the Dynkin diagram of $D_{n+1}$.
This extra particle has exclusion statistics parameter $1$, which
gives a reduction of the central charge by $\frac{1}{2}$, which is indeed the
difference of the central charge of the theories $D_{n+1,1}$ and
$B_{n,1}$. At general level we find that
$\KKqp = \DD^{-1}_{n+1} \otimes \MM^{-1}_k$,
which is characterized by
\begin{equation}  \label{pseudoso}
  \KK_{\psi\psi} = \begin{pmatrix}
  2 \DD_{n+1}^{-1} & -\DD_{n+1}^{-1}  \\
  -\DD_{n+1}^{-1} & 2\DD_{n+1}^{-1} & \ddots \\
  & \ddots & \ddots & -\DD_{n+1}^{-1}  \\
  & & -\DD_{n+1}^{-1} & 2 \DD_{n+1}^{-1} & -(\DD_{n+1}^{-1})_1 \\
  & & & -(\DD_{n+1}^{-1})^T_1 & 1\\
  \end{pmatrix} \,,
\end{equation}
where we see explicitly that there is an extra pseudoparticle
next to the $\DD_{n+1}^{-1}\otimes\AA_{k-1}$ part.

Accordingly, the matrix $\KK_{\phi\phi}$ is the inverse Cartan matrix
of $D_{n+1}$, with the first row and column omitted
(denoted by $\left. \DD_{n+1}^{-1}\right|_{\not 1}$)
\begin{equation}
  \KK_{\phi\phi} = \left. \DD_{n+1}^{-1}\right|_{\not 1} =
  \begin{pmatrix}
  2 & 2 & & 2 & 1 & 1 \\
  2 & 3 &  & 3 & \frac{3}{2} & \frac{3}{2} \\
  \vdots & & \ddots & \vdots & & \vdots \\
  2 & 3 & \cdots & n-1 & \frac{n-1}{2} & \frac{n-1}{2} \\
  1 & \frac{3}{2} & & \frac{n-1}{2}
  & \frac{n+1}{4} & \frac{n-1}{4} \\
  1 & \frac{3}{2} & \cdots & \frac{n-1}{2}
  & \frac{n-1}{4} & \frac{n+1}{4} \\
  \end{pmatrix} \,.
\end{equation}
Finally, we have
\begin{equation}
\KK_{\psi\phi} = \begin{pmatrix}
\boldsymbol{0} & \cdots & \cdots & \cdots & \boldsymbol{0} \\
-(\DD_{n+1}^{-1})_2 & \cdots & \cdots & \cdots &
-(\DD_{n+1}^{-1})_{n+1} \\
-1 & \cdots & -1 & -\frac{1}{2} & -\frac{1}{2}\\
\end{pmatrix} \ ,
\end{equation}
where $\boldsymbol{0}$ stands for the (column) vector with
all zeroes (of `length' $(n+1)(k-2)$ in this case).
Putting the parts together, we find
\begin{equation}
\KKqp = \begin{pmatrix} \KK_{\psi\psi} & \KK_{\psi\phi} \\
\KK_{\psi\phi}^T & \KK_{\phi\phi} \end{pmatrix}
= \DD_{n+1}^{-1} \otimes \MM_{k}^{-1} \ .
\end{equation}
To put emphasis on the fact that the pseudoparticle matrix is
bigger that the one for the $D_{n+1}$ CFT, we gave the matrices
$\KK_{\psi\psi}$ etc. explicitly, as we will do for all non
simply-laced cases.

\subsubsection{The case $C_{n,k}$}

In this case, we need $A_{2n-1,k}$ as the theory for the embedding.
For level $k=1$ we need the particles corresponding to the {\em even} nodes
to be pseudoparticles. These will be the extra pseudoparticles for
$k>1$, giving $n-1$ extra pseudoparticles. We again will specify the
matrix $\KKqp$ by its parts $\KK_{\psi\psi}$, etc.
\begin{equation}
  \KK_{\psi\psi} = \begin{pmatrix}
  2 \AA_{2n-1}^{-1} & -\AA_{2n-1}^{-1}  \\
  -\AA_{2n-1}^{-1} & 2\AA_{2n-1}^{-1} & \ddots \\
  & \ddots & \ddots & -\AA_{2n-1}^{-1}  \\
  & & -\AA_{2n-1}^{-1} & 2 \AA_{2n-1}^{-1} &
  \begin{matrix} -(\AA_{2n-1}^{-1})_2
  & \cdots & -(\AA_{2n-1}^{-1})_{2n-2}\end{matrix} \\
  & & &
  \begin{matrix}-(\AA_{2n-1}^{-1})^T_{2\hphantom{n-2}} \\ \vdots  \\
   -(\AA_{2n-1}^{-1})^T_{2n-2}\end{matrix} & 2 \AA^{-1}_{n-1} \\
  \end{pmatrix} \,,
\end{equation}
\begin{equation}
  \KK_{\psi\phi} = \begin{pmatrix}
  \begin{matrix} \boldsymbol{0} & \boldsymbol{0} & \cdots & \boldsymbol{0} \\
  -(\AA_{2n-1}^{-1})_1 &  -(\AA_{2n-1}^{-1})_3 & \cdots &
  -(\AA_{2n-1}^{-1})_{2n-1} \end{matrix} \\
  \KK_{\psi_{e}\phi}
  \end{pmatrix} \,,
\end{equation}
The matrix $\KK_{\psi_{e}\phi}$, which contains the coupling between
the physical particles and the {\em extra} pseudoparticles,
is described most easily by specifying its entries explicitly.
Let us first recall the elements of the inverse Cartan matrix
of $A_{2n-1}$ (compare with Appendix~\ref{cartan})
\begin{equation}
  (\AA_{2n-1}^{-1})_{i,j} =
  \min(i,j)-\frac{ij}{2n} \,,
  \qquad i,j=1,\ldots,2n-1 \,.
\end{equation}
Then we have
\begin{equation}
  (\KK_{\psi_{e}\phi})_{i,j} = \min(2i,2j-1)-\frac{2i(2j-1)}{2n} \ ,
  \qquad \begin{matrix} i=1,2,\ldots,n-1\,,\\j=1,2,\ldots,n\,.\\
  \end{matrix}
\end{equation}
For the matrix $\KK_{\phi\phi}$ we have
\begin{equation}
\begin{matrix}
  (\KK_{\phi\phi})_{i,j} = \min(2i-1,2j-1) -\frac{(2i-1)(2j-1)}{2n} \ ,
  \qquad i,j=1,2,\ldots,n \,.
\end{matrix}
\end{equation}
Note that the elements of the matrix describing the extra pseudoparticles
$(\KK_{\psi_{e}\psi_{e}})_{i,j}=\min(2i,2j)-\frac{(2i)(2j)}{2n}$, where
$i,j=1,\ldots,n-1$ is indeed equal to $2\AA_{n-1}^{-1}$.

\subsubsection{The case $D_{n,k}$.}

As we already used the matrix corresponding to $D_{n+1,k}$ in the
case of $B_{n,k}$, we will be brief here.
\begin{equation}
\KKqp = \DD^{-1}_n \otimes \MM^{-1}_k = \begin{pmatrix}
\DD^{-1}_{n} \otimes \AA_{k-1} &
\begin{matrix}
\vdots \\ \vdots  \\ \vdots & -\DD^{-1}_n \\
\end{matrix} \\
\hdotsfor{2} \\
\begin{matrix}
\hphantom{\cdots} & \hphantom{\cdots} &  -\DD_n^{-1}
\end{matrix} &
\begin{matrix}\vdots & \hp \DD_n^{-1}\end{matrix}
\end{pmatrix} \ .
\end{equation}
So we have $n (k-1)$ pseudoparticles, and $n$ physical ones.

\subsubsection{The cases $E_{n,k}$ with $n=6,7,8$.}

For $E_{n,k}$, we simply have a similar result as for
the other simply-laced cases.
\begin{equation}
\KKqp = \EE^{-1}_n \otimes \MM^{-1}_k = \begin{pmatrix}
\EE^{-1}_{n} \otimes \AA_{k-1} &
\begin{matrix}
\vdots \\ \vdots  \\ \vdots & -\EE^{-1}_n \\
\end{matrix} \\
\hdotsfor{2} \\
\begin{matrix}
\hphantom{\cdots} & \hphantom{\cdots} &  -\EE_n^{-1}
\end{matrix} &
\begin{matrix}\vdots & \hp \EE_n^{-1}\end{matrix}
\end{pmatrix} \ ,
\end{equation}
so the $n(k-1)$ pseudoparticles couple via $\EE_n\otimes\AA_{k-1}$.

\subsubsection{$F_{4,k}$} \label{secf4}

The embedding used this time is based upon $E_{6,k}$.
Now we expect to have two extra pseudoparticles, based on the level
$1$ case
(cf.\ \eqref{eqPBBp}, Section \ref{exampl}),
which turns out to be true. The couplings of these extra
pseudoparticles are related to the nodes $1$ and $5$ (see Appendix
\ref{cartan}). For general $k$, we have the pseudoparticle matrix
\begin{equation}
  \KK_{\psi\psi} = \begin{pmatrix}
  2 \EE_6^{-1} & -\EE_6^{-1}  \\
  -\EE_6^{-1} & 2\EE_6^{-1} & \ddots \\
  & \ddots & \ddots & -\EE_6^{-1}  \\
  & & -\EE_6^{-1} & 2 \EE_6^{-1} & -(\EE_6^{-1})_1 & -(\EE_6^{-1})_5 \\
  & & & -(\EE_6^{-1})^T_1 & \frac{4}{3} & \frac{2}{3} \\
  & & & -(\EE_6^{-1})^T_5 & \frac{2}{3} & \frac{4}{3} \\
  \end{pmatrix} \,,
\end{equation}
while the physical particles have
\begin{equation}
  \KK_{\phi\phi} = \begin{pmatrix}
  \frac{10}{3} & 4 & \frac{8}{3} & 2\\
  4 & 6 & 4 & 3 \\
  \frac{8}{3} & 4 & \frac{10}{3} & 2\\
  2 & 3 & 2 & 2\\
  \end{pmatrix} \,.
\end{equation}
Physical and pseudoparticles are coupled via
\begin{equation}
  \KK_{\psi\phi} = \begin{pmatrix}
  \boldsymbol{0} & \boldsymbol{0} & \boldsymbol{0} & \boldsymbol{0} \\
  -(\EE_6^{-1})_2 & -(\EE_6^{-1})_3 & -(\EE_6^{-1})_4 & -(\EE_6^{-1})_6 \\
  \frac{5}{3} & 2 & \frac{4}{3} & 1 \\
  \frac{4}{3} & 2 & \frac{5}{3} & 1 \\
  \end{pmatrix} \,.
\end{equation}
Again, if we combine the physical and extra pseudoparticles in the
right way, we find the matrix $\EE_6^{-1}$.

\subsubsection{$G_{2,k}$}
\label{g2oneqp}

Finally we come to the last case, which is $G_{2,k}$. This case is special
in the sense that if we use a similar procedure as we used in all
the other cases, we find a description in which the number of physical
particles does not equal the rank of the algebra, as was the situation
in the other cases. This will have consequences as we consider the related
parafermions in Section \ref{pfsec}.
In Appendix \ref{g2app} we will provide a
different description of $G_{2,k}$, which does have two physical particles.
For now, we will just use the description based on the K-matrices for
$D_{4,k}$, in which we embed $G_{2,k}$. It turns out that we need
three extra pseudoparticles, leaving only one physical particle.
Note that the coupling of the extra pseudoparticles is given by
Eqn.~\eqref{eqPBBq} in Section \ref{exampl}.
\begin{equation}
  \KK_{\psi\psi} = \begin{pmatrix}
  2 \DD_4^{-1} & -\DD_4^{-1}  \\
  -\DD_4^{-1} & 2\DD_4^{-1} & \ddots \\
  & \ddots & \ddots & -\DD_4^{-1}  \\
  & & -\DD_4^{-1} & 2 \DD_4^{-1} &
  -(\DD_4^{-1})_1 & -(\DD_4^{-1})_3 & -(\DD_4^{-1})_4 \\
  & & & -(\DD_4^{-1})^T_1 & 1 & \frac{1}{2} & \frac{1}{2}\\
  & & & -(\DD_4^{-1})^T_3 & \frac{1}{2} & 1 & \frac{1}{2} \\
  & & & -(\DD_4^{-1})^T_4 & \frac{1}{2} & \frac{1}{2} & 1 \\
  \end{pmatrix} \,,
\end{equation}
\begin{equation}
  \KK_{\phi\phi} = \begin{pmatrix} 2 \end{pmatrix}  \,,
\end{equation}
\begin{equation}
  \KK_{\psi\phi} = \begin{pmatrix}
  \boldsymbol{0} \\
  -(\DD_4^{-1})_2 \\
  1 \\ 1 \\ 1 \\
  \end{pmatrix} \,.
\end{equation}

\section{K-matrices for coset conformal field theories}
\label{cosets}

Having identified the K-matrices for the affine Lie algebra CFTs, one
might hope to find K-matrices for more general CFTs. An obvious class
to look at are the coset conformal field theories, as most
CFTs can be written in a coset form. In this section, we will provide
K-matrices for a class of coset CFTs. In our search for the K-matrices
for coset CFTs, we will be mainly guided by the central charge.
We can test our results by comparing to known coset K-matrices.
For diagonal cosets of simply-laced affine Lie algebras, the results
of the K-matrices are due to McCoy and co-workers. See, for
instance, \cite{BM}.

Having obtained a scheme, we will apply it to the cosets
$\mathfrak{so}(2n)_k/\mathfrak{so}(2n-1)_k$ with $k=1,2$,
where the latter is the non-trivial
one. The parafermionic cosets are dealt with in Section \ref{pfsec},
as they require a different approach. This already shows that the
scheme we found is by no means unique, but useful anyway.

\subsection{Diagonal cosets}
\label{diacos}

As said, the central charge is an important quantity to keep in mind
in determining the K-matrices for the cosets. Let us take a look at the
general coset $G/H$, where $H \subset G$ is maximal. Let us assume that
both $G$ and $H$ are of the form $\KKe\oplus\KKqp$, with equal rank $n$.
Also, both quasiparticle matrices can contain pseudoparticles.
So the central charge of these theories (denoted by $c(G)$ and $c(H)$)
is given by
\begin{align}
c (G) &= n - c(\KK_{\psi\psi}(G)) \ , &
c (H) &= n - c(\KK_{\psi\psi}(H)) \ ,
\end{align}
where $c(\KK_{\psi\psi}(G))$ denotes the central charge corresponding
to the pseudoparticle matrix of $G$. Let us further assume that all the
pseudoparticles which appear in $\KK_{\psi\psi}(G)$ also appear in
$\KK_{\psi\psi}(H)$. This restricts the applicability of the construction,
but still covers a large class of cosets. Now the argument of the
central charge suggests to take the pseudoparticle K-matrix of $H$, and
change the pseudoparticles which do {\em not} appear in the pseudoparticle
matrix of $G$ into physical particles. The central charge corresponding
to this matrix is $c(\KK_{\psi\psi}(H))-c(\KK_{\psi\psi}(G))$. This indeed
equals the central charge of the coset theory, which is given by
$c(G)-c(H)$. Note that the matrix we propose for the coset theory is not
of the form $\KK\oplus\KK^{-1}$. This is in fact consistent with
known results for K-matrices of coset conformal
field theories, as we will discuss below.
This construction does work for the cosets of the type
$X_{n,k} \oplus X_{n,l} / X_{n,k+l}$, where $X_n$ is a
simply-laced Lie algebra.
Indeed, using this, we reproduce the results of McCoy for these
diagonal cosets, see for instance \cite{BM}.

The construction above is in fact more generally
applicable as we will show in
the next subsection, where we will show a non-trivial example based on the
coset of $\mathfrak{so}(2n)_k/\mathfrak{so}(2n-1)_k$.

\subsection{$\mathfrak{so}(2n)_k/\mathfrak{so}(2n-1)_k$}

Applying the construction above to the coset
$\mathfrak{so}(2n)_k/\mathfrak{so}(2n-1)_k$
at level $k=1$, we find the K-matrix $\KK=(1)$, which is obviously the
correct result for this $c=\frac12$ CFT. Another coset with $c \leq 1$ is
the case $k=2$, which has $c=1$. We find the following K-matrix
\begin{equation} \label{socoset}
  \KK = \begin{pmatrix}
  1 & \vdots & -1 &  \ldots & -1 & -\frac12 & -\frac12 \\
  \hdotsfor{7} \\
  -1 & \vdots & \\
     & \vdots & \\
  -1 & \vdots &   & & 2 \DD_n^{-1} & \\
  -\frac12 & \vdots &  \\
  -\frac12& \vdots &
  \end{pmatrix} \,,
\end{equation}
where only the first particle is physical.  As mentioned, this matrix yields
the correct central charge $c=1$ by construction.
That it indeed describes the correct $c=1$ CFT can be seen as follows.
Applying the dual composite construction to
\begin{equation} \label{soc1}
  \KK = \begin{pmatrix}
  1 - \frac{1}{2n} & \frac{1}{2n}    \\
  \frac{1}{2n}     & 1 - \frac{1}{2n} \end{pmatrix} \,,
\end{equation}
where both particles are physical, we find
\begin{equation} \label{soc2}
  \cD_{12} \KK = \begin{pmatrix}
  1 & -\frac{1}{2} & -\frac{1}{2} \\
  -\frac{1}{2} & \frac{n}{2} & 1 - \frac{n}{2}   \\
  -\frac{1}{2} &1 - \frac{n}{2}   & \frac{n}{2} \end{pmatrix} \,.
\end{equation}
Now applying the composite construction to the two pseudoparticles
in \eqref{soc2} $(n-2)$ times we find \eqref{socoset}.
On the other hand, the UCPF based on \eqref{soc1}, summed over
$m_1+m_2 \equiv 0 \ \mod \, 2n$, equals the $c=1$ $\mathfrak{u}(1)$-character
\begin{equation}
  \frac{1}{(q)_\infty} \sum_{k\in \mathbb Z} \ q^{n(2n-1)k^2} \,,
\end{equation}
by using the Durfee square identity (see, e.g., \cite{Andr})
\begin{equation}
\frac{1}{(q)_\infty} = \sum_{m \geq 0} \frac{q^{m^2}}{(q)_m(q)_m} \ .
\end{equation}
So we indeed find that the matrix
\eqref{socoset} describes a $c=1$ conformal field theory, namely
the free boson compactified on a circle.

In addition to this non-trivial example, also the equivalence
used in the theory of $G_2$-holonomy -- namely between
$\mathfrak{so}(7)_1/G_{2,1}$ and the tricritical Ising model --
works, if we take the $G_2$ (level $k=1$)
description of Appendix \ref{g2app}. We find the K-matrix
\begin{equation}
  \KK = \begin{pmatrix} \hp 1 & -\frac{1}{2} \\ -\frac{1}{2} &
  \hp 1 \end{pmatrix} \,,
\end{equation}
with one physical and one pseudoparticle. This is indeed the K-matrix
corresponding to the minimal model with $c=\frac{7}{10}$.

\subsection{Parafermions}  \label{pfsec}

Generalized parafermionic conformal field theories were defined by
Gepner \cite{Gep} as a generalization of the $\ZZ_k$ parafermions
of \cite{ZF}. The generalized parafermion theories can be viewed as
cosets based on general affine Lie algebras (ALA's) and
$\mathfrak{u}(1)$ theories
\begin{equation}  \label{pfcoset}
  X^{\rm pf}_{n,k} = \frac{X_{n,k}}{\mathfrak{u}(1)^n} \ ,
\end{equation}
where $n$ is the rank of the Lie algebra $X_{n}$, and $k$ the level.
The central charge of the parafermion CFT \eqref{pfcoset} is given by
\begin{equation}
\label{pfcc}
  c_{\rm pf} = c_{\rm ALA} -n \, ,
\end{equation}
where $c_{\rm ALA}$ is the central charge of the corresponding affine
Lie algebra theory (see Eqn.~\eqref{eqcala}).
The parafermion cosets \eqref{pfcoset} are somewhat different
in comparison to the diagonal cosets of Section \ref{diacos},
and need to be treated differently. Before we come to the
discussion of the K-matrices, we first fix some notations
concerning the parafermion fields, following \cite{Gep}.

The primary fields of the theory $\Phi_\lambda^\Lambda$ are labeled by a
(highest) weight $\Lambda$ and a charge $\lambda$, which is also an
element of the weight lattice, and is defined modulo $k \mathcal{M}_L$,
i.e. $k$ times the long root lattice.
To obtain a complete, independent set of parafermion fields, one
has to impose the following restrictions. The charge $\lambda$ must be
`accessible' from $\Lambda$ by subtracting roots (including $\alpha_0$)
from $\Lambda$. Furthermore, the (proper) external automorphisms
$\sigma$ (see \cite{dFMS}) of the affine Lie algebra give rise to field
identifications
\begin{equation}
  \Phi^\Lambda_\lambda \equiv \Phi^{\sigma(\Lambda)}_{\lambda+\sigma(0)} \,,
\end{equation}
where $\sigma(0)$ denotes the image of the affine weight $k\Lambda_0$
under $\sigma$.

An important check on the K-matrices for the parafermionic
CFTs is based on the relation between the parafermionic
partition functions and the string functions $c^\Lambda_\lambda$
of the corresponding affine Lie algebras \cite{Gep}
\begin{equation}
\label{partstring}
  Z_{\rm pf}^{\Lambda,\lambda}= (\eta)^n c_\lambda^\Lambda \,,
\end{equation}
where $\eta=q^\frac{1}{24}\prod_{k=1}^\infty (1-q^k)$ is the
Dedekind function.  As an example, we will express
the partition function
$Z_{\rm pf}^{\Lambda,\lambda}$ with $\Lambda =(0,\ldots,0) \equiv \id$
in terms of UCPFs based on the K-matrices
for the parafermion CFTs. Using Eqn.~\eqref{partstring},
we can check our results against the known (tabulated) string
functions.

We will use the matrices $\KK_e$ of the corresponding affine Lie
algebras as a starting point for obtaining the parafermionic
matrices $\KK^{\rm pf}$.
The matrices $\KK_e$ correspond to the (elementary) electron-like
particles and composites (up to order $k$) of these elementary
particles. The operators corresponding to these (elementary)
particles have the form
\begin{equation}
  \Phi^\id_\lambda :e^{i \boldsymbol{\alpha} \cdot
 \boldsymbol{\varphi}}: \,,
\end{equation}
where $\boldsymbol{\varphi}=(\varphi_1,\dots\varphi_n)$ is a set of
bosonic fields, which correspond to the $\mathfrak{u}(1)$ degrees of
freedom  and determine the quantum numbers of the particles via the
constants $\alpha_i$.
For the order $k$ composites, the parafermion fields are
trivial, i.e.\ $\Phi^\id_{k\mu} = \id$, for $\mu\in \mathcal M_L$
($\mu$ a long root), in which case only the vertex operator part
remains.

In this section, we are interested in the K-matrices
for the parafermionic CFTs. These can be obtained
from the matrices $\KK_e$ of the corresponding affine Lie algebra
theories by subtracting from the particles which have a non-trivial
parafermion field $\Phi^\id_\lambda$ the part of the exclusion statistics
which corresponds to the vertex operator
$:e^{i \boldsymbol{\alpha} \cdot \boldsymbol{\varphi}}:$.
This can be done `by hand' by calculating the exclusion statistics
of the vertex operators. Actually, because there are
always particles which do not have a parafermion field (or equivalently,
a trivial parafermion field), this can be done by applying what we
will call an X-transformation. Such a transformation is like a
W-transformation. However, the matrices associated to an X-transformation
are not $SL(r,\ZZ)$ matrices, but rather $SL(r,\QQ)$. This is because
the quantum numbers of the largest composites (which are the particles
with trivial parafermion fields) are $k$ times the quantum numbers
of the particles in the $k=1$ formulation. In general, the non-zero
non-diagonal entries take the form $\frac{l}{k}$, with $l=1,\ldots,k-1$.
Explicitly, in the case of the
$\ZZ_k=\tfrac{\mathfrak{sl}(2)_k}{\mathfrak{u}(1)}$ parafermions we find
the following
\begin{equation}  \label{xforpf}
  \XX =
  \begin{pmatrix}
  1 &&&& -\frac{1}{k} \strt{-\frac{1}{k}}\\
  & 1 &&& -\frac{2}{k} \\
  && \ddots && \vdots \\
  &&& 1 & -\frac{k-1}{k} \strt{-\frac{k-1}{k}}\\
  &&&& \hp 1 \\
  \end{pmatrix} \,.
\end{equation}
For more general parafermions, the matrices are (a little) more complicated.
In fact, each entry of the matrix \eqref{xforpf} becomes an $n \times n$
matrix. Although fractions appear in
$\XX$, the quantum numbers of the particles after the transformation
are still integers, because the largest composite is of order $k$.
More precisely, the X-transformation is such that all the
quantum numbers of the transformed particles are in fact zero;
in a sense all the vertex operators containing the chiral boson fields are
stripped of from the parafermionic fields.
The transformed matrix $\KKe$ splits in two pieces, namely a
part containing the order $k$ composites and the part
corresponding to the parafermions $\Phi^\id_\lambda$, which is
the matrix we are looking for. We will denote this matrix by
$\KK^{\rm pf}$. In the quasiparticle sector, the pseudoparticles will
completely decouple from the physical quasiparticles and hence
the transformed matrix is of the form
$\KK_{\psi\psi} \oplus \widetilde{\KK}_{\phi\phi}$, where
$\widetilde{\KK}_{\phi\phi}$ is a deformed quasiparticle
matrix. So we conjecture that the K-matrices for parafermionic
CFTs are given by the inverse of the pseudoparticle matrix
$\KK_{\psi\psi}$, of the corresponding affine Lie algebra CFT
\begin{equation} \label{pfcon}
  \KK^{\rm pf} = \KK_{\psi\psi}^{-1} \, .
\end{equation}
A first check on the proposed matrices is the corresponding
central charge. The central charge corresponding to the
matrices $\KK_{\psi\psi}$ is given by
\begin{equation}
c_{\psi\psi} = (n+p)k - c_{\rm ALA} \ ,
\end{equation}
where $p$ is the difference in rank between the affine algebra
under consideration and the one used to `build' the K-matrices
(thus for simply-laced algebras, $p=0$).
The rank of the matrix $\KK_{\psi\psi}$ is $(k-1)(n+p)+e$, where
$e$ is the number of `extra' pseudoparticles needed for the non
simply-laced algebras. Thus we have the following result for the
central charge of matrices $\KK^{\rm pf}$
\begin{equation}
c_{\rm pf} = c_{\rm ALA} - n - (p-e) \, .
\end{equation}
For all the affine algebras, except $G_{2,k}$, the K-matrices of
Section~\ref{kqp} have $p=e$, so we obtain the correct result of
Eqn.~\eqref{pfcc}.
However, we also find that the construction above does not work for the
description of $G_{2,k}$ as given in Section \ref{g2oneqp},
because there the number of physical quasiparticles is $1$ instead
of $2$, which is the rank of $G_2$.
Luckily, there exists another way to represent the $G_{2,k}$ affine Lie
algebra, which does have two physical quasiparticles. The inverse
of the pseudoparticle matrix therefore has the correct central charge.
The corresponding K-matrices can be found in Appendix~\ref{g2app}.
It has been checked that this $G_{2,k}$ parafermion
K-matrix does give rise to the corresponding string functions (for
$k=2,3$).

\subsubsection{The case $\mathfrak{so}(5)_2$ as an example}

As an example, we will discuss the characters of the
parafermionic theory associated to $\mathfrak{so}(5)_2$.

The conjectured pseudoparticle $\KK_{\psi\psi}$ for
$\mathfrak{so}(5)_2$ is given
by Eqn.~\eqref{pseudoso} with $n=2 \, , \, k=2$
\begin{equation} \label{eqso52ps}
  \KK_{\psi\psi} =
  \begin{pmatrix}  (\DD_3^{-1})_{11} & & - (\DD_3^{-1})^T_{1} \\
   &  &  \\
  - (\DD_3^{-1})_{1} & & 2 \DD_3^{-1} \end{pmatrix} =
  \begin{pmatrix}
  \hp1 & -1 & -\frac{1}{2} & -\frac{1}{2} \\
  -1 & \hp2 & \hp1 & \hp1 \\
  -\frac{1}{2} & \hp1 & \hp\frac{3}{2} & \hp\strt{\frac{1}{2}} \frac{1}{2} \\
  -\frac{1}{2} & \hp1 & \hp\frac{1}{2} & \hp\frac{3}{2} \\
\end{pmatrix} \,.
\end{equation}
The K-matrix which is supposed to describe the
$\mathfrak{so}(5)_2$
parafermions is simply the inverse of the pseudoparticle
matrix, where it is assumed that all particles are physical
\begin{equation}
\KK^{\rm pf} =
  \begin{pmatrix}
  2 & \hp1 & \hp0 & \hp0 \\
  1 & \hp\frac{3}{2} & -\frac{1}{2} & -\frac{1}{2} \\
  0 & -\frac{1}{2} & \hp1 & \hp0 \\
  0 & -\frac{1}{2} & \hp0 & \hp1 \\
  \end{pmatrix} \,.
\end{equation}
The UCPF based on this K-matrix, namely
\begin{equation} \label{pfchar}
  Z_{\rm pf}^{\Lambda=\id} = \sum
  \frac{q^{\frac{1}{2} \bm \cdot \KK^{\rm pf} \cdot \bm }}
  {\prod_i (q)_{m_i} } \ ,
\end{equation}
with $\bm$ a $4$-dimensional vector, is the sum
over string-functions
\begin{equation}
  Z_{\rm pf}^{\Lambda=\id} = \sum_{\lambda}
  (\eta)^l c_\lambda^\Lambda \,.
\end{equation}
The sum over $\lambda$ runs over the independent parafermion fields
$\Phi^{(0,0)}_{(\lambda_1,\lambda_2)}$
(where we assume that the first root is the short root).
The various string-functions $c^{(0,0)}_{(\lambda_1,\lambda_2)}$
are obtained by restricting the sum in Eqn.~\eqref{pfchar}.
Explicitly, we have
\begin{equation}
  c^{(0,0)}_\lambda =  \frac{q^{-\frac{1}{12}}}{(q)_\infty^2}
  \sum_{res(\lambda)}
  \frac{q^{\frac{1}{2} \bm \cdot \KK^{\rm pf} \cdot \bm }}
  {\prod_i (q)_{m_i}} \,,
\end{equation}
where
\begin{equation}
  res(\lambda) = \begin{cases}
  \begin{matrix}
  2 m_1 + m_2 + 2 m_3 &= 0 \bmod 4 \\ m_3+m_4 &= 0 \bmod 2
  \end{matrix} & \text{for $\lambda = (0,0)$} \\
  \begin{matrix}
  2 m_1 + m_2 + 2 m_3 &= 0 \bmod 4 \\ m_3+m_4 &= 1 \bmod 2
  \end{matrix} & \text{for $\lambda = (2,0)$} \\
  \begin{matrix}
  2 m_1 + m_2 + 2 m_3 &= 2 \bmod 4 \\ m_3+m_4 &= 0 \bmod 2
  \end{matrix} & \text{for $\lambda = (0,2)$} \\
  \begin{matrix}
  2 m_1 + m_2 + 2 m_3 &= 1 \bmod 4 \\ m_3+m_4 &= 0 \bmod 2
  \end{matrix} & \text{for $\lambda = (0,1)$}
  \end{cases}
\end{equation}
The string functions $c^{(2,0)}_{(\lambda_1,\lambda_2)}$ can be obtained by
using a shift vector; more explicitly, by changing the power of $q$ in
Eqn.~\eqref{pfchar} to
$\frac{1}{2} \bm \cdot \KK^{\rm pf} \cdot \tilde{\bm}$,
where $\tilde{\bm}=(m_1-1,m_2,m_3,m_4)$.
We have not yet found similar expressions for the other (independent)
string functions, such as
$c^{(0,1)}_{(\lambda_1,\lambda_2)}$ and $c^{(1,0)}_{(\lambda_1,\lambda_2)}$.

\subsubsection{Cases checked}

The cases for which we checked that the conjectured matrices do give
the string functions $c^\id_{\bf \lambda}$ include all the affine
Lie algebras up to rank $n=3$ and level $k=2$.
In addition, we also checked
$\mathfrak{so}(5)_3$, $\mathfrak{so}(8)_2$, $E_{6,2}$, $E_{7,2}$,
$E_{8,2}$ and $F_{4,2}$.
The checks were performed by numerically calculating the
partition functions up to a certain order
in $q$, depending on the dimension of the K-matrix. These results
were compared to the weight-multiplicity tables of
Kass {\it et al.} \cite{KMPS}.
Note that despite the fact that for the higher rank algebras the
checks were performed to rather low order in $q$, we believe that
the formulas hold to all orders in $q$.

As an example, we give the K-matrix associated to the $F_4$ parafermions
at level $k=2$.
\begin{equation} \label{f4k2}
  \KK^{\rm pf} (F_{4,2}) =
  \begin{pmatrix}
  \hp\frac{3}{2} & -\frac{1}{2} & \hp0 & \hp0 &
          -\frac{1}{4} & \hp0 & \hp1 & -\frac{1}{2} \\
  -\frac{1}{2} & \hp1 & -\frac{1}{2} & \hp0 & \hp0 & \hp0 & \hp0 & \hp0 \\
  \hp0 & -\frac{1}{2} & \hp1 & -\frac{1}{2} & \hp0 & -\frac{1}{2}
           & \hp0 & \hp0 \\
  \hp0 & \hp0 & -\frac{1}{2} & \hp1 & -\frac{1}{2} & \hp0 & \hp0 & \hp0 \\
  -\frac{1}{4} & \hp0 & \hp0 & -\frac{1}{2} & \hp\frac{3}{2} & \hp0
            & -\frac{1}{2} & \hp1 \\
  \hp0 & \hp0 & -\frac{1}{2} & \hp0 & \hp0 & \hp1 & \hp0 & \hp0 \\
  \hp1 & \hp0 & \hp0 & \hp0 & -\frac{1}{2} & \hp0 & \hp2 & -1 \\
  -\frac{1}{2} & \hp0 & \hp0 & \hp0 & \hp 1 & \hp0 & -1 & \hp2 \\
  \end{pmatrix} \,.
\end{equation}

Explicitly, the relation between the parafermionic character based on the
matrix in Eqn.~\eqref{f4k2}, namely
\begin{equation} \label{f4k2pf}
  \sum_{\lambda} Z^\id_\lambda =  \sum_{\{ m_i \} }
  \frac{q^{(\frac{1}{2} \bm \cdot \KK^{\rm pf} \cdot \bm )}}
  {\prod_i (q)_{m_i}} \,,
\end{equation}
and the string functions
is as follows. Upon splitting the character in pieces containing powers
of $q$ which differ by integers, one finds
\begin{align}
\sideset{}{'}\sum_{\{ m_i \} } \frac{q^{(\frac{1}{2}
\bm \cdot \KK \cdot \bm)}} {\prod_i (q)_{m_i}} &=
q^\frac{1}{6} \qinf^4 (c^{\id}_{(0,0,0,0)} + 3 c^\id_{(0,0,0,2)})
&& (q^n \ ; n \in \NN ) \\
\sideset{}{'}\sum_{\{ m_i \} } \frac{q^{(\frac{1}{2}
\bm \cdot \KK \cdot \bm)}} {\prod_i (q)_{m_i}} &=
12 q^\frac{1}{6} \qinf^4 c^\id_{(1,0,0,0)}
&& (q^{n+\frac{1}{2}} \ ; n \in \NN ) \\
\sideset{}{'}\sum_{\{ m_i \} } \frac{q^{(\frac{1}{2}
\bm \cdot \KK \cdot \bm)}} {\prod_i (q)_{m_i}} &=
24 q^\frac{1}{6} \qinf^4 c^\id_{(0,0,1,0)}
&& (q^{n+\frac{1}{4}} \ ; n \in \NN ) \\
\sideset{}{'}\sum_{\{ m_i \} } \frac{q^{(\frac{1}{2}
\bm \cdot \KK \cdot \bm)}} {\prod_i (q)_{m_i}} &=
24 q^\frac{1}{6} \qinf^4 c^\id_{(0,0,0,1)}
&& (q^{n+\frac{3}{4}} \ ; n \in \NN ) \ .
\end{align}
The primes on the sums denote the restriction to the powers of
$q$ as indicated. The various numerical constants for the string functions
$c^\id_{\lambda}$ are the number of independent fields of the form
$\Phi^\id_{\lambda '}$ which have the same conformal dimension as the
field $\Phi^\id_{\lambda}$.

\section{Application to level restricted Kostka polynomials} \label{secF}

In Section \ref{pseudofus} we have argued that there exists an intimate
relation between the fusion rules of a CFT and the pseudoparticle
K-matrix as both count paths on the fusion diagram.
In fact, there exists a natural $q$-deformation
of the number of fusion paths giving rise to the so-called
level truncated Kostka polynomial.  This deformation shows up
as part of the UCPF expression for the characters of WZW models,
as conjectured in Section \ref{secbab}.  One would thus expect that the
level truncated Kostka polynomials can be expressed as UCPFs
with the K-matrices found in this paper.

Concretely, if $\phi_i=\phi_{\Lambda_i},\ i=1,\ldots r$, denotes the field
corresponding to the $i$-th fundamental weight of $\fg$, the multiplicity
of the field $\phi_\la$ in the fusion rule
\begin{equation} \label{eqPBFa}
  \phi_1^{n_1} \times \ldots\times \phi_r^{n_r}
\end{equation}
is given by $(N_1^{n_1}\ldots N_r^{n_r})_0{}^\lambda$.
By associating a power of $q$ to each path, determined through
the crystal graph of $\fg$, we obtain a $q$-deformation of this
number.  This is referred to as the (dual) level-$k$
truncated Kostka polynomial (or truncated $q$ Clebsch-Gordan coefficient)
of $\fg$ and we will denote it by
$M^{(k)}_{\lambda\mu}(q)$ where $\mu=\sum_i n_i \Lambda_i$.
An explicit expression of $M^{(k)}_{\lambda\mu}(q)$ for $k\to\infty$
is known (see e.g. \cite{BSb} and references therein) and originates
in Bethe-Ansatz techniques \cite{KR}.
Explicit UCPF type expressions for finite $k$ are
known for $\fg = \mathfrak{sl}(n)$ (see \cite{SS} for the most general
result and also \cite{BMS,DKKMM,Ki,HKKOTY}) and
$\mathfrak{so}(5)_1$ \cite{BSb}.
In \cite{HKOTY}, UCPF type expressions for Kostka polynomials
for general (non-twisted) affine Lie algebras were conjectured.
Proofs for some of these conjectures and expressions for some twised
cases can be found in, for instance, \cite{SS} and \cite{OSS}.
The relation between the K-matrices used in these expressions and the
ones brought forward in this paper is not clear at the moment.
We are gratefull to Ole Warnaar for bringing these references to
our attention.

According to the UCPF conjecture, $M^{(k)}_{\lambda\mu}(q)$
should be closely related to
\begin{equation} \label{eqPBFb}
  q^{\frac{1}{2} \bn\cdot\KK_{\phi\phi}\cdot\bn -
  \frac{1}{2} \bn'\cdot\KK_{\phi\phi}\cdot\bn'} \sideset{}{'}\sum_{\bm}
  q^{\frac{1}{2} \bm\cdot \KK_{\psi\psi} \cdot\bm + \bn
  \cdot\KK_{\phi\psi}\cdot \bm} \\
  \times
  \prod_i \qbin{ ((\II -\KK_{\psi\psi})\cdot\bm)_i - (\KK_{\psi\phi}
  \cdot \bn)_i + \bu_i }{m_i}  \,,
\end{equation}
where $\lambda=\sum n_i' \Lambda_i$ and $\mu=\sum_i n_i\Lambda_i$.
[We have set $\bQ=0$ as we are only discussing paths starting at the
identity representation.]

In the simply-laced case, it has been conjectured before (see, e.g.,
\cite{BH} and references therein) that  $M^{(k)}_{\lambda\mu}(q)$
can indeed be written in terms of the UCPF based on $\KKqp=
\XX_n^{-1} \otimes \MM_k^{-1}$.  Here we will focus on a specific
non simply-laced example, namely $\mathfrak{so}(5)$ at levels $k=1,2$.
We defer a general investigation
to future work.  An explicit recipe for computing
$M^{(k)}_{\lambda\mu}(q)$ for $\fg=\mathfrak{so}(5)$, at level $1$,
was given in \cite{Yam}.  Explicit
formulae for the level $k=1$ case were given in
\cite{BSb}.  Concretely,
\begin{align} \label{eqPBso51}
  M^{(1)}_{(0,0),(n_1,n_2)}(q) & =
  q^{\frac12(n_1^2+n_1n_2)+\frac38n_2^2}
  \sum_{m_1} q^{\frac12(m_1^2 - m_1 n_2)} \qbin{\frac12n_2}{m_1}\,,\qquad
   n_1+\frac12n_2+m_1\ \text{even}, n_2\ \text{even}\,, \nonumber\\
  M^{(1)}_{(1,0),(n_1,n_2)}(q) & = q^{\frac12(n_1^2+n_1n_2)+
  \frac38n_2^2-\frac12}
  \sum_{m_1} q^{\frac12(m_1^2 - m_1 n_2)} \qbin{\frac12n_2}{m_1}\,,\qquad
   n_1+\frac12n_2+m_1\ \text{odd}, n_2\ \text{even} \,,\nonumber\\
  M^{(1)}_{(0,1),(n_1,n_2)}(q) & = q^{\frac12(n_1^2+n_1n_2)+
  \frac38n_2^2-\frac38}
  \sum_{m_1} q^{\frac12(m_1^2 - m_1 n_2)} \qbin{\frac12(n_2+1)}{m_1}\,,\qquad
  \frac12(n_2+1)+m_1\ \text{even}, n_2\ \text{odd} \,.
\end{align}
The above formulae are of the UCPF form with
\begin{equation}  \label{eqPBso51a}
  \KK = \begin{pmatrix}
   \hp1 & \vdots & 0 & -\txtfrac12 \\
   \hdotsfor{4} \\
   \hp0 & \vdots & 1 & \hp\txtfrac12 \\
   -\txtfrac12 & \vdots & \txtfrac12 & \hp\txtfrac34 \end{pmatrix} \,,
\end{equation}
which is to be compared to the $B_{2,1}$ quasiparticle K-matrix
of Section \ref{secbnk}, given by
\begin{equation}  \label{eqPBso51b}
  \KK = \begin{pmatrix}
   1 & \vdots & \txtfrac12  & \txtfrac12 \\
  \hdotsfor{4} \\
  \txtfrac12  & \vdots &  \txtfrac34 & \txtfrac14 \\
  \txtfrac12  & \vdots &   \txtfrac14 &  \txtfrac34  \end{pmatrix} \,.
\end{equation}
While the pseudoparticle part of Eqns.~\eqref{eqPBso51a} and
\eqref{eqPBso51b} agree, the K-matrices obviously differ in the
physical particle part.  Both K-matrices are reminiscent of
$\mathfrak{so}(6)$, but while \eqref{eqPBso51b} has physical particles
inherited from the $\mathbf 4$ and $\bar{\mathbf 4}$ of $\mathfrak{so}(6)$,
Eqn.~\eqref{eqPBso51a} contains physical particles inherited
from the $\mathbf 4$ and the $\mathbf 6$ of $\mathfrak{so}(6)$.
Since $\mathbf 6 = \mathbf 5 \oplus \mathbf 1$ under $\mathfrak{so}(5)$,
the matrix \eqref{eqPBso51a} does indeed seem to be better suited to
describe general (truncated) Kostka polynomials for $\mathfrak{so}(5)$,
although we expect that the $\mathfrak{so}(5)$ Kostka polynomials
can also be expressed in terms of a UCPF based on \eqref{eqPBso51b}.
Unfortunately, it seems that Eqn.~\eqref{eqPBso51a} does not have a
straightforward higher level generalization.

Therefore, motivated by the
decomposition of finite dimensional irreducible
representations $W_{(n_1,n_2,n_3)}$ of $\mathfrak{so}(6)$ into those of
$\mathfrak{so}(5)$ under the regular embedding
$\mathfrak{so}(5) \to \mathfrak{so}(6)$, i.e.\
\begin{equation} \label{eqPBFf}
   W_{(n_1,n_2,0)} \cong \bigoplus_{l=0}^{n_1} W_{(n_1-l,n_2)} \,,
\end{equation}
we introduce
\begin{equation} \label{eqPBlast}
  \widetilde M_{(0,0),(n_1,n_2)}^{(1)}(q) =
  \sum_{k=0}^{n_1} \qbin{n_1}{k} M_{(0,0),(n_1-k,n_2)}^{(1)}(q) \,.
\end{equation}
Inserting the expression for $M_{(0,0),(n_1-k,n_2)}^{(1)}(q)$, and
changing $k \to n_1 -k$ in the summation, we find
\begin{equation}
  \widetilde M_{(0,0),(n_1,n_2)}^{(1)}(q) =
  \sum_{k,l; k+l+n_2/2\ \text{even}}
  q^{ \frac12 (k^2+kn_2)+ \frac38 n_2^2 + \frac12 l^2 - \frac12 ln_2}
  \qbin{\frac12 n_2}{l}\qbin{n_1}{k} \,.
\end{equation}
Now, let $p=k-l$, then
\begin{equation}
  \widetilde M_{(0,0),(n_1,n_2)}^{(1)}(q) =
  \sum_p \sum_{k,l; k-l=p} q^{\frac12 p^2 + \frac12 pn_2 + \frac38 n_2^2}
  q^{kl}  \qbin{\frac12 n_2}{l}\qbin{n_1}{k} =
  \sum_p q^{\frac12 p^2 + \frac12 pn_2 + \frac38 n_2^2}
  \qbin{n_1+\frac12 n_2}{n_1-p} \,,
\end{equation}
where, in the last step, we have used a finite version of the Durfee
square formula (see \cite{Bo2}).
Finally, letting $p\to n_1-p$, we find
\begin{equation} \label{eqPBso51c}
  \widetilde M_{(0,0),(n_1,n_2)}^{(1)}(q) =
  q^{\frac12(n_1^2 + n_1n_2) + \frac38 n_2^2} \sum_p
  q^{\frac12 p^2 - p n_1 - \frac12 pn_2} \qbin{n_1+\frac12 n_2}{p} \,.
\end{equation}
A similar computation can be given for the other sectors
$M_{(n_1',n_2'),(n_1,n_2)}^{(1)}(q)$ of \eqref{eqPBso51}.
Now, Eqn.~\eqref{eqPBso51c} is of the UCPF form with
\begin{equation}\label{eqPBso51d}
  \KK = \begin{pmatrix}
   \hp1 & \vdots & -1 & -\txtfrac12 \\
   \hdotsfor{4} \\
   -1 & \vdots & 1 & \hp\txtfrac12 \\
   -\txtfrac12 & \vdots & \txtfrac12 & \hp\txtfrac34 \end{pmatrix} \,,
\end{equation}
which has the same $\KK_{\psi\psi}$ and $\KK_{\phi\phi}$ parts as
\eqref{eqPBso51a}, but differs in the coupling $\KK_{\psi\phi}$.

Now consider the $\mathfrak{so}(5)$, level $k=2$ case.
As an Ansatz we take the pseudoparticle matrix of Section
\ref{secbnk} (see also Eqn.~\eqref{eqso52ps}), and the
physical particles of Eqn.~\eqref{eqPBso51d}, and adjust
the coupling between them.  Specifically, let
\begin{equation} \label{eqPBFc}
\KK = \begin{pmatrix}
1 & \vdots    & -1   & -\txtfrac12   & -\txtfrac12    &
    \vdots & 0 & 0  \\
  \hdotsfor{8} \\
  -1 & \vdots & 2 & 1   & 1     & \vdots & -1   & -\txtfrac12 \\
  -\txtfrac12 & \vdots & 1 & \txtfrac32 &  \txtfrac12  & \vdots &
  -\txtfrac12 & -\txtfrac34 \\
  -\txtfrac12 & \vdots & 1 & \txtfrac12 & \txtfrac32   & \vdots & -\txtfrac12
  & -\txtfrac14 \\
  \hdotsfor{8} \\
  0    & \vdots & -1 & -\txtfrac12   & -\txtfrac12 & \vdots & 1 & \txtfrac12 \\
  0    & \vdots & -\txtfrac12 & -\txtfrac34 & -\txtfrac14 & \vdots &
  \txtfrac12 &\txtfrac34
  \end{pmatrix} \,.
\end{equation}
Note that this matrix is not invertible, as is the case for the matrix
in Eqn.~\eqref{eqPBso51d}.
Thus, Eqn.~\eqref{eqPBFb} reads explicitly
\begin{align} \label{eqPBFca}
\widetilde{M}^{(2)}_{(n_1',n_2'),(n_1,n_2)}(q)  & =
 q^{\frac12 (n_1^2+n_1 n_2)+
  \frac38 n_2^2 - \frac12 (n_1^{\prime2}+n'_1 n'_2)-
  \frac38 n_2^{\prime2}} \nonu \\
& \times \sideset{}{'}\sum_{{\bf m}}
q^{\frac14(2m_1^2+4m_2^2+3m_3^2+3m_4^2)-\frac12 m_1(2m_2+m_3+m_4)+
m_2(m_3+m_4)+\frac12 m_3m_4} \nonu \\
&  \times q^{- \frac12 n_1(2m_2+m_3+m_4)-
\frac14 n_2(2m_2+3m_3+m_4)} \nonu \\
& \times  \qbin{\frac12(2m_2+m_3+m_4)+u_1}{m_1}
\qbin{m_1-(m_2+m_3+m_4) + n_1+\frac12 n_2+u_2 }{m_2} \nonu \\ & \times
\qbin{\frac12 (m_1-(2m_2+m_3+m_4)) +\frac12 n_1+\frac 34 n_2+u_3}{m_3}
\nonu\\ & \times
\qbin{\frac12 (m_1-(2m_2+m_3+m_4)) +\frac12 n_1+\frac14 n_2+u_4 }{m_4} \,,
\end{align}
with some appropriate restriction on the summation over $(m_1,\ldots,m_4)$.

Numerical evidence suggests the following conjecture (cf.\ \eqref{eqPBlast})
\begin{equation} \label{eqPBFd}
  \widetilde M^{(2)}_{(n_1',n_2'),(n_1,n_2)}(q)  =
  \sum_{k=0}^{n_1} \qbin{n_1}{k} \sum_{l=0}^{n_1'}
  {M}^{(2)}_{(n_1'-l,n_2'),(n_1-k,n_2)}(q)  \,,
\end{equation}
or equivalently,
\begin{equation} \label{eqPBFe}
  M^{(2)}_{(n_1',n_2'),(n_1,n_2)}(q)  =
  \sum_{k=0}^{n_1} (-1)^k q^{\frac12 k(k-1)} \qbin{n_1}{k}
  \sum_{l=0}^{n_1'} (-1)^l
  \widetilde{M}^{(2)}_{(n_1'-l,n_2'),(n_1-k,n_2)}(q)\,,
\end{equation}
where the vectors $\bu$ in \eqref{eqPBFca}, for given $(n_1',n_2')$,
are given in Table \ref{tabFa}.\footnote{We have not been able
to find the $\bu$-vectors
corresponding to the remaining integrable highest weight modules
at level $2$, i.e.\ $(n_1',n_2')=(2,0), (1,1)$ and $(0,2)$.}
\begin{table}
\begin{center}
\begin{tabular}{|c|c|}
\hline
$(n_1',n_2')$ &  $(u_1;u_2,u_3,u_4)$ \\ \hline
$(0,0)$       &  $(0;0,0,0)$ \\
$(1,0)$       &  $(0;1,\txtfrac12,\txtfrac12)$ \\
$(0,1)$       &  $(0;\txtfrac12,\txtfrac14,\txtfrac34)$ \\ \hline
\end{tabular}\medskip
\caption{The vectors $\bn'$ and $\bu$ for the $\mathfrak{so}(5)_2$
Kostka polynomials} \label{tabFa}
\end{center}
\end{table}
The summation restrictions are such that
$2m_2+m_3+3m_4\equiv 2(n_1-n_1')+(n_2-n_2')\mod4$,
and $n_1+\frac{n_2}{2}+m_1\equiv n_1'+\frac{n_2'}{2} \mod2$.

Again, the conjectured formula \eqref{eqPBFd} is strongly reminiscent
of the decomposition of finite dimensional irreducible
representations \eqref{eqPBFf}.
This suggests that while the procedure of Section \ref{secbnk}
does produce a pseudoparticle K-matrix leading to the correct
central charge, it still overcounts the number of fusion paths.
This overcounting can also be seen by applying the analysis of
Section \ref{pseudofus}, as the pseudoparticle K-matrix does
not give rise to the same recursion relations as the $\mathfrak{so}(5)_2$
fusion rules.  For this reason we also expect that the $\mathfrak{so}(5)_2$
characters, when written in UCPF form using the K-matrices of Section
\ref{secbnk}, will need alternating sign corrections.

\section{Discussion}

In this paper, we proposed a scheme to obtain the K-matrices for the
CFTs with affine Lie algebra symmetry. This construction was based on
character identities, which were applied to certain abelian covering states.
After projecting out some degrees of freedom, the K-matrices were
obtained. Subsequently, these K-matrices were used to obtain the K-matrices
of coset CFTs. Also, they appeared in some expressions for the level-k
restricted Kostka polynomials.

It would be interesting to investigate if the K-matrices obtained here
indeed are the central objects in the Kostka polynomials related to a general
affine Lie algebra. An interesting open question is whether similar K-matrices
can be used for more general CFTs, such as the twisted affine Lie algebras
(and their parafermions), which were studied in \cite{DGZa} and \cite{DGZb}.
Another interesting class of theories which might be addressed in a similar
fashion are the affine Lie superalgebras and  the related parafermions
(see, for instance, \cite{CRS} and \cite{JM} for the case
$\mathfrak{osp}(1|2)$).

Most of our consistency checks on whether we obtained the
correct K-matrices were based on the fact that the central charge worked
out correctly.  Even though this proved to be an extremely restrictive
`guide', the ultimate verification of course relies on the construction
of the CFT characters in the UCPF form using these K-matrices.  While we
have proved this in special cases, and did numerical checks in others,
a complete verification requires tools beyond the scope of this
paper, and will require proving a host of new $q$-identities.
A systematic approach towards a full proof will undoubtedly benefit
from a better algebra-geometric understanding of the role of K-matrices
(see, e.g., \cite{BH,FKLMM,FJKLM,FGT} for some initial studies).

{\em Note added.}
In an earlier version of this paper we refered 
to the Kostka polynomials of Section 6 as ``generalized
Kostka polynomials" to indicate the generalization of
the standard $A_n$ Kostka polynomials to general simple Lie
algebras.  In order to avoid confusion with the ``generalized
Kostka polynomials", introduced independently by
Schilling and Warnaar \cite{SW} and by Kirillov and
Shimozono \cite{KS} (cf. \cite{OSS} for a discussion), which are more
general than the  Kostka polynomials which are the subject of this paper,
we will simply refer to the polynomials in this paper as (level restricted)
Kostka polynomials. We thank Ole Warnaar for communication on these points.

\begin{appendix}

\section{Cartan matrices and their inverses}
\label{cartan}

In this appendix, we will list of the Cartan matrices of the simple
Lie algebras, to clarify the conventions used in this paper. In addition,
we will give some other properties, namely the dimension and the
dual Coxeter number. Other properties can be found, for instance, in
\cite{dFMS}.

In the Cartan matrices, the empty entries correspond to zeros, unless
otherwise implied by the dots. Even though we only use matrices corresponding
to simply laced
Lie algebras, we will give the Cartan matrices of all the simple
Lie algebras, for completeness.  We will denote the
Cartan matrix corresponding to the Lie algebra $X_n$ by $\mathbb X_n$. \\

\noindent{$\boldsymbol{A_n:}$}
The Cartan matrix for $A_n$ is given by
\begin{equation}
  \AA_n = \begin{pmatrix}
  2 & -1 \\
  -1 & 2 & -1 \\
  & -1 & 2 \\
  & & & \ddots & -1 \\
  & & & -1 & 2 & -1 \\
  & & & & -1 & 2 \\
  \end{pmatrix}
\end{equation}
\begin{equation}
  \AA_n^{-1} = \txtfrac{1}{n+1}\begin{pmatrix}
  n & n-1 & n-2 & \dots & 2 & 1 \\
  n-1 & 2(n-1) & 2(n-2) & & 4 & 2 \\
  n-2 & 2(n-2) & 3(n-3) & & 6 & 3 \\
  \vdots & & & \ddots & & \vdots \\
  2 & 4 & 6 & & 2(n-1) & n-1 \\
  1 & 2 & 3 & \cdots & n-1 & n \\
  \end{pmatrix}
\end{equation}

\noindent{$\boldsymbol{B_n:}$}

\begin{align}
\BB_n &=
  \begin{pmatrix}
  2 & -1 \\
  -1 & 2 & -1 \\
  & -1 & 2 \\
  & & & \ddots & -1\\
  & & & -1 & 2 & -2 \\
  & & & & -1 & 2 \\
  \end{pmatrix}
  &
  \BB_n^{-1} &=
  \begin{pmatrix}
  1 & 1 & 1 & \cdots & 1 & 1 \\
  1 & 2 & 2 & & 2 & 2 \\
  1 & 2 & 3 & & 3 & 3 \\
  \vdots & & & \ddots & & \vdots \\
  1 & 2 & 3 & & n-1 & n-1 \\
  \frac{1}{2} & 1 & \frac{3}{2} & \cdots & \frac{n-1}{2} & \frac{n}{2} \\
  \end{pmatrix}
\end{align}

\noindent{$\boldsymbol{C_n:}$}

\begin{align}
  \CC_n &=
  \begin{pmatrix}
  2 & -1 \\
  -1 & 2 & -1 \\
  & -1 & 2 & \\
  & & & \ddots & -1 \\
  & & & -1 & 2 & -1 \\
  & & & & -2 & 2 \\
  \end{pmatrix} &
  \CC_n^{-1} & =
  \begin{pmatrix}
  1 & 1 & 1 & \cdots & 1 & \frac{1}{2} \\
  1 & 2 & 2 & & 2 & 1 \\
  1 & 2 & 3 & & 3 & \frac{3}{2} \\
  \vdots & & & \ddots & & \vdots \\
  1 & 2 & 3 & & n-1 & \frac{n-1}{2} \\
  1 & 2 & 3 & \cdots & n-1 & \frac{n}{2} \\
  \end{pmatrix}
\end{align}

\noindent{$\boldsymbol{D_n:}$}

\begin{align}
  \DD_n &=
  \begin{pmatrix}
  2 & -1 \\
  -1 & 2 & -1 \\
  & -1 & 2 \\
  & & & \ddots & -1\\
  & & & -1 & 2 & -1 & -1\\
  & & & & -1 & 2 & 0\\
  & & & & -1 & 0 & 2\\
  \end{pmatrix} &
  \DD_n^{-1} & =
  \begin{pmatrix}
  1 & 1 & 1 & \cdots & 1 & \frac{1}{2} & \frac{1}{2} \\
  1 & 2 & 2 & & 2 & 1 & 1 \\
  1 & 2 & 3 &  & 3 & \frac{3}{2} & \frac{3}{2} \\
  \vdots & & & \ddots & \vdots & & \vdots \\
  1 & 2 & 3 & \cdots & n-2 & \frac{n-2}{2} & \frac{n-2}{2} \\
  \frac{1}{2} & 1 & \frac{3}{2} & & \frac{n-2}{2}
  & \frac{n}{4} & \frac{n-2}{4} \\
  \frac{1}{2} & 1 & \frac{3}{2} & \cdots & \frac{n-2}{2}
  & \frac{n-2}{4} & \frac{n}{4} \\
  \end{pmatrix}
\end{align}

\noindent{$\boldsymbol{E_6:}$}

\begin{align}
  \EE_6 &= \begin{pmatrix}
  2 & -1 & 0 & 0 & 0 & 0\\
  -1 & 2 & -1 & 0 & 0 & 0\\
  0 & -1 & 2 & -1 & 0 & -1\\
  0 & 0 & -1 & 2 & -1 & 0\\
  0 & 0 & 0 & -1 & 2 & 0\\
  0 & 0 & -1 & 0 & 0 & 2 \\
  \end{pmatrix} &
  \EE_6^{-1} &= \txtfrac{1}{3} \begin{pmatrix}
  4 & 5 & 6 & 4 & 2 & 3 \\
  5 & 10 & 12 & 8 & 4 & 6 \\
  6 & 12 & 18 & 12 & 6 & 9 \\
  4 & 8 & 12 & 10 & 5 & 6 \\
  2 & 4 & 6 & 5 & 4 & 3 \\
  3 & 6 & 9 & 6 & 3 & 6 \\
  \end{pmatrix}
\end{align}

\noindent{$\boldsymbol{E_7:}$}

\begin{align}
  \EE_7 &= \begin{pmatrix}
  2 & -1 & 0 &  0 & 0 & 0 & 0\\
  -1 & 2 & -1 & 0 & 0 & 0 & 0\\
  0 & -1 & 2 & -1 & 0 & 0 & 0\\
  0 & 0 & -1 & 2 & -1 & 0 & -1\\
  0 & 0 & 0 & -1 & 2 & -1 & 0\\
  0 & 0 & 0 & 0 & -1 & 2 & 0 \\
  0 & 0 & 0 & -1 & 0 & 0 & 2 \\
  \end{pmatrix} &
  \EE_7^{-1} &= \txtfrac{1}{2} \begin{pmatrix}
  3 & 4 & 5 & 6 & 4 & 2 & 3\\
  4 & 8 & 10 & 12 & 8 & 4 & 6\\
  5 & 10 & 15 & 18 & 12 & 6 & 9\\
  6 & 12 & 18 & 24 & 16 & 8 & 12 \\
  4 & 8 & 12 & 16 & 12 & 6 & 8\\
  2 & 4 & 6 & 8 & 6 & 4 & 4\\
  3 & 6 & 9 & 12 & 8 & 4 & 7\\
  \end{pmatrix}
\end{align}

\noindent{$\boldsymbol{E_8:}$}

\begin{align}
  \EE_8 &= \begin{pmatrix}
  2 & -1 & 0 & 0 & 0 & 0 & 0 & 0\\
  -1 & 2 & -1 & 0 & 0 & 0 & 0 & 0\\
  0 & -1 & 2 & -1 & 0 & 0 & 0 & 0\\
  0 & 0 & -1 & 2 & -1 & 0 & 0 & 0\\
  0 & 0 & 0 & -1 & 2 & -1 & 0 & -1\\
  0 & 0 & 0 & 0 & -1 & 2 & -1 & 0\\
  0 & 0 & 0 & 0 & 0 & -1 & 2 & 0 \\
  0 & 0 & 0 & 0& -1 & 0 & 0 & 2 \\
  \end{pmatrix} &
  \EE_8^{-1}&= \begin{pmatrix}
  2 & 3 & 4 & 5 & 6 & 4 & 2 & 3\\
  3 & 6 & 8 & 10 & 12 & 8 & 4 & 6\\
  4 & 8 & 12 & 15 & 18 & 12 & 6 & 9\\
  5 & 10 & 15 & 20 & 24 & 16 & 8 & 12\\
  6 & 12 & 18 & 24 & 30 & 20 & 10 & 15\\
  4 & 8 & 12 & 16 & 20 & 14 & 7 & 10\\
  2 & 4 & 6 & 8 & 10 & 7 & 4 & 5\\
  3 & 6 & 9 & 12 & 15 & 10 & 5 & 8\\
  \end{pmatrix}
\end{align}

\noindent{$\boldsymbol{F_4:}$}

\begin{align}
  \FF_4 &= \begin{pmatrix}
  \hp2 & -1 & \hp0 & \hp0\\
  -1 & \hp2 & -2 & \hp0\\
  \hp0 & -1 & \hp2 & -1\\
  \hp0 & \hp0 & -1 & \hp2\\
  \end{pmatrix} &
  \FF_4^{-1} = \begin{pmatrix}
  2 & 3 & 4 & 2\\
  3 & 6 & 8 & 4\\
  2 & 4 & 6 & 3\\
  1 & 2 & 3 & 2\\
  \end{pmatrix}
\end{align}

\noindent{$\boldsymbol{G_2:}$}

\begin{align}
\GG_2 &= \begin{pmatrix}
  \hp2 & -3 \\
  -1 & \hp2\\
  \end{pmatrix} &
  \GG_2^{-1} &= \begin{pmatrix}
  2 & 3\\
  1 & 2\\
  \end{pmatrix}
\end{align}

\setlength{\unitlength}{1 mm}
\begin{table}[h]
\begin{tabular}[h]{|c|l|c|c|}
\hline
$X_n$ & Dynkin diagram & $\text{dim}\ X_n$ & $\dCo$ \\
\hline
$A_n$ &
\begin{picture}(40,12)(0,6)
\multiput(2,10)(10,0){4}{\circle{2}}
\put(4,10){\line(1,0){6}}
\put(24,10){\line(1,0){6}}
\dottedhline(14,10){1}{7}{.2}
\put(1,5){$1$}
\put(11,5){$2$}
\put(18,5){$n-1$}
\put(31,5){$n$}
\end{picture} & $n(n+2)$ & $n+1$ \\
$B_n$ &
\begin{picture}(45,12)(0,6)
\multiput(2,10)(10,0){4}{\circle{2}}
\put(42,10){\circle*{2}}
\put(4,10){\line(1,0){6}}
\dottedhline(14,10){1}{7}{.2}
\put(24,10){\line(1,0){6}}
\multiput(34,9.5)(0,1){2}{\line(1,0){6}}
\put(1,5){$1$}
\put(11,5){$2$}
\put(17,5){$n-2$}
\put(28,5){$n-1$}
\put(41,5){$n$}
\end{picture} & $n(2n+1)$ & $2n-1$ \\
$C_n$ &
\begin{picture}(45,12)(0,6)
\multiput(2,10)(10,0){4}{\circle*{2}}
\put(42,10){\circle{2}}
\put(4,10){\line(1,0){6}}
\dottedhline(14,10){1}{7}{.2}
\put(24,10){\line(1,0){6}}
\multiput(34,9.5)(0,1){2}{\line(1,0){6}}
\put(1,5){$1$}
\put(11,5){$2$}
\put(17,5){$n-2$}
\put(28,5){$n-1$}
\put(41,5){$n$}
\end{picture} & $n(2n+1)$ & $n+1$ \\
$D_n$ &
\begin{picture}(40,17)(0,6)
\multiput(2,10)(10,0){5}{\circle{2}}
\put(32,20){\circle{2}}
\put(4,10){\line(1,0){6}}
\put(24,10){\line(1,0){6}}
\put(34,10){\line(1,0){6}}
\put(32,12){\line(0,1){6}}
\dottedhline(14,10){1}{7}{.2}
\put(1,5){$1$}
\put(11,5){$2$}
\put(17,5){$n-3$}
\put(28,5){$n-2$}
\put(33,16){$n-1$}
\put(41,5){$n$}
\end{picture} & $n(2n-1)$ & $2n-2$ \\
$E_6$ &
\begin{picture}(40,17)(0,6)
\multiput(2,10)(10,0){5}{\circle{2}}
\put(22,20){\circle{2}}
\multiput(4,10)(10,0){4}{\line(1,0){6}}
\put(22,12){\line(0,1){6}}
\put(1,5){$1$}
\put(11,5){$2$}
\put(21,5){$3$}
\put(31,5){$4$}
\put(23,16){$6$}
\put(41,5){$5$}
\end{picture}
& $78$ & $12$ \\
$E_7$ &
\begin{picture}(50,17)(0,6)
\multiput(2,10)(10,0){6}{\circle{2}}
\put(32,20){\circle{2}}
\multiput(4,10)(10,0){5}{\line(1,0){6}}
\put(32,12){\line(0,1){6}}
\put(1,5){$1$}
\put(11,5){$2$}
\put(21,5){$3$}
\put(31,5){$4$}
\put(41,5){$5$}
\put(51,5){$6$}
\put(33,16){$7$}
\end{picture} & $133$ & $18$ \\
$E_8$ &
\begin{picture}(63,17)(0,6)
\multiput(2,10)(10,0){7}{\circle{2}}
\put(42,20){\circle{2}}
\multiput(4,10)(10,0){6}{\line(1,0){6}}
\put(42,12){\line(0,1){6}}
\put(1,5){$1$}
\put(11,5){$2$}
\put(21,5){$3$}
\put(31,5){$4$}
\put(41,5){$5$}
\put(51,5){$6$}
\put(61,5){$7$}
\put(43,16){$8$}
\end{picture} & $248$ & $30$ \\
$F_4$ &
\begin{picture}(40,12)(0,6)
\multiput(2,10)(10,0){2}{\circle{2}}
\multiput(22,10)(10,0){2}{\circle*{2}}
\put(4,10){\line(1,0){6}}
\put(24,10){\line(1,0){6}}
\multiput(14,9.5)(0,1){2}{\line(1,0){6}}
\put(1,5){$1$}
\put(11,5){$2$}
\put(21,5){$3$}
\put(31,5){$4$}
\end{picture} & $52$ & $9$ \\
$G_2$ &
\begin{picture}(20,12)(0,6)
\put(2,10){\circle{2}}
\put(12,10){\circle*{2}}
\multiput(4,9)(0,1){3}{\line(1,0){6}}
\put(1,5){$1$}
\put(11,5){$2$}
\end{picture} & $14$ & $4$ \\
\hline
\end{tabular} \bigskip
\caption{Some properties of the finite dimensional simple Lie algebras}
\label{lieprop}
\end{table}

In Table \ref{lieprop} we list some of the properties of the simple
Lie algebras. The black nodes in the Dynkin diagrams correspond to
the short roots.

In addition to the Cartan matrices given above, we will frequently use
the symmetrized Cartan matrix of $B_k$, which we denote by $\MM_k^{-1}$.
Explicitly, we have
\begin{align}
\MM_k &= \begin{pmatrix}
1 & 1 & 1 & \cdots & 1 \\
1 & 2 & 2 & & 2 \\
1 & 2 & 3 & & 3 \\
\vdots & & & \ddots & \vdots \\
1 & 2 & 3 & \cdots & k \\
\end{pmatrix} \ , &
\MM_k^{-1} &= \begin{pmatrix}
2 & -1 \\
-1 & 2 & -1 \\
& -1 & 2 \\
& & & \ddots & -1\\
& & & -1 & 2 & -1\\
& & & & -1 & 1\\
\end{pmatrix} \,.
\end{align}

The simple Lie algebras are labeled by $X_n$, where $n$ is the rank,
and $X$ can be $A,B,\ldots,G$. As we will only be dealing with
the untwisted affine Lie algebras, we will use the notation
$X_{n,k}$, rather than $(X_n^{(1)})_k$, which
is more common in the literature.  Sometimes, we will use the notation
$\mathfrak{sl}(n)_k$, $\mathfrak{so}(2n-1)_k$, $\mathfrak{sp}(2n)_k$
and $\mathfrak{so}(2n)_k$ for the infinite series of untwisted affine
Lie algebras. Here, and in the rest of the paper, the level is denoted by $k$.

Blackboard bold, such as $\AA$ is used for matrices, while vectors
are in boldface, such as $\bQ$. If we want to specify a column of a
matrix, say $\AA$, we use the notation $(\AA)_c$, where the integer $c$
denotes the column we want to specify. In bilinear forms such as
$\bm^T\cdot\KK\cdot\bm$, we will frequently omit the transposition
symbol $^T$.

\section{Obtaining the $\mathfrak{so}(5)_1$ matrices}
\label{so5app}

\begin{figure}[ht]
\begin{center}
\includegraphics{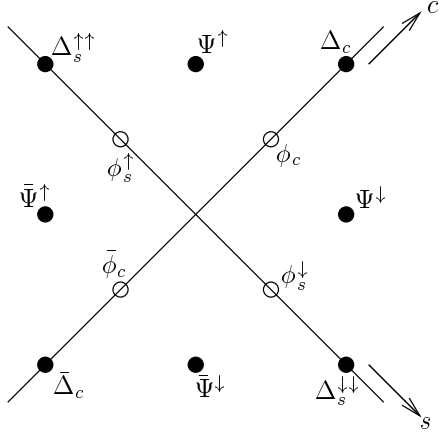}
\caption{The roots $(\bullet)$ and weights $(\circ)$ of $\mathfrak{so}(5)$.}
\label{so5diagram}
\end{center}
\end{figure}

The electron matrix for $\mathfrak{so}(5)_1$ can be obtained by
using knowledge about the root diagram and the associated parafermions
(see \cite{Gep} for general parafermion theories). We will anticipate
that it is in fact possible to use a quantum Hall type of basis for
this theory. So we define a set of electron operators, where the
vertex operator part is chosen in such a way that the spin and charge
are such that we actually have electron-like operators. The
matrix $\KK_{\text{e}}$
is obtained via the connection with the exclusion statistics, i.e.,
we calculate the associated exclusion statistics parameters of these
electron operators. From \cite{Gep} we obtain that at level $k=1$, the
short roots of $\mathfrak{so}(5)$ come with a parafermion operator, which
is in fact the Majorana fermion $\psi$, which has the same exclusion statistics
parameter as a fermion, namely $1$.
The root diagram of $\mathfrak{so}(5)$ is given in figure \ref{so5diagram}.
The electron operators we take to be part of the quantum Hall basis
correspond to $\bar{\Psi}^\downarrow$, $\Delta^{\uparrow\uparrow}_s$ and
$\Delta_c$. These operators take the form (at level $k=1$)
\begin{equation}
\bar{\Psi}^\downarrow  = \psi :e^{\frac{i}{\sqrt{2}}(\varphi_c+\varphi_s)}: \,,
  \qquad
  \Delta^{\uparrow\uparrow}_s  = \, :e^{i \sqrt{2}\varphi_s}: \,, \qquad
  \Delta_c  = \, :e^{i \sqrt{2}\varphi_c}:  \ ,
\end{equation}
where $\varphi_s$ and $\varphi_c$ are spin and charge bosons, respectively,
chosen according to the spin and charge direction indicated in Figure
\ref{so5diagram}.
{}From these operators, we infer the following exclusion statistics matrix
\begin{equation} \label{so5keapp}
  \KKe = \begin{pmatrix}
    \hp2 & -1 & -1 \\
    -1 & \hp2 & \hp0 \\
    -1 & \hp0 & \hp2 \\
  \end{pmatrix} \,,\qquad
  \bte = \begin{pmatrix}  \hp1\\ \hp0\\ -2 \end{pmatrix}\,,\quad
  \bs_{\rm e} = \begin{pmatrix} -1\\\hp2\\ \hp0 \end{pmatrix} \,.
\end{equation}
We should comment on a few things here. First of all, the matrix we found
is  equal to the Cartan matrix of
$\mathfrak{so}(6)$, which
relates to the so-called covering state of the state related to
$\mathfrak{so}(5)$.
This is analogous to the situation of the Moore-Read state, which is
related to a two-layer state. So we could have started from this
K-matrix, and performed a similar construction as was done in Section
\ref{mrcase} to find the K-matrices for the Moore-Read state. This would
lead to the same matrix \eqref{so5keapp}. In addition, in the
quasiparticle sector, there is a pseudoparticle, just as in the
Moore-Read case.
The matrix for the quasiparticle sector can simply be obtained by inverting
the matrix \eqref{so5keapp}. As said, it is important to notice that the
particle in the quasiparticle sector which has trivial quantum numbers,
is to be considered as a pseudoparticle. Otherwise, we would not obtain the
correct central charge, and hence, not the correct description. We find
\begin{equation} \label{so5kqpapp}
\KKqp = \begin{pmatrix}
  1 & \frac{1}{2} & \frac{1}{2} \\
  \frac{1}{2} & \frac{3}{4} & \frac{1}{4} \strt{\frac{1}{4}} \\
  \frac{1}{2} & \frac{1}{4} & \frac{3}{4} \\
\end{pmatrix} \,,\qquad
  \btqp = \begin{pmatrix} 0\\ 0\\ 1 \end{pmatrix}\,,\quad
  \bsqp = \begin{pmatrix} \hp0\\ -1\\ \hp0 \end{pmatrix}\,.
\end{equation}
To obtain the K-matrices for $\mathfrak{so}(5)$ at general level, we take
$k$ copies of the level-1 formulation, and do a similar construction as
described in Section \ref{compcons}. This gives the result of
Section \ref{kqp}.

\section{The case $G_{2,k}$}  \label{g2app}

In Section \ref{kqp} we found that the K-matrices for the affine
Lie algebra $G_{2,k}$ are special in the sense that the number of
physical quasiparticles is not equal to the rank of this algebra
(which is 2), if we use the standard construction of Section
\ref{construction}. Here, we will find another way of describing
this theory, which does have two physical quasiparticles.
We will start by deriving the K-matrices for level
$k=1$, in a similar way as we did for $\mathfrak{so}(5)_1$ in
Appendix \ref{so5app}. We continue by explaining how to obtain
the K-matrices for general level $k$. This is a little different from
Section~\ref{construction}, as the P-transformation which is needed is
different.

\begin{figure}[ht]
\begin{center}
\includegraphics{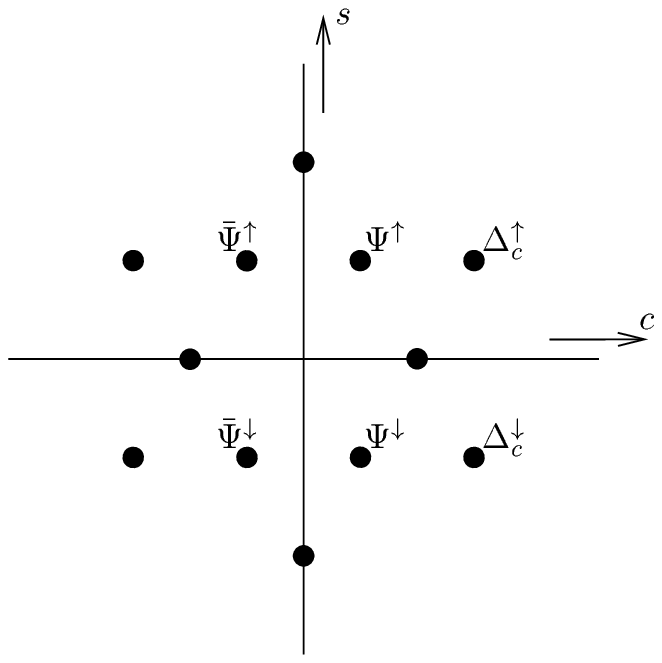}
\caption{The roots of $G_2$.}
\label{g2diagram}
\end{center}
\end{figure}

The root lattice for the Lie algebra $G_2$ is given in
Figure \ref{g2diagram}.
In fact, it is not possible to pick four electron-like operators, such that
the K-matrix is the Cartan matrix of the enveloping algebra
$\mathfrak{so}(8)$, but we will stay as close as possible.

The short roots come with two types of
parafermions, $\psi_1$ and $\psi_2$, which belong to the $\ZZ_3$
parafermion theory.
The operators needed to form the quantum Hall basis are
\begin{align}
\Psi^\uparrow & = \psi_1 : e^{i/\sqrt{6} \phi_c + i/\sqrt{2} \phi_s} : \ ,&
\bar{\Psi}^\downarrow & =
\psi_2 : e^{-i/\sqrt{6} \phi_c - i/\sqrt{2} \phi_s} :\ ,\\
\Delta_c^\uparrow & = \: : e^{i 3 /\sqrt{6} \phi_c + i/\sqrt{2} \phi_s} : \ ,&
\Delta_c^\downarrow & = \: : e^{i 3 /\sqrt{6} \phi_c - i/\sqrt{2} \phi_s} : \ ,
\end{align}
where $\phi_{c,s}$ are the charge and spin boson. As the K-matrix for the
$\ZZ_3$ parafermions is given by
\begin{equation}
\label{z3pf}
\KK_{\ZZ_3}^{\rm pf} = \begin{pmatrix}
\strt{\frac43} \frac43 & \frac23 \\ \frac23 & \frac43 \end{pmatrix} \ ,
\end{equation}
and the statistics parameters due to the vertex operators of the spin and
charge bosons are easily calculated, we find the following data for the
`electron' sector of the $G_{2,k=1}$ theory
\begin{equation} \label{g2k1e}
  \KKe = \begin{pmatrix}
   2 & \hp0 & \hp1 & 0 \\
   0 & \hp2 & -1 & 0 \\
   1 & -1 & \hp2 & 1 \\
   0 & \hp0 & \hp1 & 2 \\
  \end{pmatrix} \,, \qquad
  \bt_e = - \begin{pmatrix} \hp1 \\ -1 \\ \hp3 \\ \hp3 \end{pmatrix} \,,\quad
  \bs_e = \begin{pmatrix} \hp1 \\ -1 \\ \hp1 \\  -1 \end{pmatrix} \,.
\end{equation}
By the duality construction, we find the dual data
\begin{equation} \label{g2k1phi}
  \KKqp = \begin{pmatrix}
  \hp1 & -\frac12 & \vdots & -1 & \hp\frac12 \\
  -\frac12 & \hp1 & \vdots & \hp1 & -\frac12 \\
  \hdotsfor{5} \\
  -1 & \hp1 & \vdots & \hp2 & -1 \\
  \hp\frac12 & -\frac12 & \vdots & -1 & \hp1 \\
  \end{pmatrix} \,,\qquad
\bt_\qp = \begin{pmatrix} 0 \\ 0\\ 1\\ 1 \end{pmatrix} \,, \quad
\bs_\qp = \begin{pmatrix} \hp0 \\ \hp0\\ -1\\ \hp1 \end{pmatrix} \,,
\end{equation}
where the first two particles are pseudoparticles, which reduce the
central charge, and take care of the non-abelian statistics. Note
that we do not use the usual ordering of the Cartan matrix (compare
Appendix~\ref{cartan}), because in the quasiparticle sector, we
want the first to particles to be the pseudoparticles.

Picking the operators associated to the right
roots is crucial in finding a basis for the $G_2$ affine Lie
algebra. The way we have chosen them here gives a description which does
give the right central charge, and has two physical quasiparticles.

We would like to comment on the difference between the pseudoparticle
matrices for the two descriptions of $G_{2,1}$.
If we apply the composite construction on the
$2\times2$ pseudoparticle matrix of this appendix, we indeed find the
pseudoparticle matrix (at level-$1$) of Section~\ref{kqp}. This matrix
also appeared in Section~\ref{pseudofus}, Eqn.~\eqref{eqPBBq}. So the
pseudoparticles are equivalent in both cases.

We now proceed by constructing the matrices for level-$k$.
As usual, the covering is of the form $\KKe\otimes \II_k$. The
required P-transformation turns out to be of the form
(compare with Appendix~\ref{relbases})
\begin{equation}
\PP' = \begin{pmatrix}
\II_4 & \JJ_4^u & \cdots & \JJ_4^u \\
\JJ_4^l & \II_4 & \ddots & \vdots \\
\vdots & \ddots & \ddots & \JJ_4^u \\
\JJ_4^l & \cdots & \JJ_4^l & \II_4 \\
\end{pmatrix} \ ,
\end{equation}
where $\JJ^u_4$ and $\JJ^l_4$ are given by
\begin{align} \label{jforg2}
\JJ_4^u &= \begin{pmatrix}
1 \\ & 0 \\ & & 0 \\ & & & 1
\end{pmatrix} \ ,&
\JJ_4^l &= \begin{pmatrix}
0 \\ & 1 \\ & & 1 \\ & & & 0
\end{pmatrix} \ .
\end{align}
Because $\JJ^u_4+\JJ^l_4=\II_4$, all composites up to order $k$
are formed. To display the resulting matrix, it is most convenient to
reorder the particles in the order of increasing quantum numbers
(this is not done automatically, because of the form of the P-transformation).
To conveniently display the `permuted' K-matrix for the electron sector,
we define a  modified Cartan matrix of $\DD_4$
\begin{equation}
\cM(a,b,c) =
\begin{pmatrix}
a & 0 & b & 0\\
0 & a & c & 0\\
b & c & a & b\\
0 & 0 & b & a\\
\end{pmatrix} \ .
\end{equation}
Then, the electron K-matrix for $G_{2,k}$ can be described by
\begin{equation}
\KK^{G_{2,k}}_{\rm e} =
\begin{pmatrix}
\cM(2,0,-1) & \cM(2,0,-1) & \cdots & & \cdots & \cM(2,1,-1) \\
\cM(2,0,-1) & \cM(4,0,-2) & & & & \cM(4,2,-2) \\
\vdots &  & \ddots & & & \vdots \\
& & & \makebox[3cm]{$\cM(2\min(i,j), \max(i+j-k,0), -\min(i,j) )$}  \\
\vdots & & & & \ddots & \vdots \\
\cM(2,1,-1) & \cM(4,2,-2) & \cdots & & \cdots & \cM(2k,k,-k) \\
\end{pmatrix} \ .
\end{equation}
To make this a little more clear, we give the result for $k=2$ explicitly
\begin{equation}
\KK_e = \begin{pmatrix}
2 & \hp0 & \hp0 & 0 & \vdots & 2 & \hp0 & \hp1 & 0 \\
0 & \hp2 & -1 & 0 & \vdots & 0 & \hp2 & -1 & 0 \\
0 & -1 & \hp2 & 0 & \vdots & 1 & -1 & \hp2 & 1 \\
0 & \hp0 & \hp0 & 2 & \vdots & 0 & \hp0 & \hp1 & 2 \\
\hdotsfor{9} \\
2 & \hp0 & \hp1 & 0 & \vdots & 4 & \hp0 & \hp2 & 0 \\
0 & \hp2 & -1 & 0 & \vdots & 0 & \hp4 & -2 & 0 \\
1 & -1 & \hp2 & 1 & \vdots & 2 & -2 & \hp4 & 2 \\
0 & \hp0 & \hp1 & 2 & \vdots & 0 & \hp0 & \hp2 & 4 \\

\end{pmatrix} \ .
\end{equation}
The quasiparticle sector for $k>2$ is characterized by the following matrices
(compare with Section \ref{kmatrices})

\begin{align}
\label{pseudog2k}
\KK_{\psi\psi} & = \begin{pmatrix}
2 \DD_4^{-1} & -\DD_4^{-1} & & &  -(\DD_4^{-1})_1 \\
-\DD_4^{-1} & 2\DD_4^{-1} & \ddots \\
& \ddots & \ddots & -\DD_4^{-1}  \\
& & -\DD_4^{-1} & 2 \DD_4^{-1} & &  -(\DD_4^{-1})_3 \\
-(\DD_4^{-1})^T_1 & & & & 1 & 0\\
& & &  -(\DD_4^{-1})^T_3 & 0 & 1\\
\end{pmatrix} \ , \\
\KK_{\phi\psi} & =
\settowidth{\kolom}{$-(\DD_4^{-1})_2^T$}
\settowidth{\kolomm}{$-\DD_4^{-1}$}
\begin{pmatrix}
& \makebox[\kolomm]{} & \makebox[\kolomm]{} & -(\DD_4^{-1})_2^T
& \makebox[\kolom]{$0$} & \makebox[\kolom]{$1$} \\
 -(\DD_4^{-1})_4^T & & & & \frac12 & 0\\
\end{pmatrix} \ , \\
\KK_{\phi\phi} & = \begin{pmatrix} 2 & 0 \\ 0 & 1 \\ \end{pmatrix} \,, \\
\bt_\qp & = (0,\ldots,0;1,1) \,, \\
\bs_\qp & = (0,\ldots,0;-1,1) \,.
\end{align}
So, although the form of the K-matrix differs from the general
description, we still find that all the elements are related to the
(inverse) Cartan matrix of the Lie algebra $D_4$.

Now that we have a description of $G_2$ which
does have two quasiparticles (for every $k$), we can use the same
conjecture \eqref{pfcon} to find the K-matrices for the parafermions, namely
the parafermion theory $G_{2,k}/[\mathfrak{u}(1)]^2$. So, without giving the
explicit form, it is found that the parafermion K-matrix
$\KK_{G_2}^{\rm pf} = \KK_{\psi\psi}^{-1}$ does have the right properties.
It gives the correct central charge, and reproduces the string functions
as described in Section \ref{pfsec}.

For the case $k=1$, we indeed find that the parafermions associated
to $G_2$ are the $\ZZ_3$ parafermions. At level $k\geq 2$ we find
the K-matrices of the $G_2$ parafermions,
which for $k=2$ is given by
\begin{equation}
\label{g2pfk2}
\KK_{G_2,k=2}^{\rm pf} = \begin{pmatrix}
\strt{\frac53}\hp\frac53 & \hp\frac13 & -\frac12 & \hp0 & \frac43 & \frac32 \\
\hp\frac13 & \hp\frac53 & -\frac12 & \hp0 & \frac23 & \frac42 \\
\strt{\frac12}-\frac12 & -\frac12 & \hp1 & -\frac12 & 0 & 0 \\
\hp0 & \hp0 & -\frac12 & \hp1 & 0 & 0 \\
\strt{\frac43}\hp\frac43 & \hp\frac23 & \hp0 & \hp0 & \frac83 & \frac43 \\
\hp\frac23 & \hp\frac43 & \hp0 & \hp0 & \frac43 & \frac83 \\
\end{pmatrix} \ .
\end{equation}
Note the `asymmetry' between the parafermions 3 and 4.

\section{Relating different bases}
\label{relbases}

In Section~\ref{construction} we pointed out that the K-matrices for
$\mathfrak{sl}(3)_k$
found in \cite{ABS} differ from the ones we presented here. The reason for
this was also given. In \cite{ABS}, all the particles in the electron sector
were chosen such that their charge all had the same sign. Consequently, the
K-matrix for level-1 was based on the roots $\alpha_1$ and $-\alpha_2$.
This resulted in the following K-matrix and quantum number vectors
\begin{align}  \label{ksl3}
  \KKe^{' (k=1)} &= \begin{pmatrix} 2 & 1\\ 1 & 2\\\end{pmatrix} \,, &
  \bte' &= -\begin{pmatrix} 1 \\ 1 \end{pmatrix} \,, &
  \bse' &= \begin{pmatrix} \hp 1 \\ -1 \end{pmatrix} \,.
\end{align}
In this appendix, we will explain in detail the relation between this approach
and the one followed in this paper. The matrix \eqref{ksl3} can also be used
to obtain K-matrices for $\mathfrak{sl}(3)_k$.
This formulation is different, but
can be related to the one obtained in Section~\ref{construction}. We will first
show that we can construct the $\mathfrak{sl}(3)_k$ K-matrices found in
\cite{ABS} using the P-transformations. We then explicitly relate the two
constructions.

So, let us begin with the covering matrix based on
Eqn.~\eqref{ksl3}, which is
constructed in the usual way, by taking a direct sum of $k$ copies:
$\KKe^{'\rm cover} = \KKe^{'(k=1)} \otimes \II_k$.
Now the P-transformation is different
than the one used in Section~\ref{construction}. It will be such that all
composites up to order $k$ are formed
(for both spin up and spin down particles).
However, $\PP$ is not lower triangular, but instead we have
\begin{equation}
  \PP' = \begin{pmatrix}
  \II_2 & \JJ_2^u & \cdots & \JJ_2^u \\
  \JJ_2^l & \II_2 & \ddots & \vdots \\
  \vdots & \ddots & \ddots & \JJ_2^u \\
  \JJ_2^l & \cdots & \JJ_2^l & \II_2 \\
  \end{pmatrix}\,.
\end{equation}
Here,
$\JJ_2^u = \bigl( \begin{smallmatrix} 1 & 0 \\ 0 & 0\end{smallmatrix} \bigr)$
and
$\JJ_2^l = \bigl( \begin{smallmatrix} 0 & 0 \\ 0 & 1\end{smallmatrix} \bigr)$.
The transformed K-matrix $\PP' \cdot \KKe^{'{\rm cover}} \cdot \PP^{'T}$
is most easily described after a suitable permutation of the particles,
which
orders the particles according to their quantum numbers;
as indicated before, all composites (up to order $k$) are formed, because
$\JJ^u_2+\JJ^l_2=\II_2$. The quantum numbers after applying the
P-transformation to the covering and the permutation to order them, are
given by $\bte' =-(1,1,2,2,\ldots,k,k)$ and $\bse'=(1,-1,2,-2,\ldots,k,-k)$.
The K-matrix becomes
\begin{equation}\label{asabs}
  \KKe' =
  \begin{pmatrix}
  2 & 0 & 2 & 0 & \cdots &  & 2 & 0 & 2 & 1 \\
  0 & 2 & 0 & 2 & \cdots &  & 0 & 2 & 1 & 2 \\
  2 & 0 & 4 & 0 &        &        & 4 & 1 & 4 & 2 \\
  0 & 2 & 0 & 4 &        &        & 1 & 4 & 2 & 4 \\
  \vdots&\vdots& &&   \ddots     &        &   &   & \vdots & \vdots \\
  2 & 0 & 4 & 1 &        &        & {2(k-1)}&{k-2}&{2(k-1)} & {k-1} \\
  0 & 2 & 1 & 4 &        &        & {k-2}& {2(k-1)}&{k-1} & {2(k-1)} \\
  2 & 1 & 4 & 2 & \cdots &  & {2(k-1)} & {k-1} & 2k & k \\
  1 & 2 & 2 & 4 & \cdots &  &{k-1} & {2(k-1)} & k & 2k
  \end{pmatrix} \,.
\end{equation}
This matrix is to be compared with $\KKe$ of Eqn.~\eqref{kesl3}.
The diagonal part of the
$2\times2$ blocks is the same, namely $2\min(i,j)$, where $i,j$ label the
blocks. The off-diagonal parts are given by $\max(k-i-j,0)$. The inverse is
found to be (again, after a suitable permutation of the particles)
\begin{equation}
\KK'_{\rm qp} =
\begin{pmatrix}
\AA^{-1}_{2} \otimes \AA_{k-1} &
\begin{matrix}
\vdots \\ \vdots \\ \vdots \\ \vdots \\ \vdots \\
\end{matrix} &
\begin{matrix}
\vphantom{\vdots} -(\AA_2^{-1})_1 & 0 \\
\vphantom{\vdots} 0 & 0  \\
\vphantom{\vdots} \vdots & \vdots \\
\vphantom{\vdots} 0 & 0 \\
\vphantom{\vdots} 0 & -(\AA_2^{-1})_2 \\
\end{matrix} \\
\hdotsfor{3} \\
\begin{matrix}
\vphantom{\vdots} -(\AA_2^{-1})_1^T & 0 & \cdots & 0 & 0 & \\
\vphantom{\vdots} 0 & 0 & \cdots & 0 & -(\AA_2^{-1})_2^T & \\
\end{matrix} &
\begin{matrix}
\vdots \\ \vdots \\
\end{matrix} &
\settowidth{\kolom}{$-(\AA_2^{-1})_1$}
\begin{matrix} \makebox[\kolom]{$\frac23$} & \makebox[\kolom]{$0$} \\
0 & \frac23\\\end{matrix}
\end{pmatrix} \,,
\end{equation}
which is to be compared with $\KKqp$ of Eqn.~\eqref{kqpsl3}.
To relate the two descriptions, we make use of the fact that we know
how to relate the matrices for $k=1$. The difference
is the use of $\alpha_2$ in the description detailed in
Section~\ref{construction}
and $-\alpha_2$ in the description of this appendix and \cite{ABS}.
Recall that the K-matrix for level $k=1$ from Section~\ref{construction}
is given by
$\KKe^{(k=1)}=\bigl(\begin{smallmatrix}
\hp 2 & -1 \\ -1 & \hp 2 \end{smallmatrix}\bigr)$.
So we find that we can relate the two K-matrices for level-1 by a
W-transformation, which is  given by
$\KKe^{'(k=1)} = \WW \cdot \KKe^{(k=1)} \cdot \WW^T$, where
$\WW = \bigl( \begin{smallmatrix} 1 & \hp0 \\ 0 & -1 \end{smallmatrix} \bigr)$.
Because we also know how to transform the coverings into the corresponding
K-matrices for $\mathfrak{sl}(3)_k$, we can relate the two descriptions
in terms of a W-transformation. Apart from the extra permutations which
are involved, the calculation is straighforward, and we find the relation
$\KKe'=\WWe \cdot \KKe \cdot \WWe^T$,
with (dropping the subscript $2$)
\begin{equation}
  \WWe = \begin{pmatrix}
  -\JJ^l & & & & -\JJ^u & \JJ^u\\
  & \ddots & & \dddots & & \JJ^u\\
  & & \ddots \hspace{-16pt} \dddots & & & \vdots\\
  & \dddots & & \ddots & & \JJ^u\\
  -\JJ^u & & & & -\JJ^l & \JJ^u\\
  & & & & & \JJ^u-\JJ^l\\
  \end{pmatrix} \,,
\end{equation}
where $\ddots \hspace{-16pt} \dddots$ stands for
$\bigl(\begin{smallmatrix}-\JJ^l&-\JJ^u\\-\JJ^u&-\JJ^l\end{smallmatrix}\bigr)$
if $k$ is odd and for
$\Bigl(\begin{smallmatrix}-\JJ^l&&-\JJ^u\\&-\II&\\
-\JJ^u&&-\JJ^l\end{smallmatrix}\Bigr)$ if $k$ is even.
Note that $\WWe^{-1}=\WWe$. For the quasiparticle sector we have a similar
relation, $\KKqp'=\WWqp \cdot \KKqp \cdot \WWqp^T$. But because we needed the
extra permutations, we do not have the relation $\WWqp=(\WWe^{-1})^T$.
This only holds for the case at hand if we undo this permutation. Instead,
we have
\begin{equation}
  \WWqp =
  \begin{pmatrix}
  -\II & & & &  \\
  & -\II & & &  \\
  & & \ddots & & \\
  & & & -\II & \\
  \JJ^u & \JJ^u & \cdots & \JJ^u & \JJ^u -\JJ^l \\
  \end{pmatrix}  \,.
\end{equation}
Note that in going from the one formulation to the other, we are only
transforming the physical quasiparticles, the pseudoparticles are not
changed. This should be the case, as the pseudoparticles govern the fusion
rules and the central charge.

Let us end this discussion by mentioning that the formulation for
$\mathfrak{sl}(3)_k$ of the type of Eqn.~\eqref{asabs} can be generalized
to arbitrary affine Lie algebra CFTs. The relations between the description
in this paper is precisely analogous to the relation for $\mathfrak{sl}(3)$
as described in this appendix. The only difference would be in the form
of the matrices $\JJ^u$ and $\JJ^l$. However, they still would only have
non-zero elements on the diagonal, subject to the constraint
$\JJ^u+\JJ^l=\II$.

\end{appendix}


\begin{thebibliography}{XXXX}

\bibitem{Andr}
G.E.~Andrews,
{\it Partitions: Yesterday and today},
(New Zealand Mathematical Society, Wellington, 1979).

\bibitem{ABGS}
E.~Ardonne, P.~Bouwknegt, S.~Guruswamy and K.~Schoutens,
{\it $K$-matrices for non-abelian quantum Hall states},
Phys.\ Rev.\ {\bf B61} (2000) 10298-10302, [{\tt arXiv:cond-mat/9908285}].

\bibitem{ABS}
E.~Ardonne, P.~Bouwknegt and K.~Schoutens, {\it Non-abelian
quantum Hall states -- Exclusion statistics, $K$-matrices and duality},
J.\ Stat.\ Phys.\ {\bf 102} (2001) 421-469, [{\tt arXiv:cond-mat/0004084}].

\bibitem{ALLS}
E.~Ardonne, F.J.M.~van~Lankvelt, A.W.W.~Ludwig and K~Schoutens,
{\it Separation of spin and charge in paired spin-singlet quantum Hall
states},
Phys.\ Rev.\ {\bf B65} (2002) 041305(R), [{\tt arXiv:cond-mat/0102072}].

\bibitem{AS}
E.~Ardonne and K.~Schoutens,
{\it New class of non-abelian spin-singlet quantum Hall states},
Phys.\ Rev.\ Lett.\ {\bf 82} (1999) 5096-5099,
[{\tt arXiv:cond-mat/9811352}].

\bibitem{BM}
A.~Berkovich and B.~McCoy,
{\it The universal chiral partition function for exclusion statistics},
in ``Statistical Physics on the Eve of the 21st Century'',
Series on Adv.\ in Stat.\ Mech., Vol.\ {\bf 14}, pp 240-256,
eds.\ M.T.~Batchelor and L.T.~Wille,
(World Scientific, Singapore, 1999),
[{\tt arXiv:hep-th/9808013}].

\bibitem{BMS}
A.~Berkovich, B.~McCoy and A.~Schilling,
{\it Rogers-Schur-Ramanujan type identities for the $M(p,p')$ minimal models
of conformal field theory}, Commun.\ Math.\ Phys.\ {\bf 191} (1998) 325-395,
[{\tt arXiv:q-alg/9607020}].

\bibitem{Bo2}
P.~Bouwknegt,
{\it Multipartitions, generalized Durfee squares and affine Lie
algebra characters}, J. Aust. Math. Soc. {\bf 72} (2002) 395-408,
[{\tt arXiv:math.CO/0002223}].

\bibitem{BCR}
P.~Bouwknegt, L.-H.~Chim and D.~Ridout,
{\it Exclusion statistics in conformal field theory and the UCPF
for WZW models},
Nucl.\ Phys.\ {\bf B572} (2000) 547-573, [{\tt arXiv:hep-th/9903176}].

\bibitem{BH}
P.~Bouwknegt and N.~Halmagyi,
{\it $q$-identities and affinized projective varieties,
II.\ Flag varieties}, Commun.\ Math.\ Phys.\ {\bf 210} (2000) 663-684,
[{\tt arXiv:math-ph/9903033}].

\bibitem{BSa}
P.~Bouwknegt and K.~Schoutens,
{\it Non-abelian electrons: SO(5) superspin regimes for correlated electrons
on a two-leg ladder}, Phys. Rev. Lett. {\bf 82} (1999) 2757-2760,
[{\tt arXiv:cond-mat/9805232}].

\bibitem{BSb}
P.~Bouwknegt and K.~Schoutens,
{\it Exclusion statistics in conformal field theory -- generalized
fermions and spinons for level-1 WZW theories},
Nucl.\ Phys.\ {\bf B547} (1999) 501-537,
[{\tt arXiv:hep-th/9810113}].

\bibitem{CGT}
A.~Cappelli, L.~Georgiev and I.~Todorov, {\it Parafermion Hall states
from coset projections of abelian conformal theories},
Nucl.\ Phys.\ {\bf B599} (2001) 499-530,
[{\tt arXiv:hep-th/0009229}].

\bibitem{CRS}
J.M.~Camino, A.V.~Ramallo and J.M.~S\'anchez de Santos,
{\it Graded parafermions}, \newline
[{\tt arXiv:hep-th/9805160}].

\bibitem{DKKMM}
S.~Dasmahapatra, R.~Kedem, T.~Klassen, B.~McCoy and E.~Melzer,
{\it Quasi-particles, conformal field theory, and q-series}, Int.\ J.\ Mod.\
Phys.\ {\bf B7} (1993) 3617-3648,
[{\tt arXiv:hep-th/9303013}].

\bibitem{DGZa}
X.-M.~Ding, M.D.~Gould and Y.-Z.~Zhang,
{\it Twisted $sl(3,{\bf C})_k^{(2)}$ current algebra: free field
representation and screening currents},
Phys.\ Lett.\ {\bf B523} (2001) 367-376,
[{\tt arXiv:hep-th/0109009}].

\bibitem{DGZb}
X.-M.~Ding, M.D.~Gould and Y.-Z.~Zhang,
{\it Twisted parafermions},
Phys.\ Lett.\ {\bf 530} (2001) 197-201,
[{\tt arXiv:hep-th/0110165}].

\bibitem{ES}
R.~van Elburg and K.~Schoutens,
{\it Quasi-particles in fractional quantum Hall effect edge theories},
Phys.\ Rev.\ {\bf B58} (1998) 15704-15716,
[{\tt arXiv:cond-mat/9801272}].

\bibitem{FJKLM}
B.~Feigin, M.~Jimbo, R.~Kedem, S.~Loktev and T.~Miwa,
{\it Spaces of coinvariants and fusion products},
{\it I.\ From equivalence theorem to Kostka polynomials},
[{\tt arXiv:math.QA/0205324}]; {\it II.\ Affine $\mathfrak{sl}_2$
character formulas in terms of Kostka polynomials},
[{\tt arXiv:math.QA/0208156}].

\bibitem{FKLMM}
B.~Feigin, R.~Kedem, S.~Loktev, T.~Miwa and E.~Mukhin,
{\it Combinatorics of the $\widehat{\mathfrak{sl}}_2$ spaces of
coinvariants I, II, III},
[{\tt arXiv:math-ph/9908003}, {\tt arXiv:math.QA/0009198},
{\tt arXiv:math.QA/0012190}].

\bibitem{FGT}
S.~Fishel, I.~Grojnowksi and C.~Teleman, {\it The strong Macdonald
conjecture}, \newline
[{\tt arXiv:math.RT/0107072}].

\bibitem{FNS}
E.~Fradkin, C.~Nayak and K.~Schoutens,
{\it Landau-Ginzburg theories for non-abelian quantum Hall states},
Nucl.\ Phys.\ {\bf B546} (1999) 711-730,
[{\tt arXiv:cond-mat/9811005}].

\bibitem{dFMS}
P.~Di Francesco, P.~Mathieu and D. S\'en\'echal,
{\it Conformal Field Theory},
(Springer, New York, 1997).

\bibitem{GR}
G.~Gasper and M. Rahman,
{\it Basic hypergeometric series},
(Cambridge University Press, Cambridge, 1990).

\bibitem{Gep}
D. Gepner, {\it New conformal field theories associated with Lie
algebras and their partition functions},
Nucl.\ Phys.\ {\bf B290} (1987) 10-24.

\bibitem{GS}
S.~Guruswamy and K.~Schoutens,
{\it Non-abelian exclusion statistics},
Nucl.\ Phys.\ {\bf B556} (1999) 530-544,
[{\tt arXiv:cond-mat/9903045}].

\bibitem{Hal}
F.D.M.~Haldane,
{\it ``Fractional statistics'' in arbitrary dimensions: A generalization
of the Pauli principle},
Phys.\ Rev.\ Lett.\ {\bf 67} (1991) 937-940.

\bibitem{HKKOTY}
G.~Hatayama, A.~Kirillov, A.~Kuniba, M.~Okado, T.~Takagi and Y.~Yamada,
{\it Character formulae of $\widehat{\mathfrak{sl}_n}$-modules and
inhomogeneous paths}, Nucl.\ Phys.\ {\bf B536} (1999) 575-616,
[{\tt arXiv:math.QA/9802085}].

\bibitem{HKOTY}
G.~Hatayama, A.~Kuniba, M.~Okado, T.~Takagi and Y. Yamada,
{\it Remarks on fermionic formula},
Contemp.\ Math.\ {\bf 248} (1999) 243-291,
[{\tt arXiv:math.QA/9812022}].

\bibitem{Ho}
T.-L. Ho,
{\it The broken symmetry of two-component $\nu=1/2$ quantum Hall
states},
Phys.\ Rev.\ Lett.\ {\bf 75} (1995) 1186-1189,
[{\tt arXiv:cond-mat/9503008}].

\bibitem{IOW}
S.B.~Isakov,
{\it Generalization of statistics for several species of identical
particles},
Mod.\ Phys.\ Lett.\ {\bf B8} (1994) 319-327; \\
A.~Dasni\`eres de Veigy and S.~Ouvry,
{\it Equation of state of an anyon gas in a strong magnetic field},
Phys.\ Rev.\ Lett.\ {\bf 72} (1994) 600-603,
[{\tt arXiv:hep-th/9306039}];\\
Y.-S.~Wu,
{\it Statistical distribution for generalized ideal gas of
fractional-statistics particles},
Phys.\ Rev.\ Lett.\ {\bf 73} (1994) 922-925.

\bibitem{JM}
P.~Jacob and P.~Mathieu,
{\it Graded parafermions: standard and quasi-particle bases},
Nucl.\ Phys.\ {\bf B630} (2002) 433-452,
[{\tt arXiv:hep-th/0201156}].

\bibitem{KMPS}
S.~Kass, R.V.~Moody, J.~Patera, R.~Slansky,
{\it Affine lie algebras, weight multiplicities and
branching rules}, Vol.~2, (University of California Press,
Berkeley, 1990).

\bibitem{Ki}
A.N.~Kirillov,
{\it Dilogarithm identities},
Prog. Theor. Phys. Suppl. {\bf 118} (1995) 61-142,\newline
[{\tt arXiv:hep-th/9408113}].

\bibitem{Kib}
A.N.~Kirillov,
{\it Ubiquity of Kostka polynomials},
[{\tt arXiv:math.QA/9912094}].

\bibitem{KR}
A.N.~Kirillov and N.Yu.~Reshetikhin,
{\it The Bethe Ansatz and the combinatorics of Young tableaux},
J.\ Sov.\ Math.\ {\bf 41} (1988) 925;
{\it ibid.},
{\it Representations of Yangians
and multiplicities of occurrence of the irreducible components of the
tensor product of representations of simple Lie algebras},
J.\ Sov.\ Math.\ {\bf 52} (1990) 3156-3164.

\bibitem{KS}
A.N.~Kirillov and M.~Shimozono,
{\it A generalization of the Kostka-Foulkes polynomials},
J.\ Algebr.\ Comb.\ {\bf 5} (2002) 27-69;
[{\tt arXiv:math.QA/9803062}].

\bibitem{MR}
G.~Moore and N.~Read,
{\it Nonabelions in the fractional quantum Hall effect},
Nucl. Phys. {\bf B360} (1991) 362-396;
{\it ibid.}, {\it Fractional quantum Hall effect and nonabelian
statistics}, Prog.\ Theor.\ Phys.\ Suppl.\ {\bf 107}
(1992) 157-166,
[{\tt arXiv:hep-th/9202001}].

\bibitem{NY}
A.~Nakayashiki and Y.~Yamada,
{\it On spinon character formulas}, Frontiers in Quantum Field Theories,
(H.~Itoyama et al., eds.), (World Scientific, Singapore, 1996), pp.\
367-371.

\bibitem{OSS}
M.~Okado, A.~Schilling and M.~Schimozono,
{\it Crystal bases and $q$-identities},
Contemp.\ Math.\ {\bf 291} (2001) 29-53,
[{\tt arXiv:math.QA/0104268}].

\bibitem{RR}
N.~Read and E.~Rezayi,
{\it Beyond paired quantum Hall states: parafermions and incompressible
states in the first excited Landau level},
Phys. Rev. {\bf B59} (1999) 8084-8092,
[{\tt arXiv:cond-mat/9809384}].

\bibitem{SS}
A.~Schilling and M.~Shimozono,
{\it Fermionic formulas for level-restricted 
generalized Kostka polynomials and coset branching functions},
Commun.\ Math.\ Phys.\ {\bf 220} (2001) 105-164,
[{\tt arXiv:math.QA/0001114}].

\bibitem{SW}
A.~Schilling and S.O.~Warnaar,
{\it Inhomogeneous lattice paths, generalized Kostka polynomials
and $A_{n-1}$ supernomials},
Commun.\ Math.\ Phys.\ {\bf 202} (1999) 359-401,
[{\tt arXiv:math.QA/9802111}].

\bibitem{Sca}
K.~Schoutens,
{\it Exclusion statistics in conformal field theory spectra},
Phys.\ Rev.\ Lett.\ {\bf 79} (1997) 2608-2611,
[{\tt arXiv:cond-mat/9706166}].

\bibitem{Scb}
K.~Schoutens,
{\it Exclusion statistics for non-abelian quantum Hall states},
Phys.\ Rev.\ Lett.\ {\bf 81} (1998) 1929-1932,
[{\tt arXiv:cond-mat/9803169}].

\bibitem{Ver}
E.~Verlinde, {\it Fusion rules and modular transformations
in conformal field theory}, Nucl.\ Phys.\ {\bf B300} (1988) 360-376.

\bibitem{Wen}
X.-G.~Wen, {\it Topological orders and edge excitations in fractional
quatnum Hall states}, Adv. Phys. {\bf 44} (1995) 405,
[{\tt arXiv:cond-mat/9506066}].

\bibitem{Yam}
Y.~Yamada,
{\it On $q$-Clebsch Gordan rules and the spinon character
formulas for affine $C_2^{(1)}$ algebra},
[{\tt arXiv:q-alg/9702019}].

\bibitem{ZF}
A.B.~Zamolodchikov and V.A.~Fateev,
{\it Nonlocal (parafermion) currents in two-dimensional conformal quantum
field theory and self-dual critical points in $Z_N$-symmetric statistical
systems},
Sov.\ Phys.\ JETP {\bf 62} (1985) 215-225.

\end{thebibliography}
\end{document}